\documentclass[10pt,journal,compsoc]{IEEEtran}

\usepackage{dblfloatfix}
\usepackage{blindtext}
\usepackage{algpseudocode}
\usepackage{amsmath}
\usepackage{amssymb}
\usepackage{amsthm}
\usepackage{color}
\usepackage{cite}
\usepackage{graphicx}
\graphicspath{{./}{Figs/}}
\usepackage{subcaption}
\usepackage{mathrsfs}
\usepackage{verbatim}
\usepackage{indentfirst}
\usepackage{setspace} 
\usepackage{url}
\usepackage{epsfig,endnotes,amsthm,changepage}
\usepackage{multirow}

\usepackage{epstopdf,lipsum,etoolbox}
\usepackage{graphicx}
\usepackage{amsfonts}
\usepackage{setspace}
\usepackage{cite}
\usepackage{array}
\usepackage{mdwmath}
\usepackage{mdwtab}
\usepackage{mdwtab, stmaryrd}
\usepackage{booktabs}
\usepackage{times}
\usepackage{epsfig}
\usepackage{latexsym}
\usepackage{epstopdf}
\usepackage{verbatim}
\usepackage{units}
\usepackage{amsthm}
\usepackage{mdwlist}
\usepackage{dblfloatfix}
\usepackage{blindtext}
\usepackage{array}
\newcolumntype{P}[1]{>{\centering\arraybackslash}p{#1}}
\usepackage{tabu}

\newtheorem*{proposition*}{Proposition}

\newfont{\bbb}{msbm10 scaled 500}

\newfont{\bb}{msbm10 scaled 1100}

\newcommand{\bv}{{\bf b}}

\newcommand{\hv}{{\bf h}}

\newcommand{\ov}{{\bf o}}

\newcommand{\qv}{{\bf q}}

\newcommand{\sv}{{\bf s}}

\newcommand{\Om}{{\bf O}}

\makeatletter
\DeclareRobustCommand*\cal{\@fontswitch\relax\mathcal}
\makeatother

\begin{document}
\title{Misbehavior Detection in Wi-Fi/LTE Coexistence over Unlicensed Bands}
\author{\IEEEauthorblockN{Islam Samy$^1$, Xiao Han$^2$, Loukas Lazos$^1$, Ming Li$^1$, Yong Xiao$^3$, and Marwan Krunz$^1$}\\
	\IEEEauthorblockA{
		$^1$ The University of Arizona \\
	   $^2$ The University of South Florida\\
	   $^3$Huazhong University of Science and Technology\\
Email: islamsamy@arizona.edu, xiaoh@usf.edu, \{llazos, lim\}@arizona.edu}, yongxiao@hust.edu.cn, krunz@arizona.edu
\thanks{An abridged version of this paper appeared in the Proc. of the 11th ACM Conference on Security and Privacy in Wireless and Mobile Networks (Wisec), 2018, \cite{samy2018lte}.}
}\IEEEoverridecommandlockouts
\maketitle 

\begin{abstract}
We consider the problem of fair coexistence between LTE and Wi-Fi systems in the unlicensed 5 GHz U-NII bands. We focus on the misbehavior opportunities due to the heterogeneity in channel access mechanism and the lack of a common control plane. We define selfish misbehavior strategies for the LTE that yield an unfair share of the spectrum resources. Such strategies are based on manipulating the operational parameters of the LTE-LAA standard, namely the backoff mechanism, the traffic class parameters, the clear channel access (CCA) threshold, and others. 
Prior methods for detecting misbehavior in homogeneous settings are not applicable in a spectrum sharing scenario because the devices of one system cannot decode the transmissions of another. We develop implicit sensing techniques that can accurately estimate the operational parameters of LTE transmissions under various topological scenarios and {\em without decoding.} These techniques apply correlation-based signal detection to infer the required information.
Our techniques are validated through experiments on a USRP testbed. We further apply a statistical inference framework for determining deviations of the LTE behavior from the coexistence etiquette.  By characterizing the detection and false alarm probabilities, we show that our framework yields high detection accuracy at a very low false alarm rate. Although our methods focus on detecting misbehavior of the LTE system, they can be generalized to other coexistence scenarios. 
\end{abstract}

\section{Introduction}
The high demand for wireless services has fueled a severe shortage in radio spectrum resources. 
The regulatory approach for meeting this galloping demand is to allow the coexistence of competing wireless technologies in common bands. An example of such coexistence is that of LTE and Wi-Fi  in the 5 GHz U-NII band 
\cite{fcc2010second, flores2013ieee, qualcomm2013, 3gpp2015,fcc2016use}. However, shared spectrum introduces novel challenges for the secure, efficient, and fair channel access.
Many of these challenges arise from the heterogeneity of the coexisting systems, the system scale, and the lack of explicit coordination mechanisms between them. Such heterogeneity is manifested in different  PHY-layer capabilities, channel access dynamics (dynamic vs. fixed), schedule-based vs. random access, interference-avoiding vs. interference-mitigating, etc. Altogether, this creates a complex coexistence scenario without a unified control plane. 

Several recent efforts have addressed the problem of fair coexistence of LTE/Wi-Fi and Wi-Fi/Zigbee under benign settings (e.g.,  \cite{he2016proportional, li2015modeling,guan2016cu,sagari2015coordinated,mukherjee2015system,xiao2016adaptive, chen2016optimizing, cai2016spectrum, zinno2017fair,samy2019optimum,han2020energy,hirzallah2020intelligent,abyaneh2019intelligent,hirzallah2019harmonious,hirzallah2019matchmaker,hirzallah2018modeling}). Recent analytical and experimental studies have shown that  an LTE  system could cause serious performance degradation to a co-present Wi-Fi system, even if the LTE remains protocol-compliant \cite{sagari2015modeling,ratasuk2012license}.  The main approach to address unfair channel access is to introduce the  Licensed-Assisted Access (LAA) protocol that follows the Listen-Before-Talk (LBT) mechanism \cite{rel15}.  Tao {\em et al.} showed that dynamically adjusting the contention window (CW) size can be beneficial for fair coexistence \cite{tao2015enhanced}. Follow-up works achieved further improvements by controlling other protocol parameters and applying other enhancements, e.g.,  \cite{jeon2015lte, yin2016lbt,tan2019qos,dai2018adaptive,tan2019deep}. 

Intentional violations of the coexistence etiquette to gain an unfair spectrum share have not been studied at length. Ying {\em et al.} were among the first to consider the problem of misbehavior when cycle-based LTE-U and Wi-Fi coexist \cite{ying2017detecting}. The authors recognized that because the LTE duty cycle is unilaterally controlled by the LTE system, it can be abused to increase LTE's spectrum share. They proposed a monitoring mechanism that accurately estimates the duty cycle and allows a spectrum manager to detect any misbehavior. The proposed scheme is not applicable to LTE-LAA, which is embraced by most operators and the standardization bodies \cite{rel15}. {\em In this paper, we  focus on misbehavior detection mechanisms specific to the prevailing LTE-LAA standard.} 

Our methods build upon prior works on misbehavior detection for homogeneous networks, e.g., \cite{kyasanur2005selfish, fragkiadakis2013survey,tang2014real, li2015mac,zhang2013vuln}, with notable differences. First,  heterogeneous networks do not share common coordination channels for communicating explicit control information such as the network allocation vector (NAV) field, device IDs, reservation messages (RTS/CTS), etc. Without explicit coordination, detecting the state and monitoring the behavior of stations operating under a different technology is challenging, as the messages exchanged by one system are undecodable at the other. Relevant challenges include determining which system occupies the channel, for how long, at what locality, with what range, and which stations collided, to name a few. Moreover, although the LTE-LAA and Wi-Fi standards follow the same carrier-sense multiple access (CSMA) approach, they adopt different channel contention parameters that affect 
the overall system behavior. Determining a system's behavior requires accurate estimation of these parameters but using only implicit monitoring. 
      
In this paper, we address the problem of misbehavior at the system level when heterogeneous technologies coexist. Specifically, we consider a misbehaving LTE-LAA system that coexists with Wi-Fi. LTE devices aim at occupying the shared spectrum for a longer fraction of time by manipulating the channel access mechanism 
of LAA. 
We propose a framework that detects LTE misbehavior, taking into account the absence of any means for explicit coordination.  Our framework relies on implicit sensing mechanisms that provide  accurate approximations of the operational parameters used by the misbehaving LTE devices.  
Our contributions are  summarized as follows:
\begin{itemize}
\item We study the problem of channel access misbehavior of LTE-LAA when coexisting with Wi-Fi. Although possible misbehaving strategies bear resemblance to those in a homogeneous setting, we highlight novel challenges that stem from the technology heterogeneity and lack of explicit coordination. 
\item We introduce a new suite of monitoring mechanisms that do not rely on signal decoding for estimating relevant LTE-LAA protocol parameters.  We develop implicit sensing techniques that go beyond simple LTE transmission detection to determine the presence of hidden stations, identify retransmitted frames, and specify the  LTE priority class. These are essential parameters for accurately estimating the overall LTE behavior. 
\item We validate the effectiveness of the implicit parameter-estimation techniques in a USRP testbed. We show that these techniques are reliable and accurate.
\item We propose a misbehavior detection mechanism based on statistical inference. This mechanism allows a monitor to detect any deviations of LTE behavior from the spectrum sharing etiquette. 
\item We investigate the detection performance under different traffic loads and adapt our framework accordingly to guarantee high detection and low false alarm probabilities. 
\item We perform extensive simulations to validate the proposed misbehavior detection mechanism and show that our approach yields near-perfect detection probability and a negligible false alarm rate. 
\end{itemize}

The remainder of this paper is organized as follows. We discuss related works in Section \ref{RelW}. The system and misbehavior models are introduced in Section \ref{SysMod}. 
The adopted implicit techniques for monitoring LTE activities are detailed in Section \ref{Implicit}. 
In Section \ref{Esti}, we demonstrate how the LTE channel access behavior can be accurately evaluated. 
We validate the performance of the proposed implicit techniques in Section \ref{Exp_val}. We analyze the detection scheme performance  in Section \ref{perf} and summarize the main contributions of this work in Section \ref{Conc}.

\section{Background and Related work}
\label{RelW}
\subsection{LTE-LAA Release 15}
\label{LAAs}

We consider an LTE system that follows the LAA specification, as described in  Release 15 \cite{rel15}. The standard defines four traffic priority classes. The first two classes are suitable for transmitting control messages and short frames, whereas classes $C_3$ and $C_4$ accommodate longer LTE frames. 

\begin{figure}
\begin{center}
\includegraphics[width=1\linewidth]{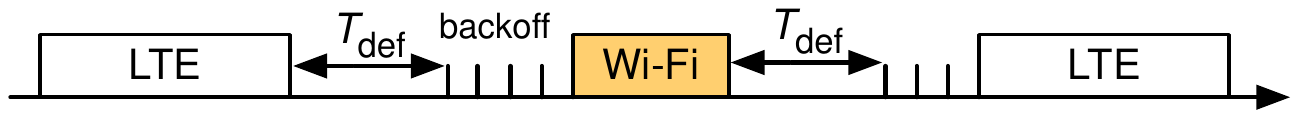}
\end{center}
\vspace{-0.15in}
\caption{Backoff between two consecutive transmissions.} \label{fig:LAA}
\vspace{-0.15in}
\end{figure}

\noindent {\bf Downlink channel access:} The downlink channel access mechanism of LTE-LAA is shown in Fig.~\ref{fig:LAA}. Channel access  parameters are listed in Table \ref{table:T1} and the channel access steps are as follows.
\begin{enumerate}
\item Before transmitting a frame, the eNB freezes for an initial time $T_{init}$ consisting of a defer time $T_{def}=16\mu s$ plus $p$ observation slots, each of length $T_s=9\mu s$. The parameter $p$ takes larger values for lower priority classes to compensate for the longer frame size. If the channel stays idle during $T_{init}$, the eNB proceeds to the backoff phase described in Step 2, otherwise it repeats Step 1. The channel state (busy or idle) is determined by sensing the power on a given channel. If the power is less than the CCA threshold ($P_{th}\approx -73$ dBm according to \cite{rel15}), for at least $4\mu s$, the channel is inferred to be idle and it is busy, otherwise. 

\item The eNB initializes the backoff counter to a value $b$ uniformly selected in $\{0,1,\dots,q-1\}$, where  $q$ is the contention window (CW) size, initially set to  a minimum value $q_{\min}$. 

\item The eNB decrements its backoff counter by one with every idle slot. If a slot is sensed busy, the eNB freezes its backoff counter until the channel becomes idle. The channel must remain idle for  $T_{init}$ before the backoff countdown can be resumed.

\item When the backoff counter becomes zero, the eNB transmits a frame with a maximum duration of $T_{\mbox{MCOP}}$. The eNB then waits for an ACK/NACK. If it receives an ACK, the transmission round is completed. Otherwise, the process is repeated from Step 1 by doubling the CW size, up to a $q_{\max}$. 
\end{enumerate}

{\bf Uplink channel access:} To make an uplink transmission, a UE must receive an uplink (UL) grant permission from the eNB. The UL grant permission determines the access type that should be used by the  UE. The standard specifies two candidate types.
\begin{itemize}
    \item Type 1: The UE follows the same backoff process described for the downlink, but with slightly different parameters as shown in Table \ref{table:T1}.
    \item Type 2: The UE waits for the channel to be idle for 25 $\mu$s and then accesses the channel without further contention. 
\end{itemize}

From both the DL and the UL procedures, we note that the priority classes differ in both the defer time and allowed CW sizes. As will be shown later, these differences can be exploited by LTE devices to shorten the time between consecutive transmissions.

\begin{table}
\caption{LTE parameters for different priority classes.}
\vspace{-0.1in}
\centering
\begin{tabular}{ | P{0.85cm} | P{0.85cm}| P{0.85cm} |   P{1.7cm} |  P{1.7cm} |} 
\hline
Class & $p$ (slots) & $q_{\min}$ (slots) & $T_{\mbox{MCOP}}$ (ms) & $q$ (slots) \\ 
\hline
\multicolumn{5}{|c|}{Downlink channel access}\\
\hline
$C_1$ & $1$ & $4$ &  $2$ & $\lbrace 4,8 \rbrace$  \\ 
\hline
$C_2$ & $1$ & $8$  & $3$ & $\lbrace 8,16 \rbrace$  \\ 
\hline
$C_3$ & $3$ & $16$  & $8$ or $10$  & $\lbrace 16-64 \rbrace$ \\ 
\hline
$C_4$ & $7$ & $16$  & $8$ or $10$ & $\lbrace 16-1024 \rbrace$\\ 
\hline
\multicolumn{5}{|c|}{Uplink channel access}\\
\hline
$C_1$ & $2$ & $4$ &  $2$ & $\lbrace 4,8 \rbrace$  \\ 
\hline
$C_2$ & $2$ & $8$  & $4$ & $\lbrace 8,16 \rbrace$  \\ 
\hline
$C_3$ & $3$ & $16$  & $6$ or $10$  & $\lbrace 16-1024 \rbrace$ \\ 
\hline
$C_4$ & $7$ & $16$  & $6$ or $10$ & $\lbrace 16-1024 \rbrace$\\ 
\hline
\end{tabular}
\label{table:T1}
\vspace{-0.1in}
\end{table}



\subsection{Related Work}

LTE/Wi-Fi coexistence in a benign setting has been studied extensively  \cite{samy2019optimum,han2020energy,hirzallah2020intelligent,abyaneh2019intelligent,hirzallah2019harmonious,hirzallah2019matchmaker,hirzallah2018modeling} as well as \cite{chen2017coexistence} and the references therein. 
Ratasuk {\em et al.}  \cite{ratasuk2012license} showed that LTE  outperforms  Wi-Fi by replacing one of the Wi-Fi deployments with an LTE cell and comparing the respective throughput. Hirzallah {\em et al.} \cite{hirzallah2016full}  showed that  different access protocols for Wi-Fi and LTE can cause an increased collision rate and latency for both systems. They suggested a CCA threshold adaptation mechanism to promote fairness. 
In \cite{abyaneh2019intelligent}, authors proposed a framework that allows nodes to adapt their CW sizes
based on observed transmissions, ensuring they receive
equal airtime. 
The idea of adapting the backoff parameters of  LTE/Wi-Fi to achieve a fair coexistence  was also studied in \cite{jeon2015lte,tao2015enhanced,yin2016lbt,tan2019qos,dai2018adaptive,tan2019deep}.    
However, these works assumed that all stations are trustful and protocol-compliant.

Misbehavior detection for channel access in homogeneous networks has been extensively studied, especially for IEEE 802.11  protocols (e.g.,  \cite{fragkiadakis2013survey,tang2014real,toledo2007robust, kyasanur2005selfish,li2015mac,zhang2013countering}). 
In \cite{kyasanur2005selfish}, the authors introduced modifications to the IEEE 802.11 protocol to simplify misbehavior detection and presented a penalty scheme for punishing selfish users.
 Li {\em et al.} used multiple backoff counter observations to calculate the probability that a monitored station remains protocol-compliant \cite{li2015mac}. Misbehavior was detected by comparing this probability to a threshold.  
 Considering a fair channel allocation \cite{tang2014real}, Tang {\em et al.} proposed a real-time misbehavior detection mechanism, which relies on an indicator function that represents the difference between the number of successful transmissions and the number of expected transmissions.   Toledo {\em et al.} applied the Kolmogorov-Smirnov test to detect misbehavior from the  number of idle slots between two transmissions \cite{toledo2007robust}. As all stations follow the same protocol, misbehavior is detected if the idle slot distribution of a station differs from that of others. 
 
Whereas there is a wealth of interest in channel access misbehavior for homogeneous networks, misbehavior between coexisting technologies is relatively new. The work closest to ours is reported in  \cite{ying2017detecting}. However, the authors considered  the misbehavior in the CSAT-based LTE-U protocol, not LAA. They  developed a method for estimating the LTE  duty cycle by tracking LTE transmissions. The latter are identified based on the frame length, as LTE frames are typically longer than Wi-Fi frames. Possible LTE misbehavior is detected by a central node called the spectrum manager, which has prior knowledge  of the permitted duty cycle for LTE. In this paper, we consider misbehavior under the LTE-LAA standard that implements a CSMA-like (LBT) channel access model and involves drastically different misbehavior actions and remedies.  Our work is similar to that of \cite{ying2017detecting} in that we also employ a central node, which we call as the \textit{hub}, to analyze the LTE behavior. 
The pivotal difference between our work and misbehavior detection in homogeneous networks lies in the monitoring mechanisms for obtaining samples of behavior. All prior works rely on frame decoding to attribute transmissions to their originators. This is not generally possible between different technologies. Moreover, LTE and Wi-Fi systems execute channel access protocols with different parameters. For instance, the LTE-LAA backoff parameters  depend on the priority class. Accurate estimation of the LTE behavior requires mechanisms for classifying frames according to their respective classes. Additional challenges stem from the heterogeneity in transmission and interference ranges. A Wi-Fi station may backoff in the presence of an LTE transmission, but the converse may not occur. 


\section{Models and  Framework Overview} 
\label{SysMod}

\subsection{ System Model}

\begin{figure}[t]
\begin{center}
\includegraphics[width=0.8\linewidth]{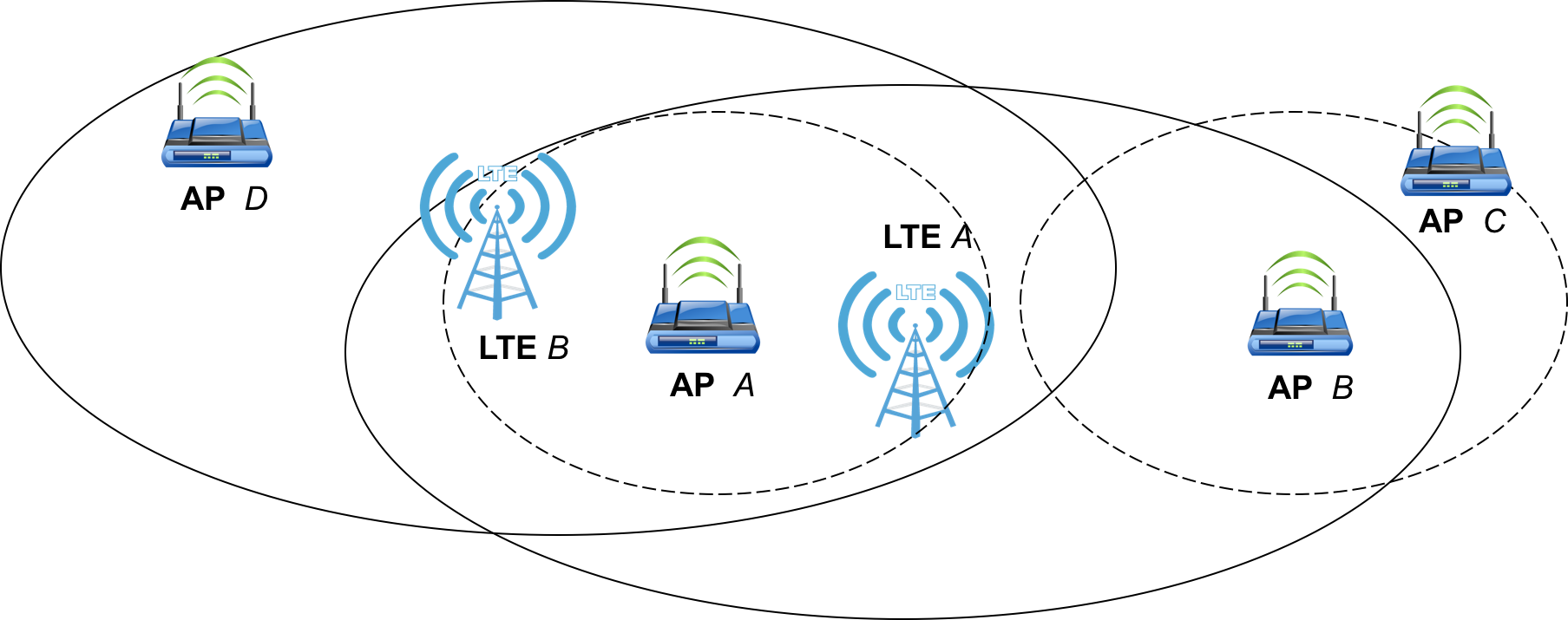}
\end{center}
\vspace{-0.1in}
\caption{Coexistence between LTE and Wi-Fi. Wi-Fi and LTE stations have different interference ranges.}
\label{fig:sys}
\vspace{-0.2in}
\end{figure}

We consider  $N_L$ LTE devices that coexist with $N_{W}$ Wi-Fi access points (APs) over the 5 GHz unlicensed band. 
The one-hop neighborhood set (other devices inside the interference range) of device $X$ is denoted by $\mathcal{N}_{X}^{(1)}$. LTE devices and Wi-Fi APs may transmit at different powers, so $Y \in 
\mathcal{N}_{X}^{(1)}$ does not imply that $X \in \mathcal{N}_{Y}^{(1)}$. As an example, in  Fig.~\ref{fig:sys},  AP $B$ is in the interference range of LTE $A$ (solid line), but LTE $A$ is not in the interference range of $B$ (dashed line). The transmission powers of the LTE and Wi-Fi are denoted by $P_\ell$ and $P_w$, respectively.
 LTE  devices  and  Wi-Fi  terminals are considered to follow  the  LTE-LAA  and  IEEE  802.11ac standards, respectively. 
 
 We consider the misbehavior of one or more LTE devices which are monitored by any  AP  in their vicinity. The monitoring APs are  capable of performing simple signal operations on RF signals like sampling and correlation.
 LTE-related observations are collected by APs and analyzed at a central hub. This could be achieved through a cloud service that allows the uploading of all observations made by the APs to a central repository. Such scenario is relevant in enterprise networks where multiple APs are under a single administrative control or can be offered as an overlay service to which APs subscribe. This assumption also helps us identify if misbehavior is detectable at the system level, given all distributed observations.
 
 We initially focus on detecting the LTE misbehavior in backlogged conditions. Under such conditions, gains in performance due to LTE misbehavior occur at the expense of the Wi-Fi system. We later consider  the  detection of LTE misbehavior  under unsaturated traffic conditions.

\subsection{Misbehavior Model}

The goal of a misbehaving LTE is to capture the channel more frequently and for a longer time than competing APs. This can be achieved by manipulating the LAA protocol parameters in the uplink or downlink direction.

{\bf Misbehavior in the downlink direction.} This is the most beneficial type of misbehavior because the downlink direction carries a far higher traffic volume than the uplink one and the eNB nodes transmit to a large number of LTE devices. An eNB node can misbehave in the following ways.

{\em 1) Decrease the defer time $p$:} An LTE device can reduce the defer time to initiate the backoff countdown process faster. It can select a defer time that belongs to a high priority class and transmit a frame of low priority class with longer duration. Alternatively, the LTE can completely ignore the defer time and initiate the backoff countdown immediately within a transmission round.

{\em 2) Increase the CCA threshold:} Another manipulation strategy is to avoid freezing the backoff counter in the presence of active Wi-Fi APs. This leads to  faster acquisition of the medium, which can be beneficial if a high-power eNB transmission overshadows Wi-Fi transmissions. Note that avoiding backoff freezing can occur in benign settings due to hidden terminals or due to power asymmetry. This is illustrated in Fig.~\ref{fig:sys}. Assume that Wi-Fi AP $B$ acts as a monitor for the behavior of LTE $A$, which is outside the interference range of AP $C$ (hidden terminal) and therefore does not freeze its backoff counter when AP $C$ transmits. This behavior may be perceived by AP $B$, which is within the interference range of LTE $A$, as misbehavior. In another scenario, AP $B$  observes that LTE $A$ does not freeze its backoff window when AP $B$ is active. This could be due to the transmission power asymmetry or due to misbehavior. 
 
{\em 3) Reduce the backoff window:} An LTE system can manipulate the LAA backoff process  by selecting its backoff counter from a smaller window range $b \in \{0,1,\dots,q_m-1\}$, where $q_m < q$. 
The value of $q_m$ may be selected from a high-priority class so that the LTE appears to be protocol-compliant. Moreover, LTE can avoid increasing its CW size after a collision, to reduce the delay between two consecutive channel access attempts. 
This can be done by simply ignoring any mandated doubling of the CW size after a collision or by taking advantage of the low CW sizes allowed for high-priority classes. Here, we consider a general model in which the LTE remains protocol-compliant for  a fraction of time $0 \leq \alpha \leq 1$, while it uses a smaller CW of size $q_m$ for the remaining time. 
As an example of backoff  manipulation, the LTE device can consistently select backoff values in the range $\{0,1,2,3\}$, irrespective of the priority class. Moreover, in the event of successive collisions, it can maintain $q_m =4$. Essentially, all priority classes are treated as if they were of $C_1$. 
We emphasize that there is an inherent difficulty in attributing collisions to a transmitting device because: (a) collisions are receiver-dependent, and (b) in a heterogeneous setting, one system cannot decode the transmissions of other systems. Hence, detecting misbehavior that involves collisions is challenging. 

\begin{figure}[t]
\begin{center}
\includegraphics[width=1.0\linewidth]{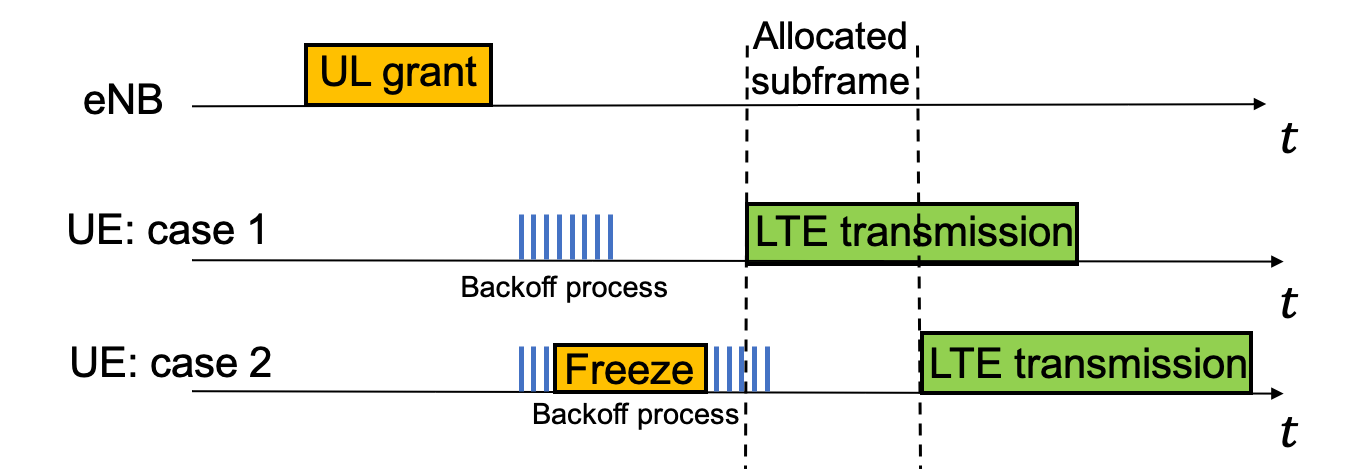}
\end{center}
\vspace{-0.1in}
\caption{Uplink channel access procedures.} 
\vspace{-0.2in}
\label{fig:LTE_UL}
\end{figure}

{\bf Misbehavior in the uplink direction.}
Unlike DL transmissions, any UL transmission must be preceded by an UL grant permission from the eNB. This grant specifies the subframe where the UL transmission should start. Once the UL grant is received, the UE starts the backoff process. Here, the misbehavior strategies described for the DL direction are possible. However, they have limited benefit for the UE because a failure to seize the channel for a Type 1 transmission will lead to a much more aggressive channel access strategy for the next subframe (Type 2).  

If the UE cannot complete the backoff process before the beginning of the allocated subframe due to contention (case 2 in Fig. \ref{fig:LTE_UL}), the UE is allowed to access the channel in the following subframe by only deferring for an initial time $T_{init}=T_{def}+p$, where $p=1.$ That is, the UE senses the channel for minimal time and does not follow a backoff process, thus increasing the chances of capturing the channel substantially. 

Given the significantly higher volume of DL traffic, the requirement for a DL frame to schedule an UL transmission, and the aggressive channel access strategy of UL Type 2 transmissions, we only focus on detecting LTE misbehavior in the DL direction.

\subsection{Misbehavior Detection Framework Overview}

\begin{figure}[t]
\begin{center}
\includegraphics[width=0.9\linewidth]{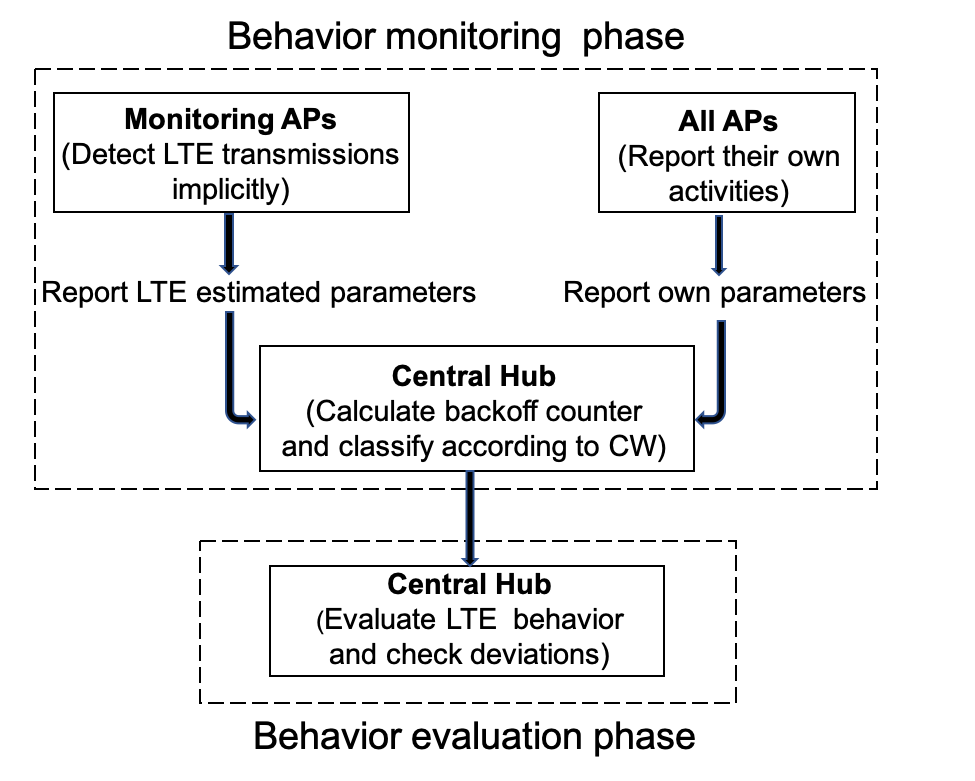}
\end{center}
\vspace{-0.1in}
\caption{Overview of the misbehavior detection mechanism.} 
\vspace{-0.2in}
\label{fig:Overview}
\end{figure}

To detect misbehaving eNBs, we propose a detection framework which  consists of a behavior monitoring phase and a behavior evaluation  phase, as shown in Fig. \ref{fig:Overview}. During the behavior monitoring phase, monitoring APs listen to the wireless medium when they do not transmit. Each monitoring AP overhears LTE frames and infers behavior-related parameters such as the start and end times of the LTE frame, the transmitting eNB, the retransmission round, the traffic class, and the topological relation of the AP to the LTE (whether the AP is a hidden  terminal to the transmitting LTE or not). {\em All parameters are implicitly estimated without decoding LTE frames.} Monitoring APs periodically report a time series of observations along with a time series of their own activity to a central hub for further processing.   

In the behavior evaluation phase, the hub processes the information reported by the distributed network of APs to derive the channel access pattern of each monitored eNB. If the access pattern is deemed to deviate from the LAA specifications, the LTE system is considered misbehaving. 

\section{Behavior Monitoring Phase}
\label{Implicit} 

The key challenge in monitoring the LTE behavior is the system heterogeneity. The monitoring APs cannot decode LTE transmissions as they may not be equipped with LTE receivers. In this section, we present several techniques for the implicit estimation of the LTE operating parameters. Specifically, each monitoring AP listens to the wireless medium when it is not active. Upon detection of channel activity that is not Wi-Fi decodable, it processes the signal without decoding to determine if it belongs to an LTE. For the $i^{th}$ detected LTE transmission, the AP estimates an observation vector $\ov(i)$ of six parameters 
\begin{equation}
\ov(i):=<t_s(i), t_e(i), ID(i), C(i), r(i), h(i)>,
\end{equation}
where $t_s(i)$ and $t_e(i)$ denote the start and end times of the $i^{th}$ transmission, respectively, $ID(i)$ denotes an eNB identity, $C(i)$ denotes the LTE traffic class, $r(i)$ denotes the retransmission round, and $h(i)$ is a flag that denotes if the monitoring AP belongs to the one-hop neighborhood of the transmitting LTE. In the remainder of the section, we describe this parameter estimation.

\subsection{Detecting LTE Transmissions}
\label{Det.CP}

\begin{figure}[t]
\begin{center}
\includegraphics[height=0.17\textheight]{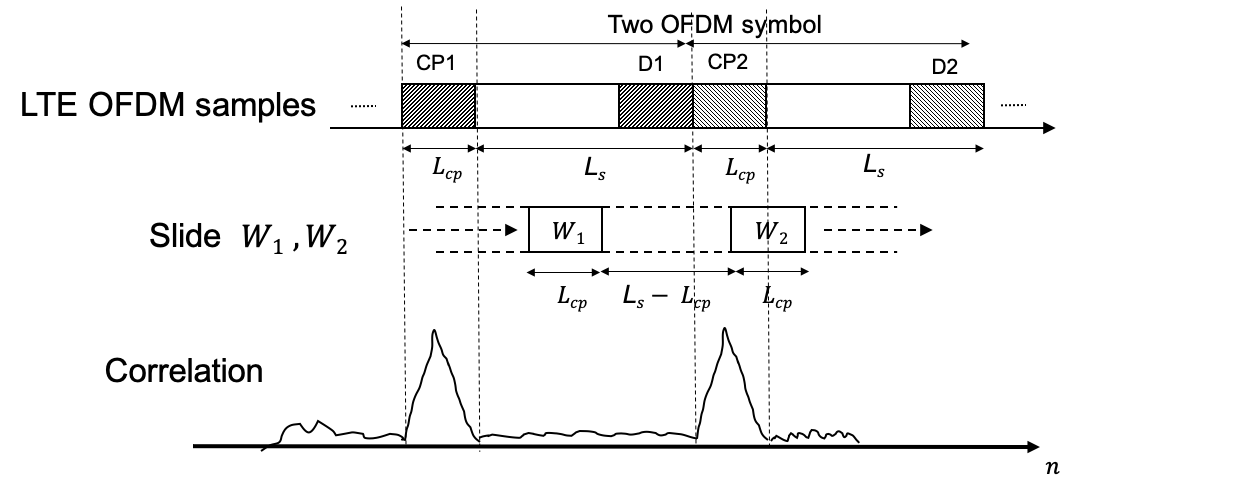}
\end{center}
\vspace{-0.2in}
\caption{Detecting LTE transmissions using CP correlation.} 
\vspace{-0.2in}
\label{fig:LTE_CP}
\end{figure}

 The first step for estimating the LTE operating parameters is to determine when and for how long eNBs transmit. This allows the estimation of $t_s(i)$ and $t_e(i).$ To detect LTE transmissions, we adopt the cyclic prefix (CP)-based method proposed in \cite{hirzallah2016full}. Like any OFDM modulated signal,  LTE transmissions utilize the CP concept to mitigate inter-symbol interference. The end of an OFDM symbol is appended at the beginning, forming the CP. In Fig. \ref{fig:LTE_CP}, CP1 is equal to D1, CP2 is equal to D2, etc. 

A Wi-Fi AP can attribute a signal to an eNB by verifying that the CP and its copy are $L_{S}-L_{CP}$ samples away, where $L_{S}$ and $L_{CP}$ denote the lengths, in samples, of the LTE symbol and the CP, respectively. The duration of the LTE OFDM symbol, and consequently the appended CP, are fixed to unique values in the LTE standard \cite{rel15}. Based on the fixed duration and the sampling rate, the AP determines the  values of $L_{S}$ and $L_{CP}$.  
The main idea of this method is to detect high signal  correlation when the CP samples are correlated to the end of the LTE symbol.

\medskip

\noindent \underline{\bf Algorithm 1: LTE transmission detection}

 {\bf Step 1:} The AP samples the received signal.
 
 {\bf Step 2:} The AP fixes two time windows $W_1$ and $W_2$ of length $L_{CP}$, separated by $L_S-L_{CP}$ samples. Then, it shifts the two windows simultaneously by one sample at a time while keeping the window separation fixed to $L_{S}-L_{cp}$. 
 
 {\bf Step 3:} For each shift $n$, the AP obtains the vectors of signal samples $\sv_1(n)$ and $\sv_2(n)$ that correspond to windows $W_1$ and $W_2$ (each of length $L_{CP}$) and computes  
\begin{equation}
\rho(n)=\frac{|A(n)|^2}{(\max(E_{\sv_1}(n),E_{\sv_2}(n)))^2},
\label{rho}
\end{equation}
where $A(n)$ is the correlation between $\sv_1(n)$ and $\sv_2(n)$, 
\begin{equation}
A(n)=\sum_{k=0}^{L_{CP}-1}\sv_1(n-k)\sv_2^*(n-k-L_S).
\label{rho1}
\end{equation}
Here, $\sv^{\ast}$ is the complex conjugate of $\sv$. The energies $E_{\sv_1}(n)$ and $E_{s_2}(n)$ are computed as 
\begin{equation}
E_{\sv_1}(n)=\sum_{k=0}^{L_{CP}-1}\sv_1(n-k)\sv_1^*(n-k),
\label{rho2}
\end{equation}
\begin{equation}
E_{\sv_2}(n)=\sum_{k=0}^{L_{CP}-1}\sv_2(n-k-L_S)\sv_2^*(n-k-L_S).
\label{rho3}
\end{equation}
We use the max in the denominator to ensure that $\rho(n)$ always stays within $[ 0,1]$ and to help minimize the value of  $\rho(n)$ when $\sv_1(n)$ and $\sv_2(n)$ are different.

{\bf Step 4:} If $\sv_1(n) \approx \sv_2(n)$, the correlation spikes indicating that $\sv_{1}(n)$ is the CP of $\sv_2(n)$. The correlation spike is recognized if $\rho(n)\geq\gamma_{LTE}$
where $\gamma_{LTE}$ is a minimum correlation threshold that defines a signal match. We discuss the selection of the threshold $\gamma_{LTE}$ in Section~\ref{ssec:lte_det}.

{\bf Step 5:} The AP sets $t_s(i)$ to the time of the first local maximum (correlation spike) that exceeds $\gamma_{LTE}$  and $t_e(i)$ to the time of the last local maximum that exceeds $\gamma_{LTE}$. 
\medskip

\subsection{ Differentiating Between eNBs}

Attributing transmissions to individual eNBs is necessary for building the behavioral profile of each eNB. However, this requires: (a) to distinguish downlink LTE transmissions from uplink ones and (b) differentiate between eNBs in the downlink. 

\begin{figure}[h]
\begin{center}
\includegraphics[width=1\linewidth,height=0.1\textheight]{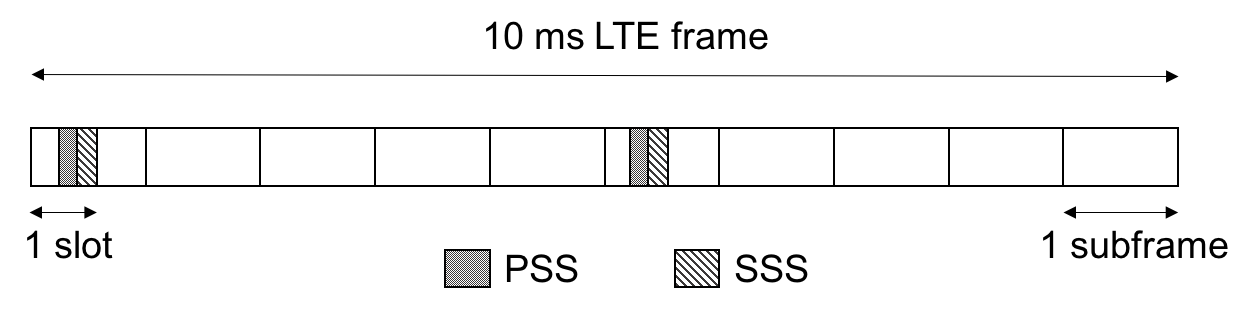}
\end{center}
\vspace{-0.2in}
\caption{The PSS and SSS fields in LTE frames. } 
\vspace{-0.1in}
\label{fig:ID}
\end{figure}

To perform these two operations, we propose that monitoring APs use two distinct frame fields included only in the DL direction. Those are the primary synchronization signal (PSS) and the secondary synchronization signal (SSS), which  are used for synchronization and carry information about the transmitting eNB's identity. As shown in Fig.~\ref{fig:ID}, the PSS and SSS fields are repeated twice at fixed locations in an LTE DL frame. Samples of DL LTE signals at those fixed locations are identical. This gives the opportunity to a monitoring AP to identify DL frames. 

Moreover, the identity of an eNB  is calculated as  ID $= ID_{1}+3ID_{2}$, where  $ID_1$ and $ID_{2}$ define the physical-layer cell identity group and physical layer identity of the LTE, respectively. The $ID_{1}$ and $ID_{2}$ values are part of the PSS and SSS fields. The pair $(ID_1, ID_2)$ is unique for every eNB, however, both can only be obtained by decoding the PSS and SSS fields.

Monitoring APs can exploit the known locations for $ID_1$ and $ID_2$ to attribute LTE transmissions to different eNBs. Note that we are not interested in extracting the ID value, but to identify LTE frames with the same ID. We use the ``signal signature'' of the static PSS and SSS fields for this attribution. 
 The main idea is to detect the unique  fields $(ID_1, ID_2)$ by sampling the LTE transmission at the PSS and SSS locations and correlating the signal samples with previously recorded samples. Two transmissions from the same eNB will exhibit a high correlation on the ID fields, if the channel effect is neutralized. A monitoring AP can identify DL frames and differentiate between different eNBs by executing the following LTE frame attribution algorithm. 
\medskip

\noindent  \underline{\bf Algorithm 2: LTE Frame Attribution}

{\bf Step 1:} For the $i^{th}$ LTE frame, the AP applies the CP-based LTE detection algorithm and synchronizes with the frame start time $t_s(i).$ 

{\bf Step 2:} The AP collects two sets of samples $\sv^{(i)}_{ID}$ and $\tilde{\sv}^{(i)}_{ID}$ of length $L_{ID}$, at the two locations of the PSS and SSS fields. 

{\bf Step 3:} The AP computes the signal correlation $ \rho_{DL}{(i)}$ between these two sets of samples as follows
\begin{equation}
 \rho_{DL}{(i)}=\frac{|\sum_{k=1}^{L_{ID}}{\sv^\ast}^{(i)}_{ID}(k) \ \tilde{\sv}^{(i)}_{ID}(k)|^2}{(\max(E_{{\sv}^{(i)}_{ID}},E_{\tilde{\sv}^{(i)}_{ID}}))^2}, 
 \end{equation}
where $E_{{\sv}^{(i)}_{ID}}$ and $E_{\tilde{\sv}^{(i)}_{ID}}$  are the energies of $\sv^{(i)}_{ID}$ and $\tilde{\sv}^{(i)}_{ID}$, respectively, calculated in a similar way to \eqref{rho2}.
A downlink transmission is  inferred  if the
correlation exceeds a threshold value that defines a signal match (see Section~\ref{subsec:LTE-ID}). If the correlation is below the threshold, the current LTE frame is ignored (uplink transmission). Otherwise, the AP proceeds to the following steps.

{\bf Step 4:} The AP maintains a signature database that includes the observed LTE signatures up to the current observation. The signature $\sv_{ID_j}$ of the $j^{th}$ LTE represents the samples carrying $ID_1$ and $ID_2$, collected from previous transmissions. The database is assumed to be initially empty and is updated gradually according to the collected observations.



{\bf Step 5:} Due to the change in the channel impulse response over time, the AP  adjusts the samples $\sv^{(i)}_{ID}$  by a fixed phase to compensate for the channel effect. The AP observes the phases of the complex samples $s^{(i)}_{ID}$ collected over the $i^{th}$ frame, denoted by vector $\mathbf{\theta^{(i)}_{ID}}$. The AP recovers the phases  $\mathbf{\theta_{ID_j}}$ of the complex samples  $\sv_{ID_j}$ stored in the signature database, for each $ID_j$. The AP computes the average phase shift between the two sample vectors $s^{(i)}_{ID}$ and $\sv_{ID_j}$ as,
\begin{equation}
  \bar{\theta}(i,j)=\frac{1}{L_{ID}} \sum_{k=1}^{L_{ID}} |\mathbf{\theta^{(i)}_{ID}}(k)-\mathbf{\theta_{ID_j}}(k)|. 
\end{equation}
The AP updates the phase part of  $\sv^{(i)}_{ID}$ as $
 \mathbf{\theta^{(i)}_{ID}}=\Big(\mathbf{\theta^{(i)}_{ID}}+ \bar{\theta}(i,j)\Big)\mod \pi.
$ The phase compensation method is explained in detail in the experimental validation section (Section \ref{subsec:LTE-ID}). 

{\bf Step 6:} The AP computes the signal correlation between $\sv^{(i)}_{ID}$ and every signature in the database, 
 \begin{equation}
 \label{sigcorr}
 \rho_{ID}{(i,j)}=\frac{|\sum_{k=1}^{L_{ID}}\sv^{\ast}_{ID_j}(k) \ \sv^{(i)}_{ID}(k)|^2}{(\max(E_{\sv_{ID_j}},E_{\sv^{(i)}_{ID}}))^2}, \forall j,
 \end{equation}
where $E_{\sv_{ID_j}}$ and $E_{\sv^{(i)}_{ID}}$  are the energies of $\sv_{ID_j}$ and $\sv^{(i)}_{ID}$, respectively, calculated in a similar way to \eqref{rho2}.

{\bf Step 7:} The AP attributes the $i^{th}$ LTE transmission to LTE $ID_j$ that yields the maximum $\rho_{ID}{(i,j)}$,
\begin{equation}
    \text{ID}=\underset{ID_j}{\text{arg}\max}\{ \ \rho_{ID}{(i,j)}| \ \rho_{ID}{(i,j)}\geq \gamma_{ID}\}.
\end{equation}

Here $\gamma_{ID}$ is a minimum correlation threshold that defines a signal match. If a match is found, the AP also replaces $\sv_{ID_j}$, the current signature of LTE $ID_j$, with $\sv^{(i)}_{ID}$.

{\bf Step 8:} If no correlation value exceeds $\gamma_{ID}$, the AP adds $\sv^{(i)}_{ID}$ to the database as a new eNB signature.
\medskip





A challenge for this method is the attribution of an LTE transmission when it collides with another transmission. Although performing such classification via signal correlation in the presence of collisions is possible \cite{gollakota2008zigzag}, we leverage the distributed nature of the monitoring operation to resolve colliding transmissions. As collisions are receiver-dependent, not all monitoring APs experience collisions. Those APs that do not experience a collision correctly classify the LTE transmission. As an example, AP $A$ in Fig.~\ref{fig:sys} is in the interference range of LTE $A$ and LTE $B$ thus being unable to classify frames of $A$ and $B$ that collide. Such frames are correctly monitored by AP $B$ and $D$.  Finally, even if colliding frames fail to be correctly classified, they only represent a small fraction of the transmitted frames. 



\subsection{Priority Class Estimation}
\label{PC}

The channel access parameters of LTE transmissions depend on the priority class. Lower priority classes utilize longer frames and thus are designed to access the channel less frequently, whereas higher classes accommodate shorter frames, shorter defer times, and smaller contention windows. 

To evaluate the compliance of an eNB with the class parameters, the APs classify frames to one of the four classes of Table \ref{table:T1} using the transmission duration. By measuring the length of the $i^{th}$ frame as $T_{\mbox{MCOP}} =t_e(i)-t_s(i)$, the AP can classify the frame to classes $C_1$, $C_2$, and $C_3/C_4$. 
Note that the $T_{\mbox{MCOP}}$ values for $C_3$ and $C_4$ are equal. However, $C_3$ has shorter defer time allowing for faster medium access and a better choice for misbehavior. Thus, for all practical purposes, we air on the conservative side and assume that any frame of length 8ms or 10ms belongs to class $C_3$. 


\subsection{Contention Window Size Estimation} 
\label{CW_est}
Another important behavior parameter is the CW used at every LTE transmission. Maintaining a small CW improves the channel access opportunities for the LTE. Monitoring APs can estimate the CW size  of an eNB by tracking the retransmission round $r(i)$ of a frame. The CW size $q(i)$ at the $i^{th}$ transmission is given by 
\begin{equation}
q(i) = \min\{2^{r(i)}q_{\min}, q_{\max}\},
\label{CWcalc}
\end{equation}
where $q_{\min}$ and $q_{\max}$ are the minimum and maximum allowed CW sizes, as listed in Table \ref{table:T1}.
Following the LTE protocol specifications, the monitoring AP sets $r(i)$ to zero after a successful transmission by the eNB and increments it by one with every retransmission attempt. Note that the AP needs to keep track of $r(i)$ individually for each eNB. 

Parameter $r(i)$ is difficult to infer in practice via overhearing because collisions are receiver-dependent. Rather than attempting to directly infer collisions, APs rely on identifying retransmissions to estimate $r(i).$ Specifically, a monitoring AP utilizes the signal correlation method to detect if the same frame is retransmitted by an LTE. The AP  exploits the fact that the payload and most fields in the header of a retransmitted frame remain identical to the original transmission. Therefore, the sampled signal of two identical transmissions should exhibit high signal correlation, even if one is corrupted by a colliding signal. 
The main challenge in performing signal correlation is identifying the start and end times of the LTE frame, along with the ID field of the collided eNB in the presence of a  collision.

 {\bf Collision between Wi-Fi and LTE:} We first consider the case of an LTE colliding with an AP. This is the most common case, as eNBs are typically deployed to minimize collisions and are usually assigned different operating frequencies. A monitoring AP can estimate $r(i)$ through the following steps. 
\begin{figure}[t]
\begin{center}
\includegraphics[width=1.0\linewidth]{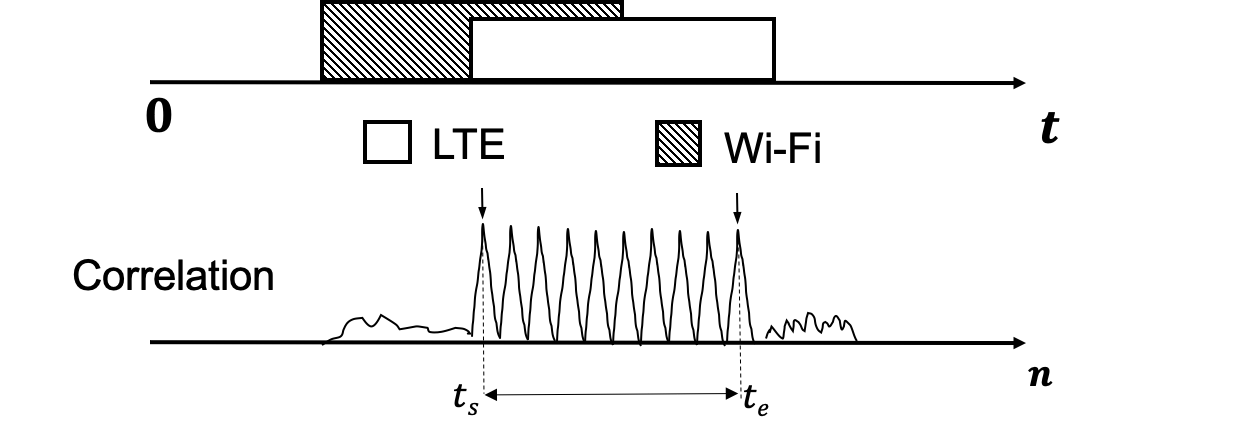}
\end{center}
\vspace{-0.1in}
\caption{Example of applying the CP-based LTE detection method in case of collisions with a Wi-Fi frame.}
\vspace{-0.2in}
\label{fig:RetCol}
\end{figure}
\medskip

\noindent \underline{\bf Algorithm 3: Transmission Round Estimation}

{\bf Step 1:} The AP applies the CP-based LTE detection method described in Algorithm 1 to determine the start time $t_s(i)$ and end time $t_e(i)$ of the $i^{th}$ LTE frame.  These times are identified by the first and last correlation peaks of the CP with the end of the symbols, respectively, as shown in  Fig.~\ref{fig:RetCol}.

{\bf Step 2:} Using the start time $t_s(i)$ as a time reference, the AP extracts the samples carrying the LTE ID. As the collision does not necessarily corrupt all samples (e.g., only half of the samples are involved in the collision in Fig.~\ref{fig:RetCol}), 
the samples carrying the LTE ID may be clean or corrupted.  If the samples are clean, the AP identifies the LTE ID field by performing Algorithm 2 and it proceeds to the following step. Otherwise, if the samples are corrupted, it proceeds to Step 5.

{\bf Step 3:} The AP buffers the samples of the $i^{th}$ and $(i+1)^{st}$ eNB transmission denoted by $\sv(i)$ and $\sv(i+1)$, respectively.

{\bf Step 4:} 
The AP correlates $\sv(i)$ with  $\sv(i+1)$  using  the correlation function in  \eqref{sigcorr} and computes the correlation value $\rho(i,i+1)$. If $\rho(i,i+1)\geq \gamma_{rt}$, where $\gamma_{rt}$ is a correlation threshold that defines a signal match, the AP identifies the $(i+1)^{st}$ frame as a retransmission and sets $r(i+1)=r(i)+1$. 

{\bf Step 5:} If the samples carrying the ID field are corrupted (no match in Step 4), the AP determines the length of the $i^{th}$ LTE  frame  as the difference between the start and end times identified in Step 1. The AP buffers the samples $\sv(i)$ of the $i^{th}$ LTE transmission.

{\bf Step 6:} The AP tracks subsequent frames transmitted by the eNB that have the same length as the  $i^{th}$ frame. For each of these frames, the AP buffers the related samples.

{\bf Step 7:} Let $\sv(j)$ be the buffered samples of  a subsequent eNB transmission. The AP correlates $\sv(i)$ with $\sv(j)$  using  the correlation function in  \eqref{sigcorr} and computes the correlation value $\rho(i,j)$. If $\rho(i,j)\geq \gamma_{rt}$, the AP identifies the  frame $j$ as a retransmission of the $i^{th}$ frame, sets $r(j)=r(i)+1$. It further identifies the ID of the $i^{th}$ frame to be the same as the ID carried in the $j^{th}$ frame.

{\bf Step 8:} If no  frame is  found to exceed $\gamma_{rt}$, the AP ignores the particular transmission. As our behavior estimation depends on many observations, we can tolerate ignoring a small percentage of collisions that remain unidentifiable. 

\medskip

\begin{figure}[t]
\begin{center}
\includegraphics[width=1.0\linewidth]{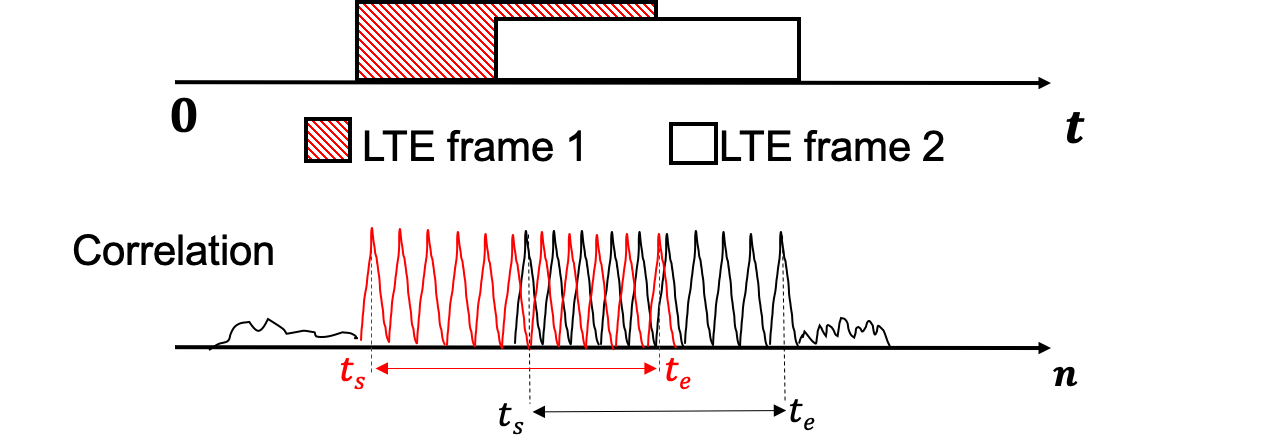}
\end{center}
\vspace{-0.2in}
\caption{Example of applying the CP-based LTE detection method in case of collisions between LTE frames.} 
\label{fig:LTECol}
\vspace{-0.1in}
\end{figure}

{\bf Collision between two eNBs:}  If the eNB transmissions collide, Algorithm 3 can be applied to each of the colliding frames separately, with a modification to Step 1. In this case, symbols in both frames cause a peak in signal correlation.  The AP observes two groups of peaks as shown in Fig.~\ref{fig:LTECol}. The peaks in each group are separated by a period equal to the duration of an LTE OFDM symbol.  Each group of peaks identifies the start and end times of the related LTE frame.
  The start time $t_s$ of the first frame (group) is identified with the first peak. The end time $t_e$ of the first frame is identified with the last peak that is multiple symbol periods away from the first peak (last peak in the group). The start of the second frame is identified as the first peak that is not periodic to the OFDM symbol length. The end of the second LTE frame is identified as the last peak that is multiple symbol periods away from the first peak of the second frame.  

From parameter $r(i)$ and the class priority $C(i)$, the  hub can infer $q(i)$ using \eqref{CWcalc}. The class $C(i)$ is used to determine both $q_{\min}$ and  $q_{\max}$. For instance, the $q(i)$ of a Class 3 frame with $r(i)=1$ should be equal to 32 according to Table 1.

\subsection{Inferring Neighbor APs}
\label{HTD}

\begin{figure}[t]
\begin{center}
\includegraphics[width=0.7\linewidth]{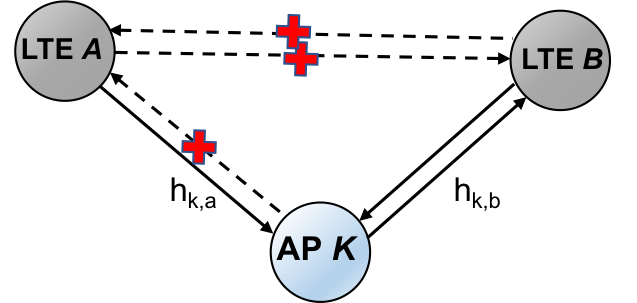}
\end{center}
\caption{AP $K$ is a hidden terminal to $A$ but not $B$. eNBs $A$ and $B$ are hidden terminals.} 
\vspace{-0.2in}
\label{fig:HT1}
\end{figure}

The final parameter to be estimated is the value of the flag $h(i)$ that defines if a monitoring AP belongs to the one-hop neighborhood of a transmitting LTE. Although $h(i)$ may be fixed  over all $i$'s for static topologies, we update it with every transmission to reflect channel fluctuations. The importance of $h(i)$ is shown
in Fig.~\ref{fig:HT1}. AP $K$ can overhear LTE $A$ when $A$ is active. Contrary, the received power at $A$ falls below $A$'s CCA threshold when $K$ is transmitting because $K$ transmits at lower power than $A$. Moreover, although $A$ and $B$ are both overheard at $K,$ they are hidden terminals to each other. These topological configurations impact how $K$ estimates the freeze time of a monitored LTE.   Let LTE $A$ be monitored by AP $K$. To determine if $K$ is overheard by $A$, i.e., $K \in \mathcal{N}_{A}^{(1)}$, the AP executes  the following algorithm.  

\medskip
\noindent \underline{\bf Algorithm 4: Inferring Neighbor APs}

\medskip
 {\bf Step 1:} A monitoring AP $K$ keeps track of the average received power over the last $z$ transmissions by eNB $A$. Let this series be represented by a $z\times 1$ vector $\mathbf{P}_{r}^{(K)}$. The $j^{th}$ element $P_{r}^{(K)}(j)$ of $\mathbf{P}_{r}^{(K)}$ is equal to 
\begin{equation}
P_{r}^{(K)}(j) = P_{\ell} |h_{A,K}(i)|^2+\sigma^2,
\end{equation}
where $h_{A,K}(j)$ denotes the channel impulse response, $P_{\ell}$ is the transmission power of the eNB, and $\sigma^2$ is the noise power.

 {\bf Step 2:} AP $K$ exploits the channel reciprocity principle $(h_{K,A}(j) = h_{A,K}(j))$ to estimate the received power at $A$ when $K$ transmits. AP $K$ generates a $z\times 1$ vector $\mathbf{P}_{r}^{(A)}$ that represents the received power at $A$, if $K$ were to transmit using power $P_{w}$ over the same channel. The $j^{th}$ average received power value at eNB $A$ is computed as
\begin{equation}
P_{r}^{(A)}(j) = P_{w} |h_{K,A}(j)|^2 +\sigma^2 = \frac{P_{w} (P_{r}^{(K)}(j)-\sigma^2)}{P_{\ell}}+ \sigma^2.
\label{pra}
\end{equation}
The number of observations $z$ is chosen based on the channel fading conditions. For slow fading channels, $z$ should take small values, whereas longer observation times are needed if the AP experiences a fast fading channel.

{\bf Step 3:} If the majority of the power samples in $\mathbf{P}_{r}^{(A)}$ exceed the CCA threshold, AP $K$ considers itself a member of the one-hop neighborhood   $\mathcal{N}_{\text{$A$}}^{(1)}$ of $A$ and sets $h(i)$ to zero. Otherwise, $K$ is a hidden terminal to  $A$, and sets flag $h(i)$  to one.  


\section{Behavior Evaluation Phase}
\label{Esti}

\begin{figure}[t]
\begin{center}
\includegraphics[width=1.0\linewidth]{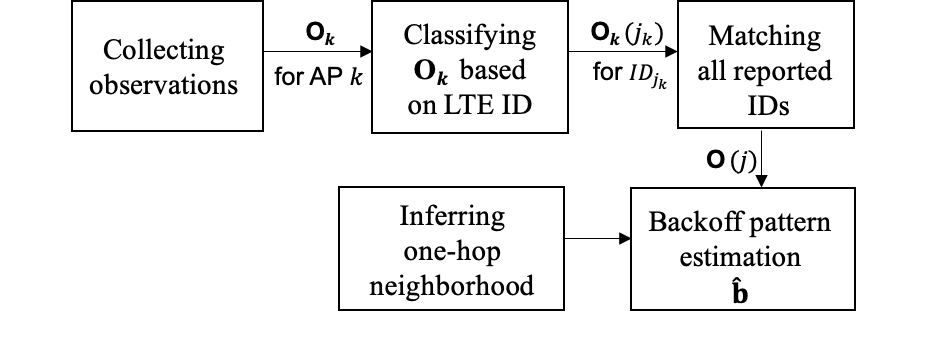}
\end{center}
\vspace{-0.2in}
\caption{Overview of the behavior evaluation phase.}
\vspace{-0.2in}
\label{fig:flow2}
\end{figure}

The behavior monitoring phase is followed by the behavior evaluation phase where the central hub models and analyzes the behavior of each eNB based on the collected observations. An overview of the behavior evaluation phase is shown in Fig.~\ref{fig:flow2}.
The hub first integrates the data from the different APs into a single observation set. Subsequently, the observation set is analyzed to  detect any LTE misbehavior.

\subsection{Integration of the AP Observations}

The first task in the  evaluation phase is to attribute the reported observations by the multiple APs to eNBs. Each AP independently associates the observations with unique  ID fields. Recall that no LTE frame decoding takes place, so the real LTE ID is not recorded in the observation sets reported by the APs. To match the unique IDs, the central hub exploits the timing reported with each observation. The intuition here is that LTE transmissions recorded by multiple APs will share common start and end times. The hub matches the different LTE ID versions using the following steps.

\medskip
\noindent \underline{\bf Algorithm 5: Matching the Reported LTE ID Fields}

{\bf Step 1:}
Let $\Om_K$ be the set of observations reported by a monitoring AP $K$ to the hub. The hub partitions $\Om_K$  based on the reported ID field, such that the observations tagged with $ID_{j_K}$ are included in subset $\Om_K(j_K)\subset \Om_K$.

{\bf Step 2:} The hub utilizes the start and end times within observation subsets from different APs to match the LTE IDs. These times are almost synchronous when the reported observations from different APs represent the same eNB. For any two LTE IDs $ID_{j_{K}}$ and $ID_{j_{L}}$ reported by monitoring APs $K$ and $L$, respectively, if the start and end times of most observations within subsets $\Om_K(j_K)$ and $\Om_L(j_L)$ are identical, then $ID_{j_{K}}$ (reported by AP $K$) and $ID_{j_{L}}$ (reported by AP $L$) represent the same eNB. The $i^{th}$ observation from $K$ and $\ell^{th}$ observation from $L$ shall satisfy the following conditions:
\begin{equation}
\label{eq:t_s_mat}
    |t_s(i)-t_s(\ell)|\leq \epsilon,~~~t_e(i)- t_s(i) = t_e(\ell) - t_s(\ell).
\end{equation}
That is, the frame start times recorded by the two monitoring APs should not differ by more than $\epsilon$ and the frame length should be the same. Parameter $\epsilon$ represents the synchronization error due to differences in propagation delay, and clock offsets. As the two APs do not necessarily collect the same number of observations, the previous relation should be true for some fraction of observations in $\Om_{K}(j_{k})$ and $\Om_{L}(j_{\ell})$.

{\bf Step 3:} The hub merges the observation sets that correspond to the same LTE ID. Specifically, if  subsets $\Om_{K}(j_{k})$ and $\Om_{L}(j_{\ell})$ are attributed to the same ID based on Step 2, the two sets are merged into a single one as follows: (a) any unique observation is retained intact and (b) for a duplicate observation $\ov(i)$ only one copy is retained, except for the hidden terminal flag $h(i)$. The flag $h(i)$ is extended to a vector $\hv(i)$ that includes the different $h(i)$'s reported by APs. 
\begin{table}
\caption{Observations reported by  AP 1 and AP 2.}
\centering
\begin{tabular}{ | P{0.7cm} | P{7.5cm}|} 
\hline
$i$& Observations reported by AP 1 ($\Om_1$) \\ 
\hline
$1$ & $<10, 150, ID_{1}, C_1(1), r_1(1), h_1(1)>$\\ 
\hline
$2$ & $<160, 300, ID_{2}, C_1(2), r_1(2), h_1(2)>$   \\ 
\hline
$3$ & $<320, 500, ID_{2}, C_1(3), r_1(3), h_1(3)>$   \\ 
\hline
$4$ & $<510, 650, ID_{1}, C_1(4), r_1(4), h_1(4)>$  \\ 
\hline
$\ell$ & Observations reported by AP 2 ($\Om_2$)\\ 
\hline
$1$  & $<160.1, 300.1, ID_{3}, C_2(1), r_2(1), h_2(1)>$   \\ 
\hline
$2$ &  $<320.1, 500.1, ID_{3}, C_2(2), r_2(2), h_2(2)>$   \\ 
\hline
$3$ & $<550.1, 700.2, ID_{4}, C_2(3), r_2(3), h_2(3)>$  \\ 
\hline
$4$ & $<720.1, 820.1, ID_{4}, C_2(4), r_2(4), h_2(4)>$  \\ 
\hline
\end{tabular}
\label{table:T2}
\vspace{-0.1in}
\end{table}

\medskip
To illustrate Algorithm 5, consider the following example. Let AP 1 and AP 2 report the observation sets shown in Table \ref{table:T2}. First, the hub separates the observations of each AP into subsets based on the ID field. This creates subsets $\Om_1(1)=\{\ov_1(1),\ov_1(4)\}$ and  $\Om_1(2)=\{\ov_1(2),\ov_1(3)\}$ for AP 1 and subsets $\Om_2(3)=\{\ov_2(1),\ov_2(2)\}$ and $\Om_2(4)=\{\ov_2(3),\ov_2(4)\}$ for AP 2. Next, the hub checks if there is any matching between the four reported ID fields by applying the checks in \eqref{eq:t_s_mat} on the observations of each subset. Based on the reported timings, the hub matches subset $\Om_1(2)$ with $\Om_2(3)$. This is because the respective observations have almost identical start and end times. Also, we observe no matching for  $\Om_1(1)$ and $\Om_2(2)$. The hub concludes that there are three different ID fields reported by the two APs, namely  $ID_1, ID_2$, and $ID_4$. The last step is to merge the observations within $\Om_1(1)$ and $\Om_2(2)$ in a new observation set. This is done by keeping only one copy for each repeated observation and expanding the hidden terminal flag to a vector. For example, $\ov_1(2)$ and $\ov_2(1)$ are merged into following observation:
\[<160T_s, 300T_s, ID_2, C_1(2), r_1(2), \hv(i)=\{h_1(2),h_2(1)\}.
\]

We emphasize that $C_1(2)$ and $r_1(2)$ should be the same as $C_2(1)$ and $r_2(1)$, respectively. Once the hub matches all reported IDs, it analyzes the behavior of each eNB individually.  Without loss of generality, we focus on the behavior evaluation of a single eNB. The same process is repeated for other eNBs. 

Let the observation set for the eNB be $ \Om = \{\ov(1), \ov(2),\ldots, \ov(n)\}$.   One vital step in evaluating the LTE behavior is the identification of the one-hop neighborhood for each eNB. This is important to estimate when a given eNB should freeze its contention process relative to other active eNBs and APs.  The topological information is inferred at the central hub using the following process.

\medskip
\noindent \underline{\bf Algorithm 6: Inferring the One-hop Neighborhood of an eNB}

{\bf Step 1:} For the $i^{th}$ observation, the hub uses the reported  vector $\hv(i)$ to identify the APs that belong to the one-hop neighborhood of the eNB that is analyzed. Let $\hv(i,k)$ be the flag reported by AP $K$. The hub places $K$ to the one-hop neighborhood of the eNB if $\hv(i,k)=0$.

 {\bf Step 2:} For eNBs, the hub utilizes the reported start and end times  for each monitored LTE to identify the intervals in which each accesses the channel. If the number of overlapping transmissions between two eNBs exceeds a threshold $\gamma_{int}$,   the hub concludes that the involved eNBs are not in range of each other.
 
{\bf Step 3:} If the number of  overlapping transmissions is below $\gamma_{int}$, the eNBs are considered to be within interference range and are placed in the one-hop neighborhood of each other. 
\medskip

\begin{figure}[t]
\begin{center}
\includegraphics[width=1.0\linewidth]{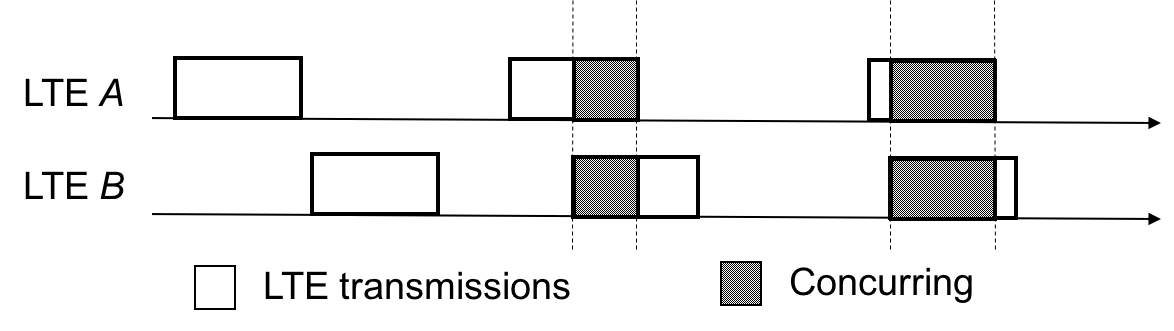}
\end{center}
\vspace{-0.1in}
\caption{Transmissions of two eNBs which are hidden terminals.} 
\vspace{-0.2in}
\label{fig:HT2}
\end{figure}

 To demonstrate Algorithm 6, consider the transmission timeline shown in Fig.~\ref{fig:HT2}. The  two transmitting eNBs are not within range and therefore several frames overlap in time, indicating that one is not aware of the other's transmissions.

\subsection{Backoff Pattern Estimation}

Consider the behavior analysis of an eNB $A$. After the observation set $\mathbf{O}$ and the one-hop neighborhood $\mathcal{N}_{A}^{(1)}$ of $A$ have been determined, the hub performs the following steps to estimate the backoff pattern $\widehat{\bv}$ of the eNB. 

\medskip
\noindent \underline{\bf Algorithm 7: Backoff Pattern Estimation}

 {\bf Step 1:}  The hub computes the inter-transmission time between two successive transmissions $\ov(i-1)$ and $\ov(i)$ as:
\begin{equation}
\label{boest}
    T(i) = t_s{(i)}-t_e{(i-1)},
\end{equation}
where $t_e{(i-1)}$ and $t_s{(i)}$  are the end and start times reported during the $\ov(i-1)$ and $\ov(i)$ observations, respectively.

 {\bf Step 2:}  Let $v_i$ denote the number of all intermediate transmissions that occur between $\ov(i-1)$ and $\ov(i)$, from stations in the one-hop neighborhood $\mathcal{N}_{A}^{(1)}$.  The hub computes $v_i$ by tracking all observations that have a starting time $t_s$ such that $t_e{(i-1)}<t_s<t_s{(i)}$ and belong to  $\mathcal{N}_{A}^{(1)}$.

  {\bf Step 3:} According to the LAA-LTE backoff process, the time $T(i)$ between two successive transmissions consists of defer, freeze, and backoff times and can be expressed as:
\begin{equation}
\label{bo0}
{T}(i) =\underbrace{ \sum_{j=1}^{v_i+1} (T_{def} + p_j\cdot T_s )}_{\text{defer time}}+ \underbrace{\sum_{j=1}^{v_i} L_j{(i)}}_{\text{freeze time}} +\underbrace{{b}(i)\cdot T_s}_{\text{backoff time}}.
\end{equation}
In eq. \eqref{bo0}, $T_{def}$ is the default defer time followed after every transmission,  $p_j$ is the number of defer slots  before the $j^{th}$ intermediate transmission and $L_j{(i)}$ is the length of the $j^{th}$ intermediate transmission. Recall that $v_i$ is the number of intermediate transmissions and $T_s$ is the slot duration. We emphasize that  the collision of more than one transmissions is registered as only one intermediate transmission whose length is the combined interval of the colliding transmissions. 

  {\bf Step 4:} Let $p(i)$ be the number of observation slots related to the class $C(i)$ (see Table~\ref{table:T1}) reported within the observation $\ov(i)$. The hub computes the defer slots $p_j$ as
  \begin{equation}
        p_j=\min\{p(i),\frac{T_j-T_{def}}{T_s}\},
\end{equation} 
    where $T_j$ is the idle time before the $j^{th}$ intermediate transmission. This means that $p_j$ is calculated based on one of the following two scenarios
\begin{itemize}
\item If the channel stays idle until the  $p(i)$ observation slots have passed $\left(\frac{T_j-T_{def}}{T_s}\geq p(i)\right)$, then $p_j=p(i)$ .
    \item When another device starts transmitting before $p(i)$ slots are observed $\left(\frac{T_j-T_{def}}{T_s}<p(i)\right)$, then $p_j=\frac{T_j-T_{def}}{T_s}<p(i)$.
    \end{itemize}
    
  {\bf Step 5:} The hub estimates the backoff counter $b(i)$ from \eqref{bo0}:
 \begin{equation}
 \label{eq:NfromT}
{b}(i)=\frac{{T}{(i)}-(v_i+1) T_{def}-\sum_{j=1}^{v_i+1} p_j\cdot T_s - \sum_{j=1}^{v_i} L_j{(i)}}{T_s}.
\end{equation}
Intuitively, eq. \eqref{eq:NfromT} states that the backoff counter selected by an eNB is equal to the time between two successive transmissions from that eNB minus all the defer time, and minus all the freeze time (normalized over the slot duration to convert it to slots). 
The correct estimation of ${b}(i)$ requires the knowledge of $p(i)$, which is determined according to the reported priority class $C(i)$ during the observation $\ov(i)$.

\begin{figure}[t]
\begin{center}
\includegraphics[width=1.0\linewidth]{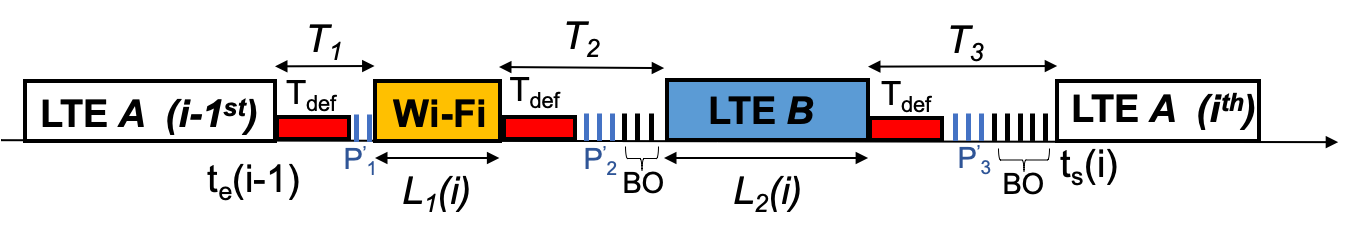}
\end{center}
\vspace{-0.1in}
\caption{Estimation of the $i^{\text{th}}$ backoff counter between two successive transmissions from LTE $A$.} 
\vspace{-0.2in}
\label{fig:f1}
\end{figure}

An example of all the timings involved in the estimation of $b(i)$ is shown in Fig.~\ref{fig:f1}. The time between two successive transmissions from LTE $A$ is $T(i) = t_s(i) - t_e(i-1)$. Two intermediate transmissions occurred during $T(i)$ from devices in the one-hop neighborhood of $A$. The first transmission was from a Wi-Fi station and the second from another LTE, so $v_i$ is set to two. The backoff is computed by reducing $T(i)$ by the duration of the two intermediate transmissions (freeze time) and the defer time before each transmission ($T_{def}$+observation slots). The  Wi-Fi transmission duration $L_1(i)$ is known because Wi-Fi APs report their own activity to the hub (see Fig.~\ref{fig:LTE_UL}), whereas the duration $L_2(i)$ for LTE $B$ is implicitly sensed and reported by APs. Finally, $p_j$ is inferred using Step 4.

{\bf Unsaturated LTE Traffic:} It is worth noting that the backoff counter in \eqref{eq:NfromT} is accurate under saturation conditions where idle slots only exist due to the backoff and channel sensing processes. However, if traffic is not saturated, idle slots can inflate the backoff estimation. To avoid misdetection due to backoff inflation, the hub excludes all observations that definitively include idle slots outside those related to the backoff process. To identify these observations, we rely on the estimated backoff counter values from \eqref{eq:NfromT}. Let $q(i)$ and $b(i)$ be the estimated CW size and backoff counter for the $i^{th}$ observation. The hub excludes $\ov(i)$ if the estimated backoff exceeds the contention window (i.e., ${b}(i)>q(i)-1.$) This is because it is expected that the idle slots due to an empty transmission queue will far exceed the small values taken by the backoff counter.  

This process will filter most of the unwanted observations where a correct backoff estimate cannot be made, especially for low levels of saturation. We emphasize that eliminating the observations that belong to unsaturated conditions provides a misbehavior opportunity to the LTE, which can reduce its defer and backoff time, once a frame arrives at its transmission queue. However, as we show through simulations, this misbehavior has a limited impact on the Wi-Fi performance due to low contention.

\subsection{LTE Misbehavior detection}
\label{Detec}

By processing each observation in set $\Om$ using Algorithm 7, the hub recovers
the estimated backoff pattern $\widehat{\bv}$ for a monitored eNB. This pattern is used to evaluate  the LTE behavior as follows. The hub creates two distributions $\mathbf{M}$ and $\mathbf{W}$ representing the observed and the expected backoff counter distributions, respectively. Distribution $\mathbf{M}$ is the empirical distribution obtained from the appearance frequency of each backoff counter value in $\widehat{\bv}$. The density function of $\mathbf{M}$ is expressed as:
\begin{equation}
\label{eq:RT_dis}
P_{\mathbf{M}}(x)=\frac{\sum_{i=1}^{n} I({b}(i)=x)}{n}, \quad x\in \{b_{\min},\dots, b_{\max}\},
\end{equation}
where $I(\cdot)$ is the indicator function, and $b_{\min}$ and $b_{\max}$ are the minimum and maximum backoff counters found in $\widehat{\mathbf{b}}$.

The expected backoff counter distribution for a protocol-compliant node is then calculated from the contention window values used in every transmission. Those can be extracted from observation set $\Om$, which contains the retransmission round $r(i)$ and class $C(i)$ for each of the $n$ observations. The contention window for the $i^{th}$ transmission   is:
\begin{equation}
\label{cw}
q(i) = \min\{2^{r(i)}q_{\min}, q_{\max}\},
\end{equation}
where $q_{\min}$ and $q_{\max}$ are the minimum and maximum allowed CW sizes for the reported class $C(i)$. Using \eqref{cw}, the hub estimates  vector  $\widehat{\qv}$ of all contention windows. A protocol-compliant eNB should choose each backoff counter at random within each contention window, leading to a density function for the following backoff counter distribution. 
\begin{equation}
\begin{split}
P_{\mathbf{W}}(x)&=\sum_{k\in\mathcal{N}_{cw}} \Pr(q=k)\cdot \Pr(x|q=k),
\end{split}
\end{equation}
where $\mathcal{N}_{cw}=\{4,8,\dots,1024\}$ is the set of all possible contention window sizes and
$\Pr(q=k)$ is the probability that the LTE uses a CW of size  $k$. Taking into account that the backoff counter selection  is uniform regardless of the CW size, we get,
\begin{equation}
\begin{split}
P_{\mathbf{W}}(x)&=\sum_{k\in\mathcal{N}_{cw}} \Pr(q=k)\cdot \frac{1}{k}.
\end{split}
\end{equation}
To determine the probability $\Pr(q=k)$, the hub relies on the frequency of appearance of value $k$ in vector $\widehat{\qv}$, which can be written as:
\begin{equation}
\label{eq:RT_dis2}
\Pr(q=k)=\frac{\sum_{i=1}^{n} I(q(i) = k)}{n}.
\end{equation}

To detect a deviation from the expected behavior, the hub measures the distance between the observed backoff counter distribution $M$ and the expected distribution based on the CW sizes $W$.  The distance is measured through the Jensen-Shannon divergence defined as
\begin{equation}
D_{JS}(\mathbf{M} ||  \mathbf{W})\triangleq \frac{1}{2}D(\mathbf{M}  ||  \mathbf{C})+\frac{1}{2}D(\mathbf{W}  ||  \mathbf{C}),
\end{equation}
where $D(\cdot||\cdot)$ is the Kullback-Leibler divergence, and 
 $\mathbf{C}=\nicefrac{1}{2}( \mathbf{M}+ \mathbf{W}).$
An eNB is suspected of misbehavior if $
D_{JS}(\mathbf{W} ||  \mathbf{M})>\delta,
$
where $\delta$ is a threshold specified by the hub.





\section{Validation of Implicit Techniques}
\label{Exp_val}

To evaluate the performance of our proposed misbehavior detection framework, we first evaluate the accuracy of the implicit  LTE monitoring techniques proposed in Section~\ref{Implicit}. For this part, we performed experimentation using the USRP platform and measured the efficacy of extracting various LTE operation parameters without decoding using signal correlation.

\subsection{Experimental Setup} 

\begin{figure*}
\begin{center}
\setlength{\tabcolsep}{-3pt}
\begin{adjustwidth}{+0.1in}{}
\begin{tabular}{@{}cccc}
\includegraphics[height=1.4in]{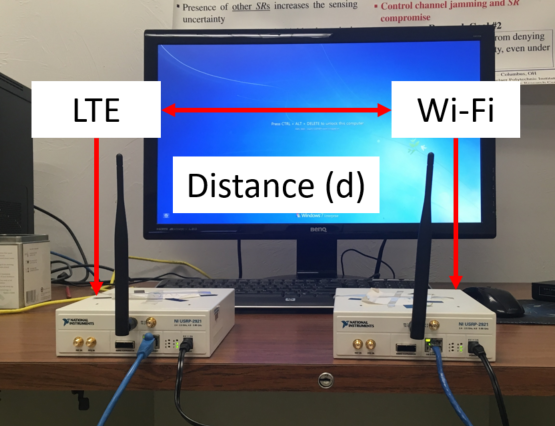} & 
\includegraphics[height=1.4in]{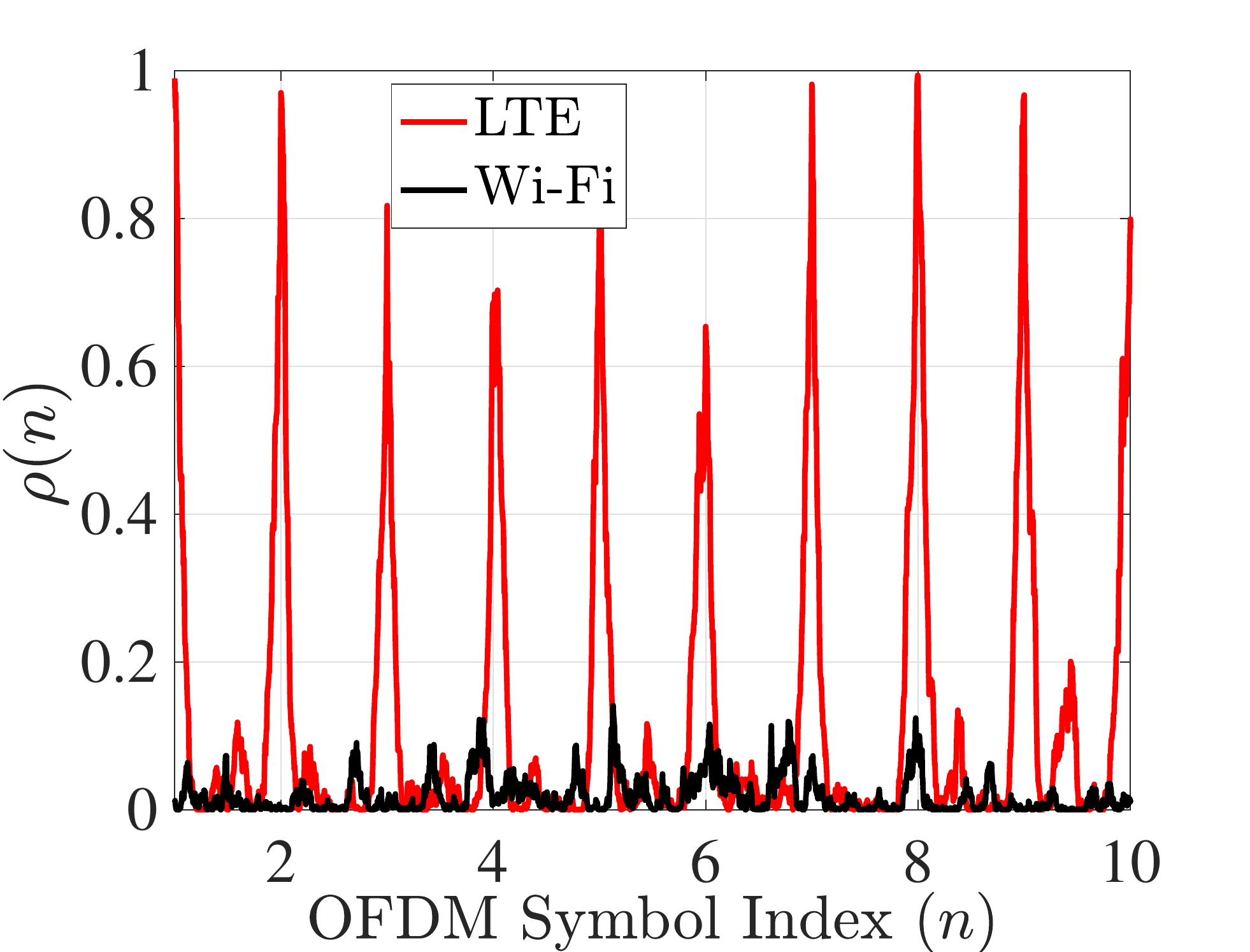} & \includegraphics[height=1.4in]{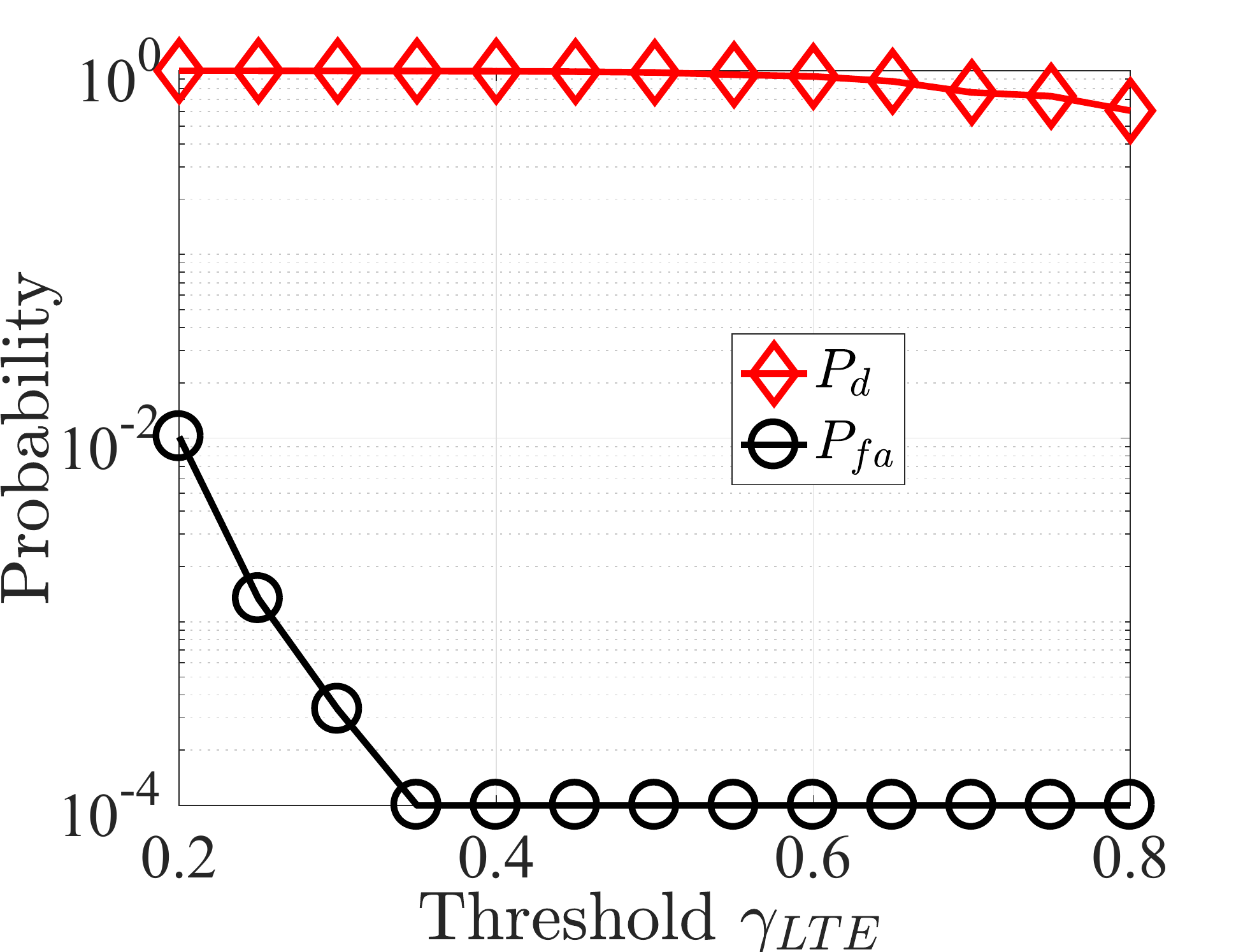} & \includegraphics[height=1.4in]{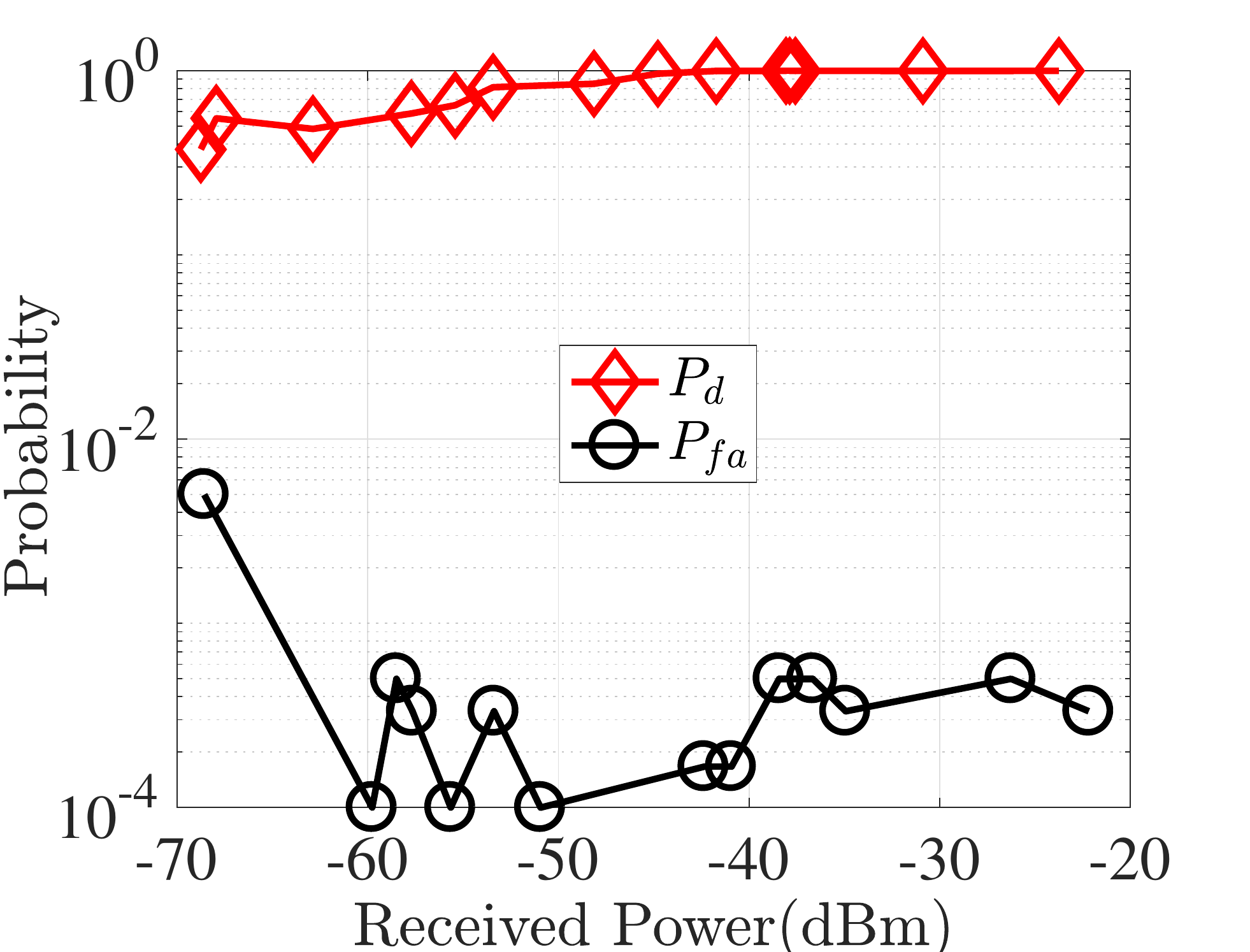} \\
(a) ~&~ (b) ~&~ (c) ~&~ (d)
\end{tabular}
\end{adjustwidth}
\end{center}
\caption{(a) Experimental setup, (b) $\rho$(n) vs. OFDM symbol index, (c) detection and false alarm probabilities as a function of the threshold $\gamma_{LTE}$, and (d) detection and false alarm probabilities  as a function of the input power at the Wi-Fi AP.}
\label{fig:SetupLTEdetection}
\end{figure*}

We set up two NI-USRP 2921 devices as a transmitting LTE and an overhearing AP, respectively. The devices were tuned to the 5 GHz U-NII band. The physical layer of the LTE device was programmed to operate according to the LTE standard.  The transmission bandwidth was set to 20MHz, whereas the IQ rate was set to 1.92MHz. The LTE frame duration was set to 10ms. Each frame consisted of 10 subframes occupying 2 slots. Each slot had a duration of 0.5ms and allowed for the transmission of 6 OFDM symbols. The duration of each OFDM symbol was set to 83.4$\mu$s of which $16.7\mu$s corresponded to the extended CP. The AP sampled the LTE signal on the same band, without implementing any further decoding. The experiment setup is shown in Fig. \ref{fig:SetupLTEdetection}(a).
\par

\subsection{Detecting LTE Transmissions}
\label{ssec:lte_det}
In the first set of experiments, we evaluated the Wi-Fi's ability in identifying the LTE signals  using the CP detection method proposed in Section \ref{Det.CP}. For LTE signals, the number of samples per data symbol was set to $L_S = 256$ and for the CP to $L_{CP}= 64$. 
The eNB continuously transmitted 6,000 OFDM symbols, repeating the sequence $\{0,1,1,0\}.$ To further  ensure that other signals are not misclassified as LTE, we repeated the experiments but configured the transmitting USRP device to send Wi-Fi OFDM symbols. Each Wi-Fi OFDM symbol had a duration of 4$\mu$s (3.2$\mu$s for the data symbol and 0.8$\mu$s for the CP). 
\begin{figure*}
\begin{center}
\center
\setlength{\tabcolsep}{-1pt}
 \begin{tabular}{ccc}
\includegraphics[height=1.4in]{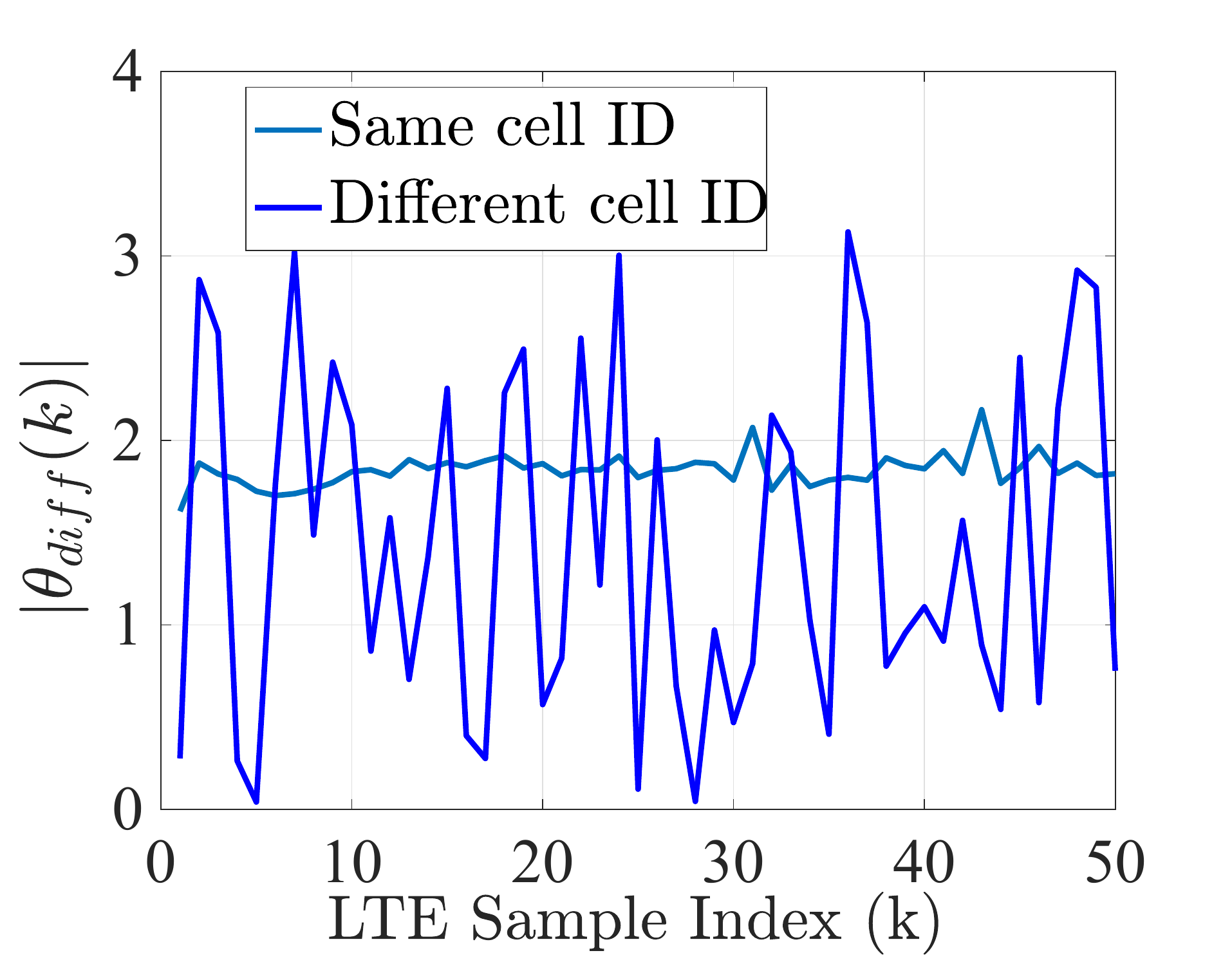} & 
\includegraphics[height=1.4in]{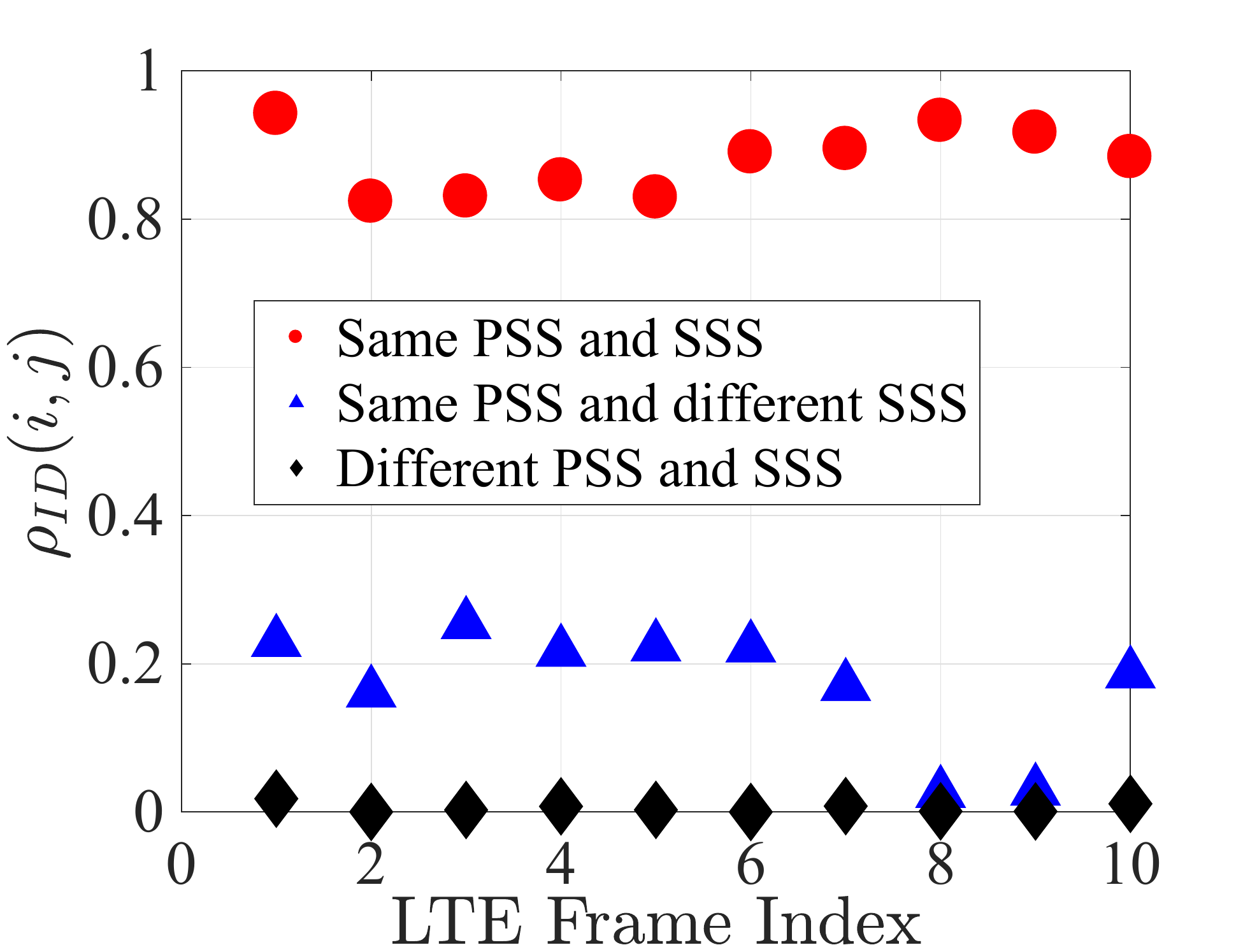} & \includegraphics[height=1.4in]{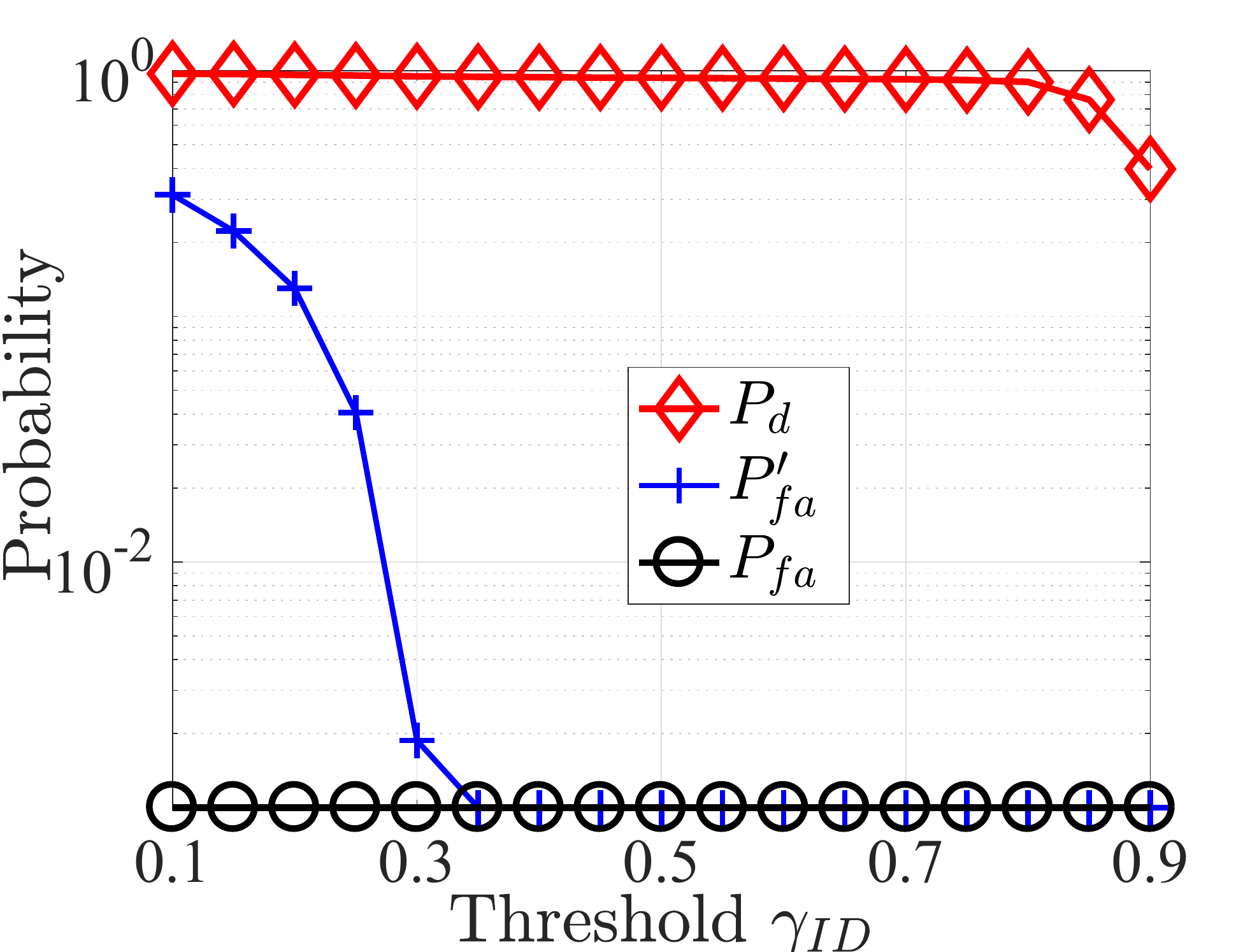} \\
(a) ~&~ (b) ~&~ (c)
\end{tabular}
\end{center}
\caption{(a) The absolute phase difference as a function of the LTE sample index, (b) correlation $\rho(i,j)$ as a function of the LTE frame index, and (c) detection and false alarm probabilities vs. the threshold $\gamma_{ID}$.}
\label{fig:LTEdetection}
\vspace{-0.1in}
\end{figure*}

The AP sampled the LTE signal and applied Algorithm 1 to compute the correlation $\rho(n)$ as a function of the shift $n$. Figure~\ref{fig:SetupLTEdetection}(b) shows sample values of $\rho(n)$ for the duration of ten LTE and Wi-Fi symbols. We observe that when an LTE transmits and the correlation windows $W_1$ and $W_2$ align with the CP and its copy, the correlation $\rho(n)$ peaks to values higher than 0.6. The peaks also denote the start time of OFDM symbols. Using the LTE CP detection parameters on Wi-Fi transmissions yields correlation values of almost zero. This is due to the different OFDM symbol length and CP length in Wi-Fi transmissions. 

In Fig.~\ref{fig:SetupLTEdetection}(c), we show the detection probability $P_{d}$ and false alarm probability $P_{fa}$ for the CP-based approach, computed over 6,000 OFDM symbols, as a function of the correlation threshold $\gamma_{LTE}$.  The $P_{d}$ was computed as the fraction of LTE symbols that were correctly classified, whereas $P_{fa}$ was calculated as the fraction of Wi-Fi OFDM symbols that were falsely classified to belong to LTE. Intuitively, a greater threshold would lower the false alarm rate but will decrease the detection probability. We observe that for thresholds in the 0.4-0.6 range, $P_{d}\approx 1$, whereas $P_{fa}=0$.  Following these experiments, we set the detection threshold to 0.4.

In the next experiment, we investigated the effect of the received signal strength (RSS) on the detection probability. We repeated the LTE signal detection experiment while moving the eNB away from the AP. In Fig. \ref{fig:SetupLTEdetection}(d), we show $P_{d}$ and $P_{fa}$ as a function of the RSS at the AP. The signal correlation threshold was set to 0.4. Even at low power levels, $P_d$ remains high whereas $P_{fa}$ remains quite low with the exception of -70dBm, which is close to the CCA threshold for detecting any activity.

\subsection{Differentiating between eNBs}
\label{subsec:LTE-ID}


In this set of experiments, we evaluated the LTE frame attribution algorithm  for classifying LTE frames to different eNBs.  In the first part of the experiments, one USRP transmitted LTE frames with the same primary and secondary synchronization signal (SSS/PSS) while the second USRP sampled the signal. The signal correlation $\rho_{ID}(i,j)$ was calculated over a total of 640 samples per frame, which is equal to the combined length of the two OFDM symbols carrying the SSS and PSS fields. 

One practical issue in correlating the sampled PSS and SSS fields over different frames is the fact that the channel changes over time. Whereas the channel attenuation could remain relatively constant for a static LTE-AP distance, the phase of the impulse response could vary more rapidly. Indeed during our experiment, we noted an almost constant phase difference between samples of the same fields  that belong to different frames of the same LTE ID.  Figure~\ref{fig:LTEdetection}(a) shows the absolute phase difference $|\theta_{diff}(k)|$ between the samples carrying the PSS and SSS fields in two consecutive frames as a function of the OFDM sample index $k$, 
\begin{equation}
 |\theta_{diff}(k)|=|\theta^{(i)}_{ID}(k)-\theta^{(i-1)}_{ID}(k)|, \quad\forall k\in[1:L_{ID}],   
\end{equation}
where $\theta^{(i-1)}_{ID}(k)$ and $\theta^{(i)}_{ID}(k)$ are the phases of the  $k^{th}$ sample in the $L_{ID}$ samples carrying both SSS and PSS fields in the $i-1^{st}$ and $i^{th}$ LTE frames, respectively. We observe a fixed shift when both frames belong to the same LTE ID. This fixed shift is due to the coherence time of the channel. The channel remains relatively constant over the transmission of one frame, but changes over multiple frames.  
On the other hand, when two frames carry different PSS and SSS fields, the phase difference is random.  To improve the eNB identification method, we apply a compensation technique for the channel response phase in Step 4 of the frame attribution algorithm, which operates as follows. 

\begin{figure*}
\begin{center}
\center
\setlength{\tabcolsep}{-1pt}
 \begin{tabular}{ccc}
\includegraphics[height=1.6in]{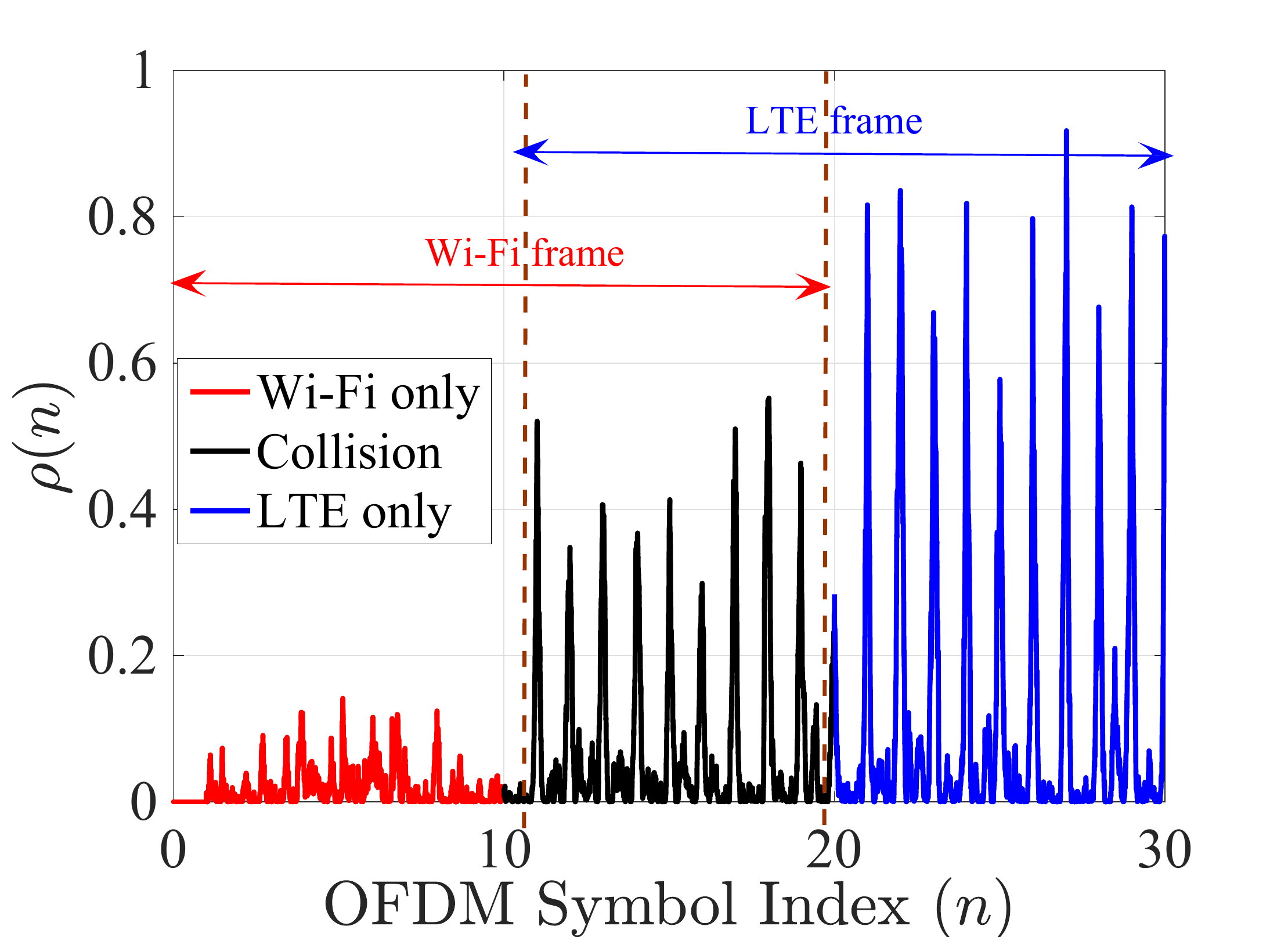} &
\includegraphics[height=1.6in]{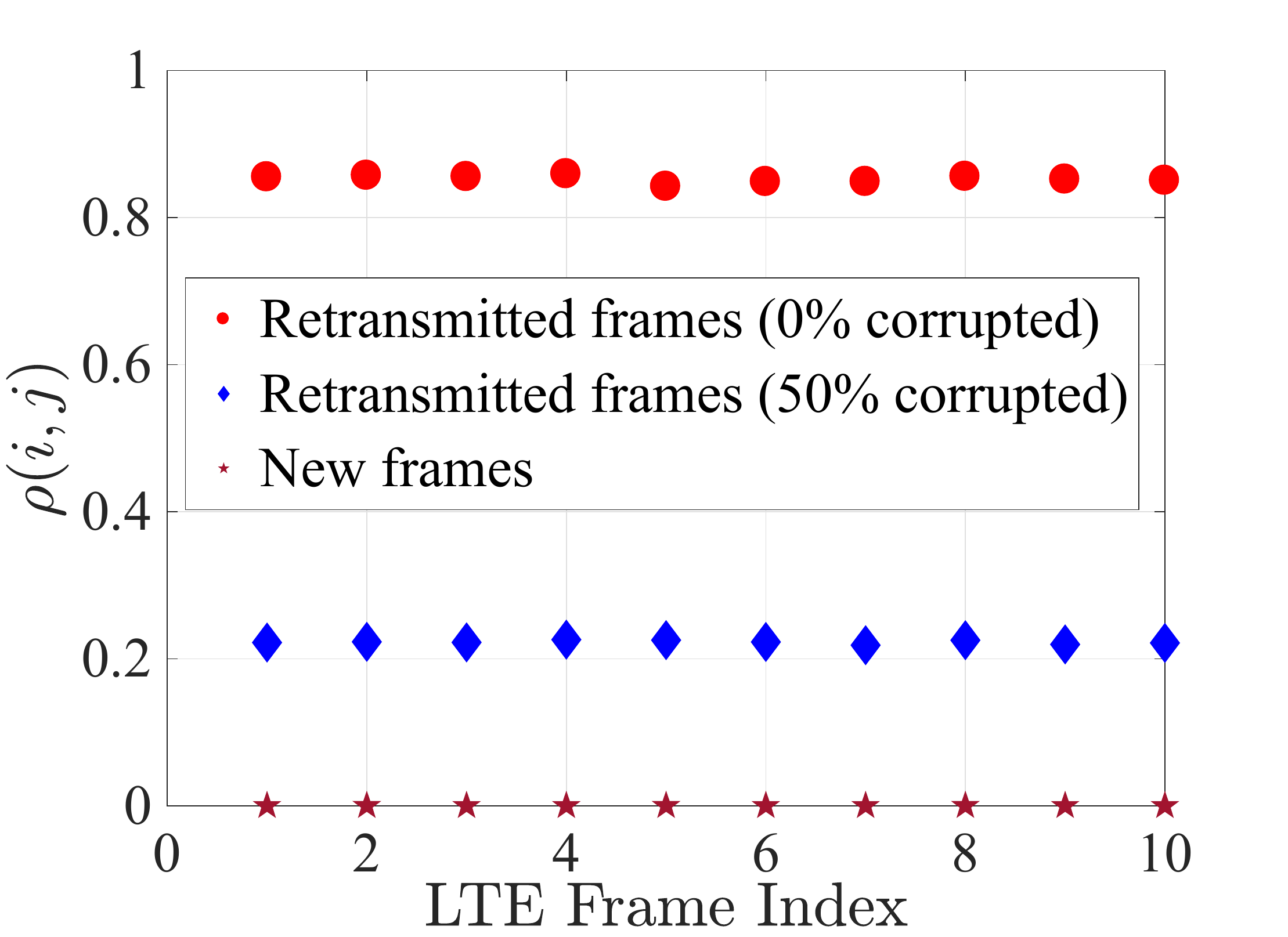} & \includegraphics[height=1.6in]{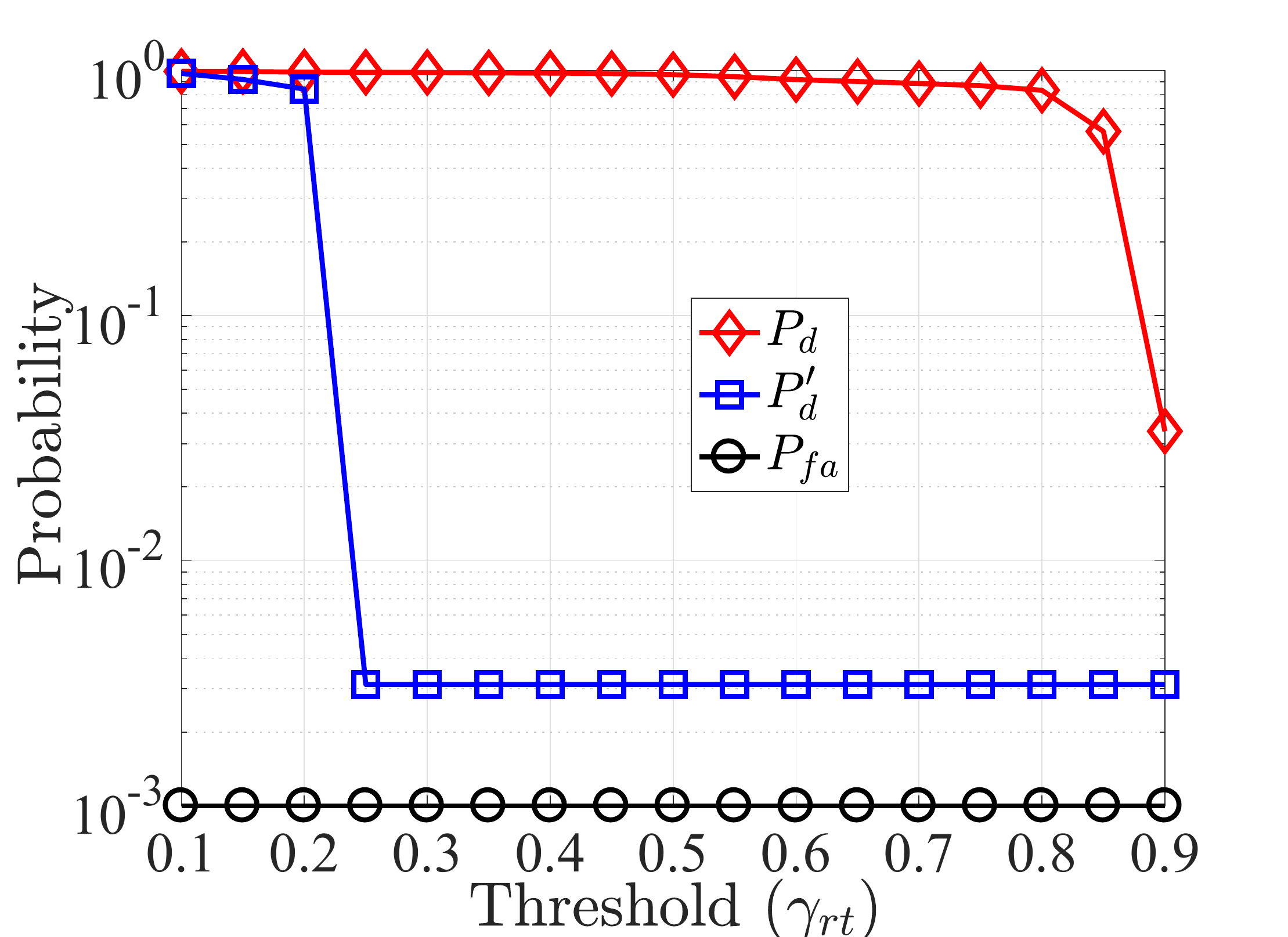} \\
(a) ~&~ (b) ~&~ (c)
\end{tabular}
\end{center}
\caption{(a) $\rho(n)$ for a collision between LTE and Wi-Fi frames, (b) correlation $\rho(i,j)$ as a function of the LTE frame index and (c) detection and false alarm probabilities as a function of the threshold $\gamma_{rt}$.}
\label{fig:CellIDdetectionandretransmissiondetection}
\vspace{-0.1in}
\end{figure*} 

\begin{itemize}
\item The AP extracts $L_{ID}=640$ samples representing $\sv^{(i)}_{ID}$, i.e., the SSS and PSS fields of the $i^{th}$ LTE frame. Denote the phases of the $L_{ID}$ samples in $\sv^{(i)}_{ID}$ by vector $\mathbf{\theta}^{(i)}_{ID}$. 
\item For each signature $\sv_{ID_{j}}$ stored in its database, the AP calculates the mean difference $\bar{\theta}(i,j)$ between $\theta^{(i)}_{ID}$ and the vector ${\theta_{ID_j}}$ denoting the phases of the corresponding  samples in  $\sv_{ID_{j}}$,
\begin{equation}
  \bar{\theta}(i,j)=\frac{1}{L_{ID}} \sum_{k=1}^{L_{ID}} |\theta^{(i)}_{ID}(k)-{\theta_{ID_j}}(k)|. 
\end{equation}
\item The  phase part of  $\sv^{(i)}_{ID}$ is compensated by $\bar{\theta}(i,j)$ as follows
\begin{equation}
\label{eq:26'}
 \theta^{(i)}_{ID}=\Big(\theta^{(i)}_{ID}+ \bar{\theta}(i,j)\Big)\mod \ \pi.
\end{equation}
\item The AP computes the correlation $\rho_{ID}(i,j)$  between $\sv^{(i)}_{ID}$ and $\sv_{ID_j}$ after the former has been phase-compensated, using equation \eqref{eq:26'}. 
\end{itemize} 
We emphasize that the proposed phase compensation method does not require decoding LTE transmissions at the AP, as all operations occur on signal samples. Fig. \ref{fig:LTEdetection}(b) shows $\rho_{ID}(i,j)$ for 10 LTE frames in the following cases: (1) frames from the same eNB, (2) frames with the same PSS (cell ID) but different SSS (eNB ID), and (3) frames with different PSS and SSS fields. For the first case, we always have a high correlation as all frames belong to the same LTE. For the second and third cases, frames belong to different eNBs, thus the correlation is much lower. In Fig. \ref{fig:LTEdetection}(c), we plot the detection and false alarm probabilities as a function of the detection threshold, we vary the threshold from 0.1 to 0.9. The false alarm $P^{\prime}_{fa}$ represents the case of the same cell ID but different SSS, whereas  $P_{fa}$ represents the case of the different PSS and SSS. 
 The value of $P_{fa}$ is almost zero even for very low thresholds. For the second case, $P^{\prime}_{fa}$ becomes almost zero when the detection threshold is selected to be higher than 0.35, whereas $P_{d}$ remains close to 1. The experiments confirm that applying the correlation technique on the PSS and SSS fields can successfully attribute LTE signals to the transmitting eNB.

 \begin{figure*}[t]
\begin{center}
 \setlength{\tabcolsep}{-0.037in}
 \begin{tabular}{ccc}
\includegraphics[trim=0.3in 0in 0.0in 0in, width=2in]{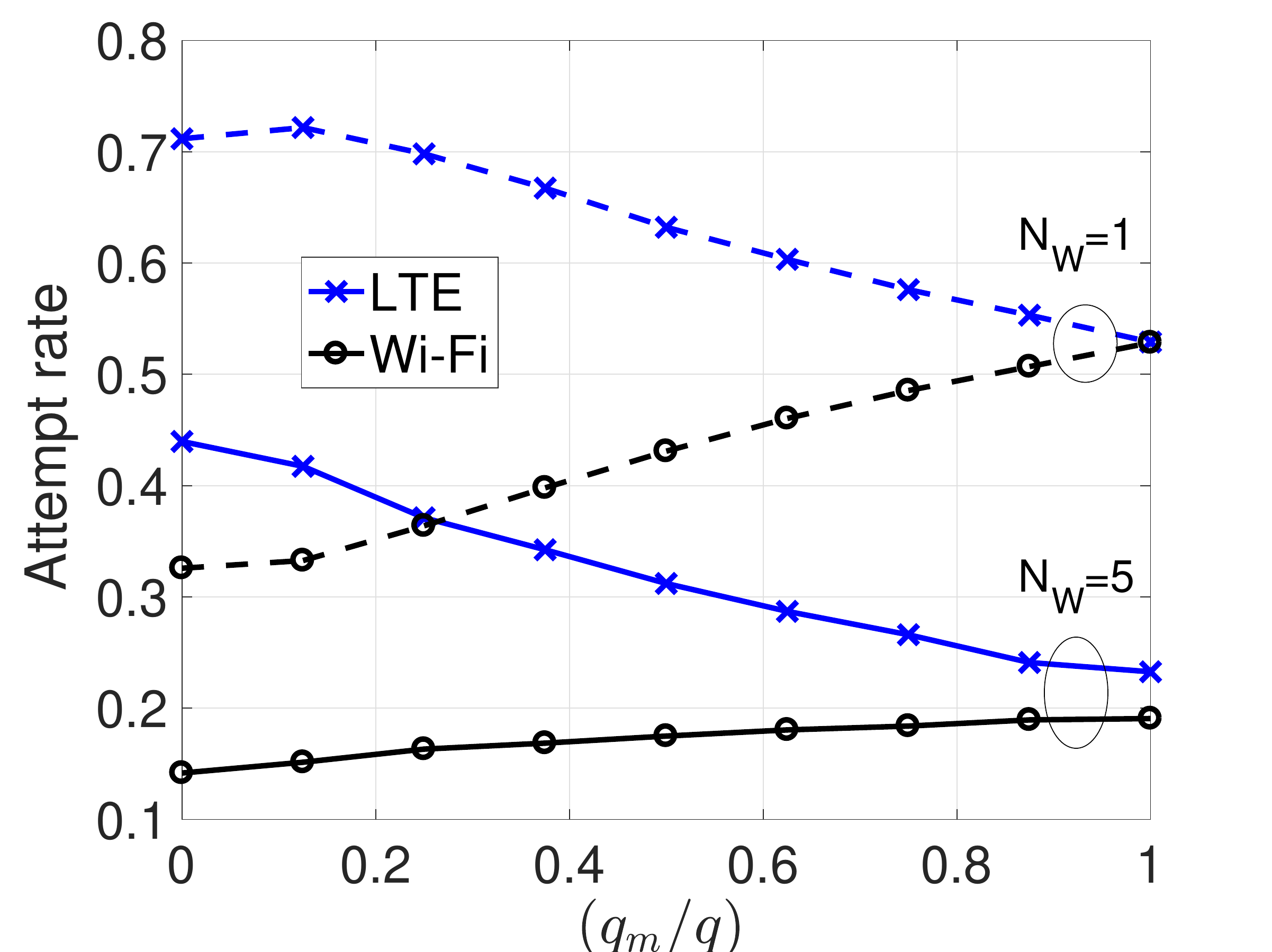} ~&\includegraphics[trim=0.3in 0in 0.0in 0in, width=2in]{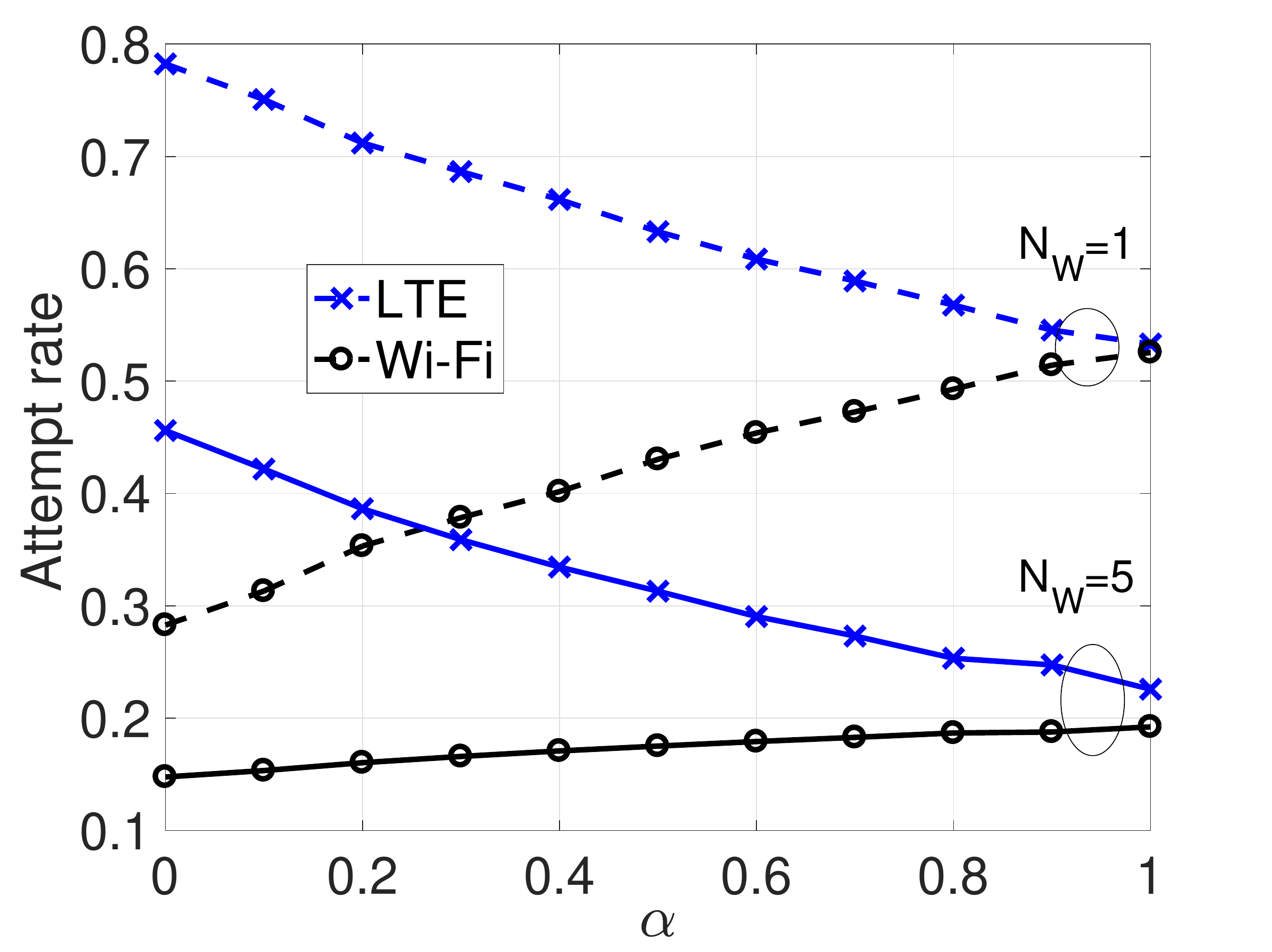} ~&\includegraphics[trim=0.3in 0in 0.0in 0in, width=2in]{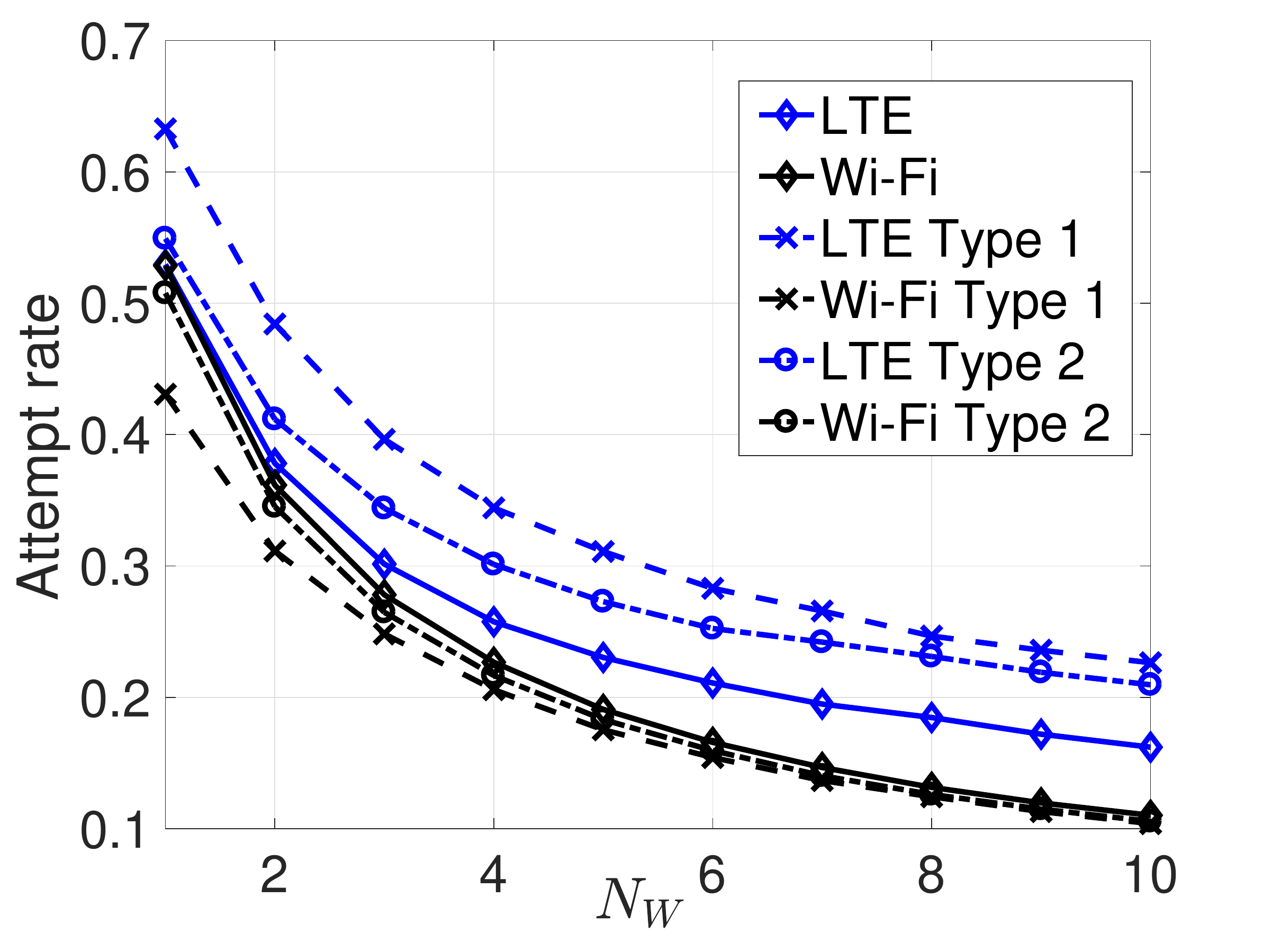} \\
(a) & (b) & (c) 
\end{tabular}  
\end{center} 
\vspace{-0.15in}
\caption{Attempt rate for LTE and Wi-Fi systems: (a)  vs.  $q_m/q$, with $N_{W}=1,5$, and $\alpha=0.5$, (b) vs. $\alpha$, with $N_{W}=1,5$, and $q_m=0.5q$, and (c) vs. number of Wi-Fi terminals, for class 3 LTE and class 3 Wi-Fi with $q_m=0.5q$, and $\alpha=0.5$. }
\label{fig:eval_perf}
\vspace{-0.15in}
\end{figure*}

\begin{figure}[t]
\begin{center}
 \setlength{\tabcolsep}{-0.037in}
 \begin{tabular}{cc}
 \includegraphics[trim=.3in 0in 0.0in 0in, width=1.82in]{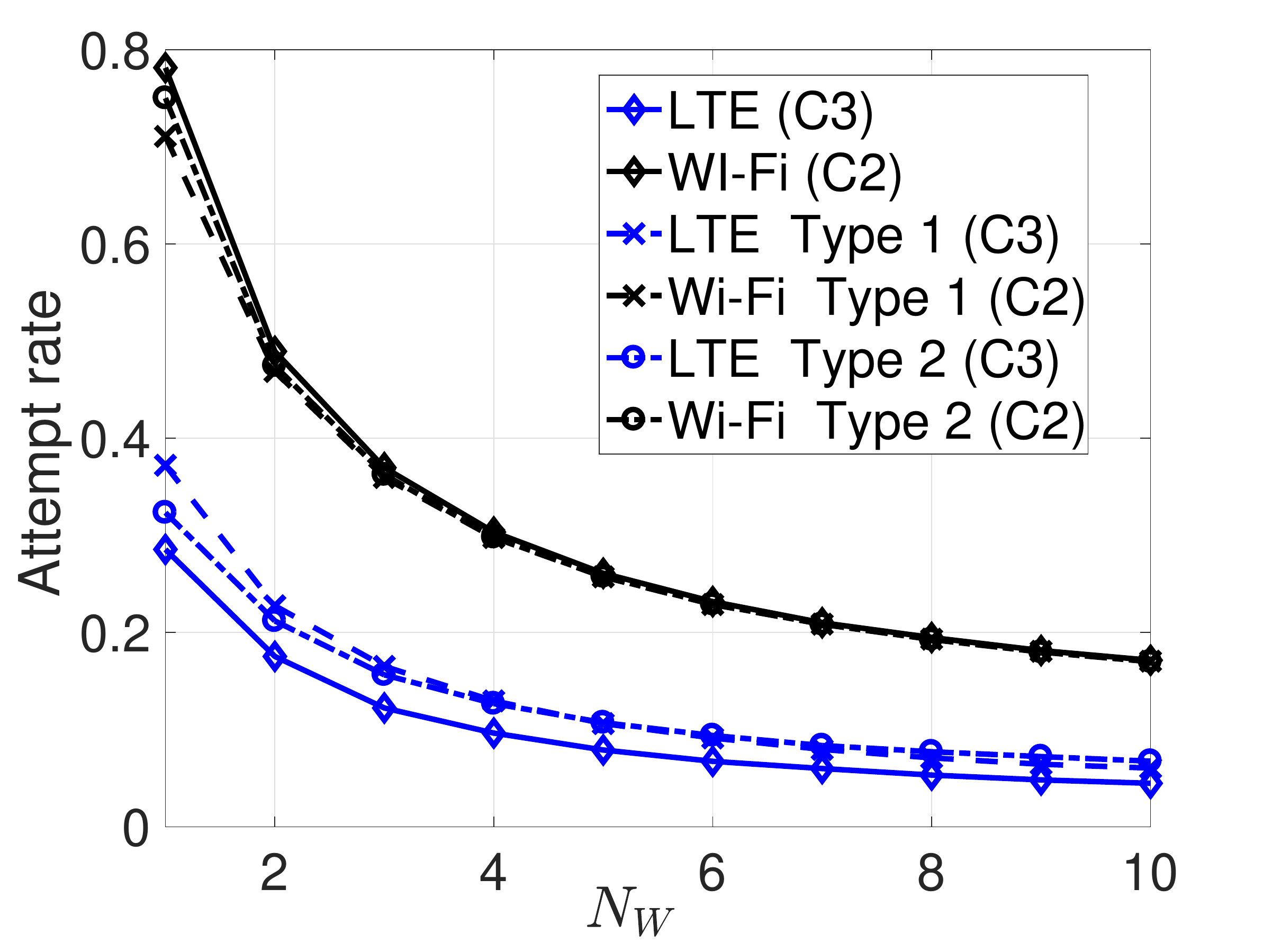} ~& \includegraphics[trim=.3in 0in 0.0in 0in, width=1.82in]{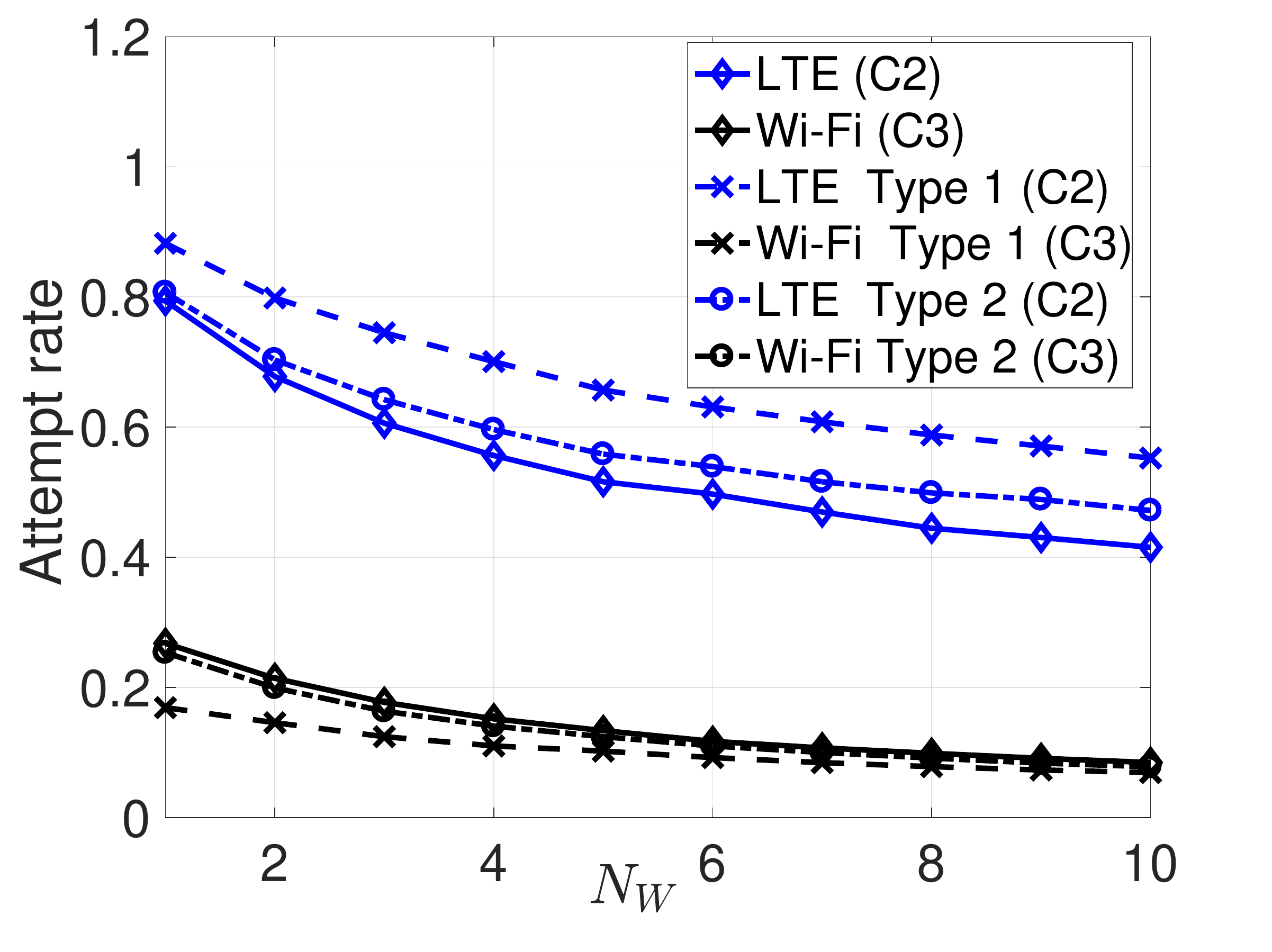} \\
(a) & (b) 
\end{tabular}  
\end{center} 
\vspace{-0.15in}
\caption{Attempt rate for LTE and Wi-Fi systems vs. number of Wi-Fi terminals for: (a) class 3 LTE and class 2 Wi-Fi with $q_m=0.5q$, and $\alpha=0.5$, and (b) class 2 LTE and class 3 Wi-Fi with $q_m=0.5q$, and $\alpha=0.5$. }
\label{fig:eval_perf2}
\vspace{-0.15in}
\end{figure}
 
\subsection{Transmission Round Estimation}

In the final set of experiments, we evaluated the transmission round estimation algorithm discussed in Section \ref{CW_est}. First, we evaluated Step 1 in Algorithm 3 by implementing a collision between LTE and Wi-Fi frames and applying the CP-based correlation method. Figure~\ref{fig:CellIDdetectionandretransmissiondetection}(a) shows the correlation $\rho(n)$ as a function of the OFDM symbol index for a sample LTE collision with Wi-Fi. We observe that once the LTE frame starts, correlation peaks start to appear. Although the correlation is not as high as the case when the samples are interference-free, it is still sufficiently high to indicate the start of the LTE frame. 

Furthermore, we implemented retransmissions of the same LTE frame and computed the signal correlation over a window of  38,400 samples, which is the length of one LTE frame. The phase compensation mechanism was also used here to account for the variations in the CIR.  The phase difference was computed over the entire frame.
Figure~\ref{fig:CellIDdetectionandretransmissiondetection}(b) shows the correlation between 10 pairs of frames when the frames in each pair are identical (retransmission due to channel impairments), identical but one is corrupted by another transmission (collision), and when they differ (not a retransmission). In the collision case, half of the samples representing the initial LTE frame are corrupted. We observe that the signal correlation between two identical transmissions is substantial enough to distinguish it from two different transmissions, even if some of the samples are corrupted. 

In Fig. \ref{fig:CellIDdetectionandretransmissiondetection}(c), we show the probability $P_{d}$ of detecting a retransmission when both the original frame and the retransmission do not collide with other frames. Moreover, we show the detection probability $P'_{d}$ of a retransmission when the original frame collided with another frame and the false alarm probability $P_{fa}$ as a function of the threshold $\gamma_{rt}$. 
The false alarm $P_{fa}$ is evaluated by changing the payload of consecutively transmitted LTE frames.  We observe that the correlation technique yields a nearly perfect detection for any threshold  less than 0.8 when the frames are not corrupted and 0.2 when the frames are corrupted. The false alarm, on the other hand, is close to zero for any threshold greater than 0.1. Selecting a threshold value equal to 0.2 allows the identification of retransmissions for both clean and corrupted frames.  The high accuracy is attributed to the large number of samples used in the computation of the correlation relative to the prior correlation mechanisms that use fewer samples. 



 \section{Performance Evaluation}
 \label{perf}
 
To validate the proposed misbehavior detection framework, we further implemented an event-based simulation for the LTE/Wi-Fi coexistence. Specifically, we deployed a  set of LTE and Wi-Fi devices in the same collision domain so that activity from every device affects the behavior of others. The eNBs followed  the LTE-LAA  standard whereas the  APs implemented the IEEE 802.11ac protocol.  LTE protocol parameters were considered  perfectly detected using the proposed implicit techniques. To isolate the impact of misbehavior, frame losses occurred only due to collisions (perfect channel conditions). 
Each experiment was run for 100,000 events, where each event corresponds to a transmission attempt by any device.  For each device, we evaluated the  transmission attempt rate defined  as the number of times a device tried to transmit (backoff reached zero) including collisions,  over the total number of attempts by any device. This metric indicates the success rate in  seizing the common medium. We further evaluated the detection and false alarm probabilities, $P_d$ and $P_{fa}$, under different misbehavior scenarios. 

\begin{figure*}[t]
\begin{center}
 \setlength{\tabcolsep}{+0.015in}
 \begin{tabular}{cccc}
\includegraphics[trim=0.3in 0in 0.0in 0in, width=2in]{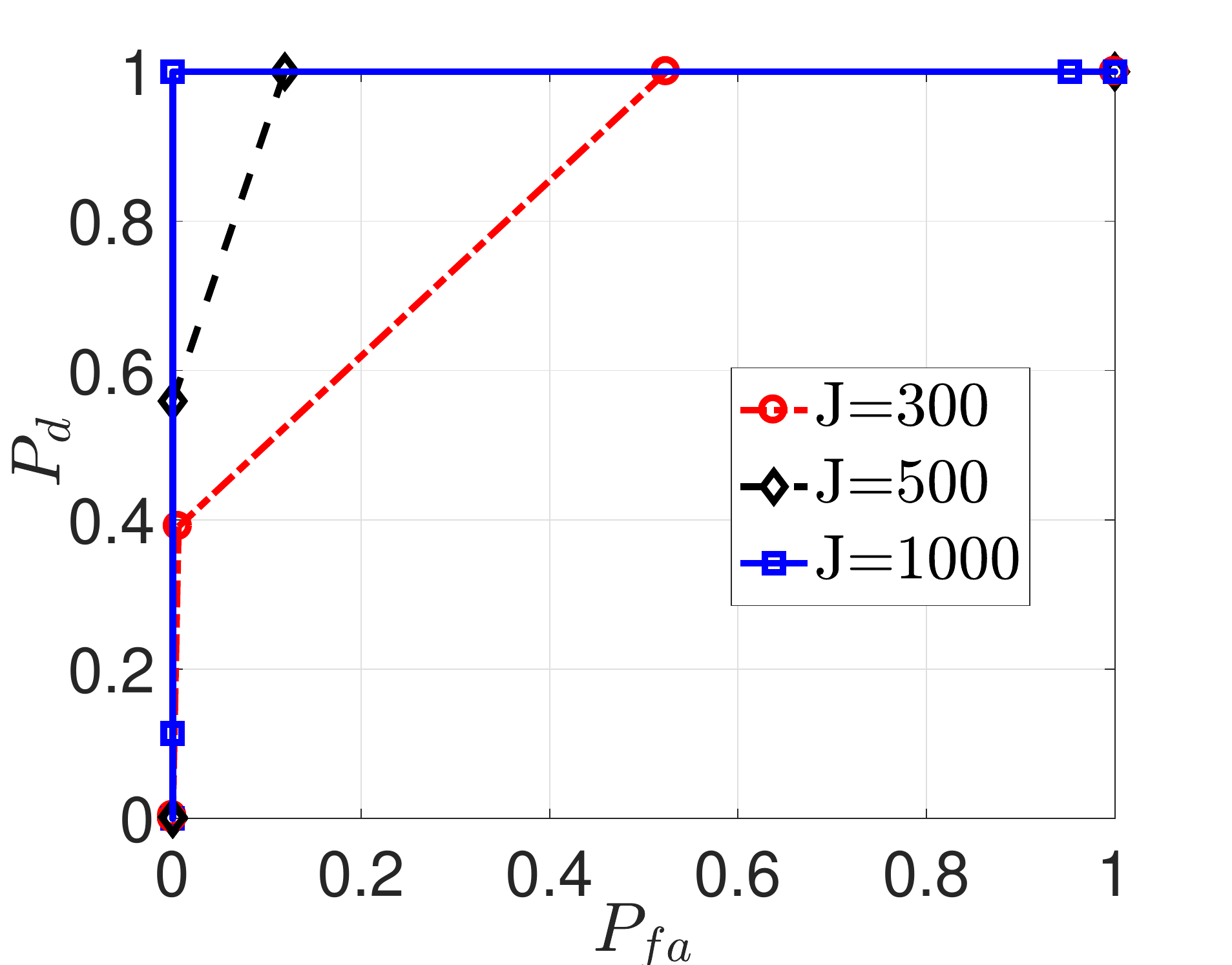} ~&\includegraphics[trim=0.3in 0in 0.0in 0in, width=2in]{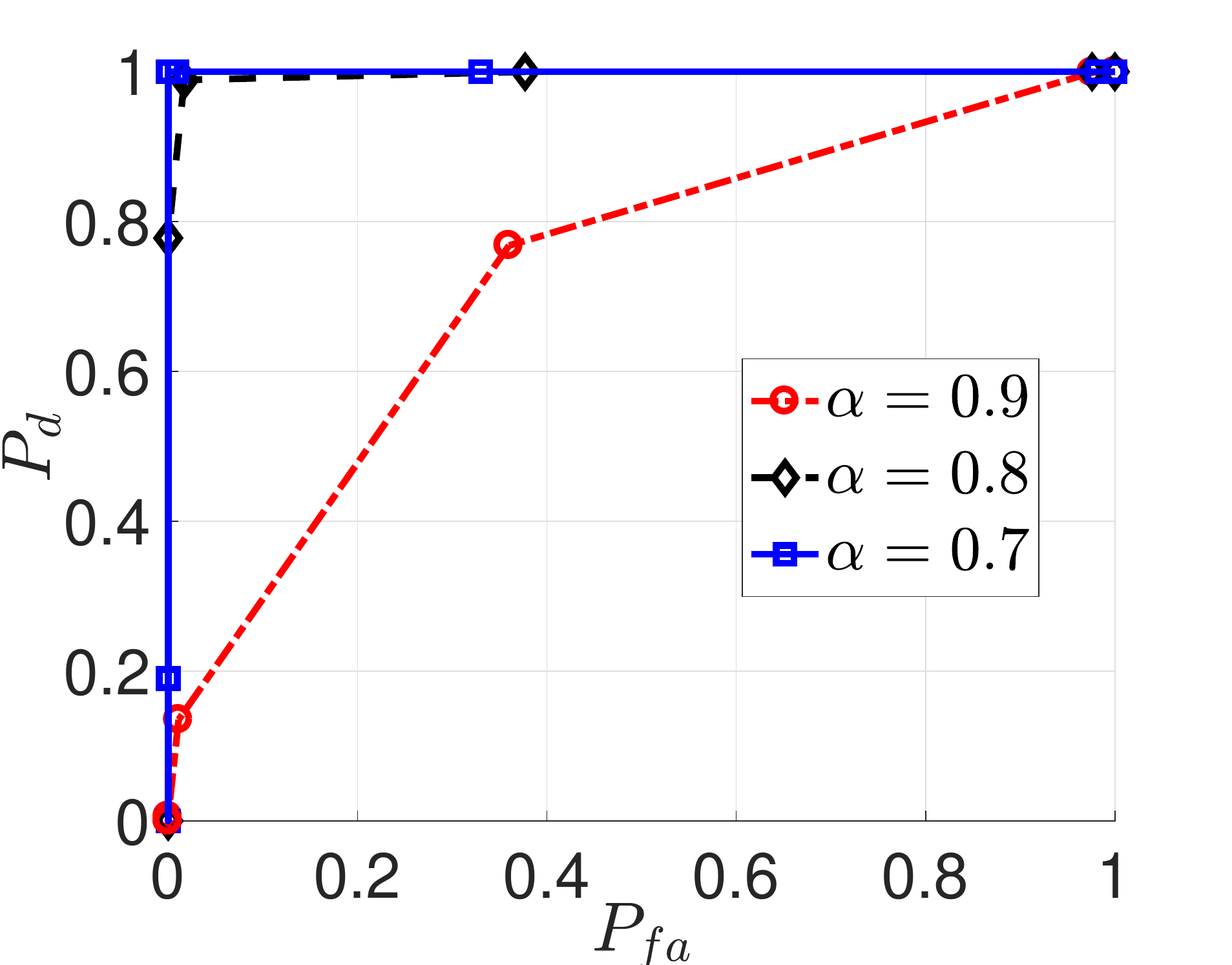}
\hspace{+0.015in}~&\includegraphics[trim=0.3in 0in 0.0in 0in, width=2in]{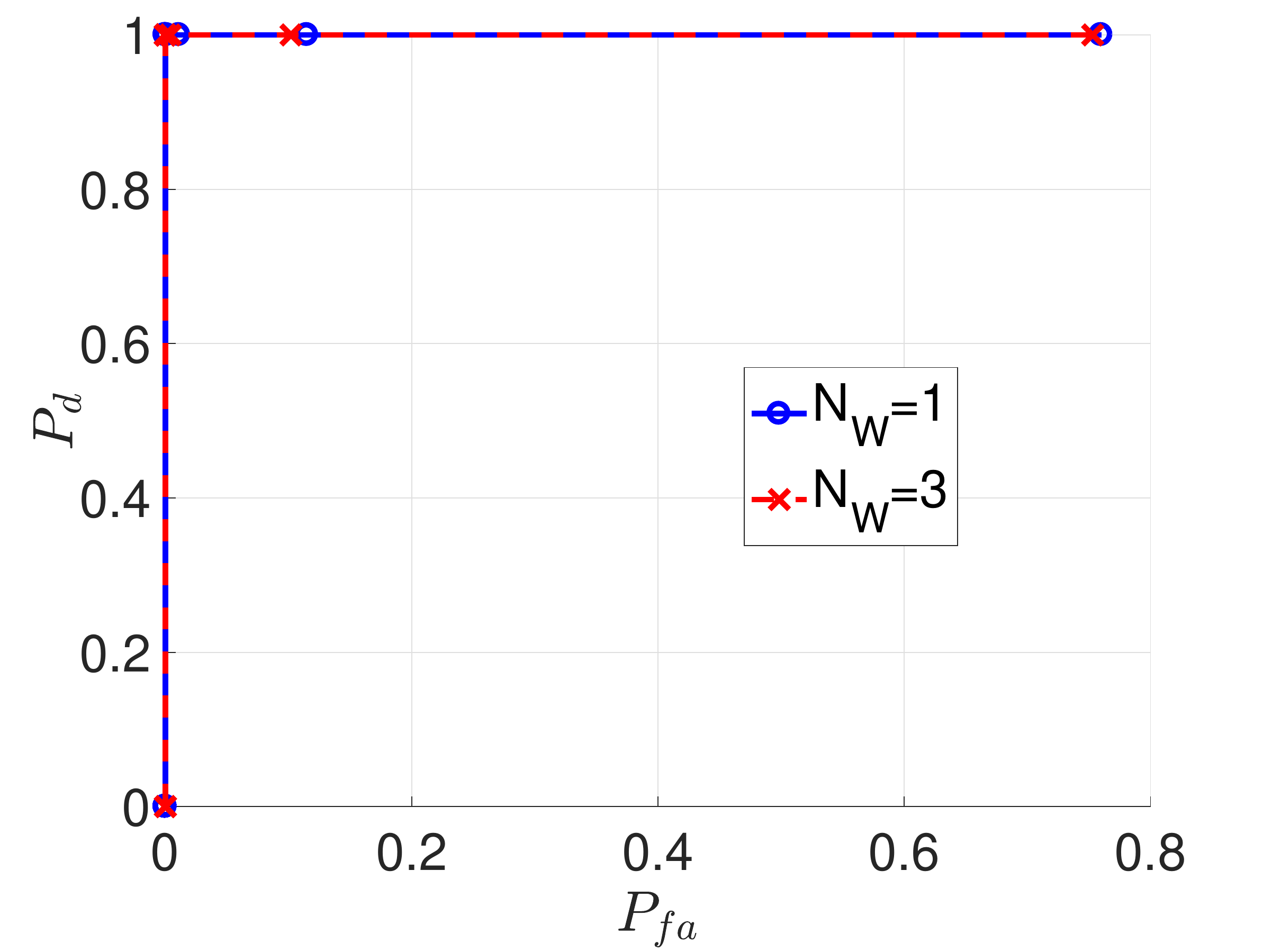}\\
(a) & (b) & (c) 
\end{tabular}  
\end{center} 
\vspace{-0.2in}
\caption{ROC curves: (a) $q_m=0.5q$, and $\alpha=0.5$, (b)  $J=1000$, and $q_m=0.5q$, and (c)  $q=16$,  and defer ($p=1$).} 
\vspace{-0.1in}
\label{fig:eval_det2}
\end{figure*}

\begin{figure}[t]
\begin{center}
 \setlength{\tabcolsep}{+0.015in}
 \begin{tabular}{cc}
\includegraphics[trim=0.3in 0in 0.0in 0in, width=1.82in]{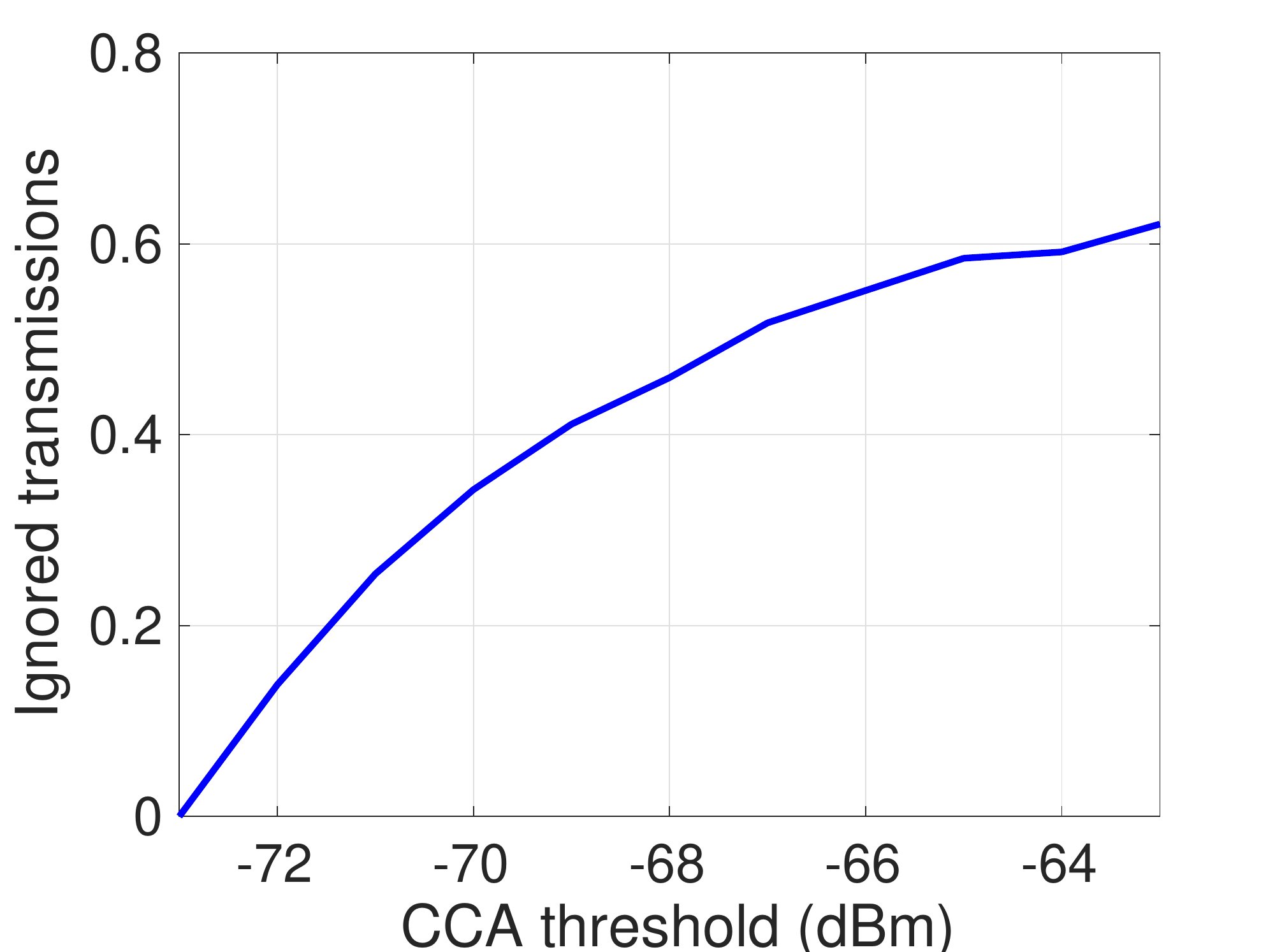} ~& \includegraphics[trim=0.3in 0in 0.0in 0in, width=1.82in]{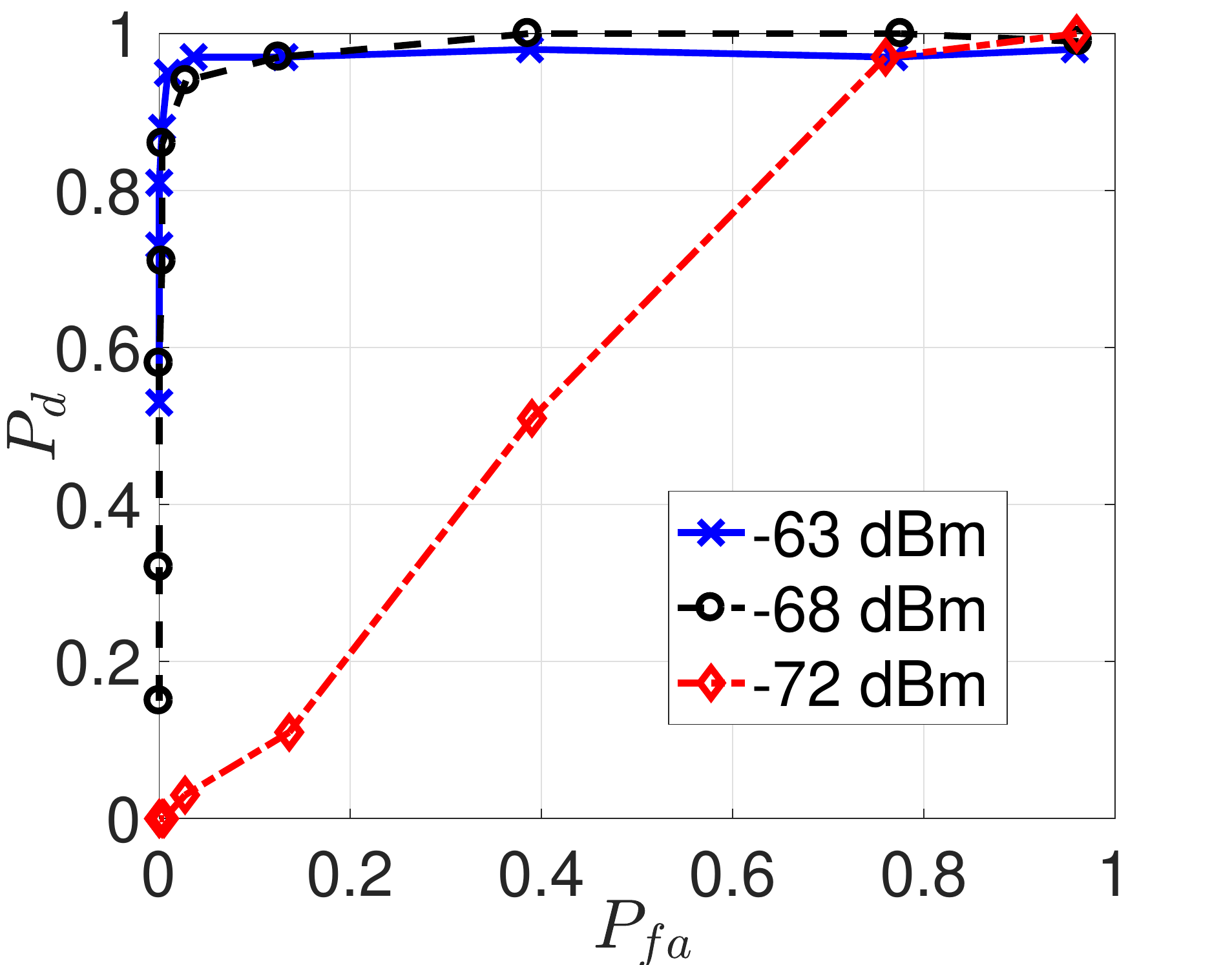} \\
(a) & (b) 
\end{tabular}  
\end{center} 
\vspace{-0.2in}
\caption{(a) Average number of ignored Wi-Fi transmissions per channel access attempt, (b) ROC curve: $q=16$, and $N_W=200$.} 
\label{fig:eval_det_CCA}
\vspace{-0.1in}
\end{figure}

\begin{figure*}[ht]
\begin{center}
 \setlength{\tabcolsep}{-0.025in}
 \begin{tabular}{cccc}
\includegraphics[trim=0.3in 0in 0.0in 0in, width=1.8in]{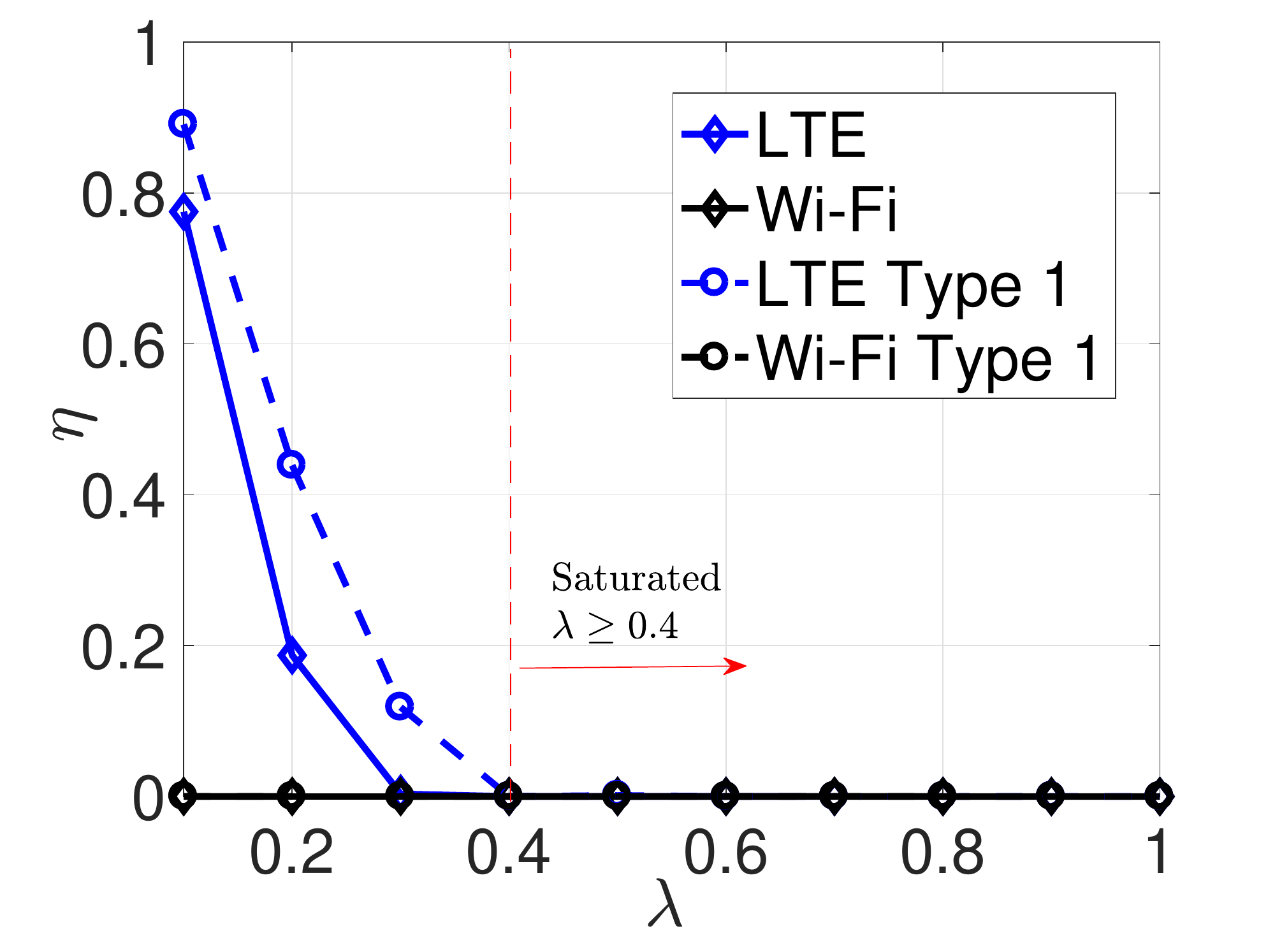}
~&\includegraphics[trim=0.3in 0in 0.0in 0in, width=1.9in]{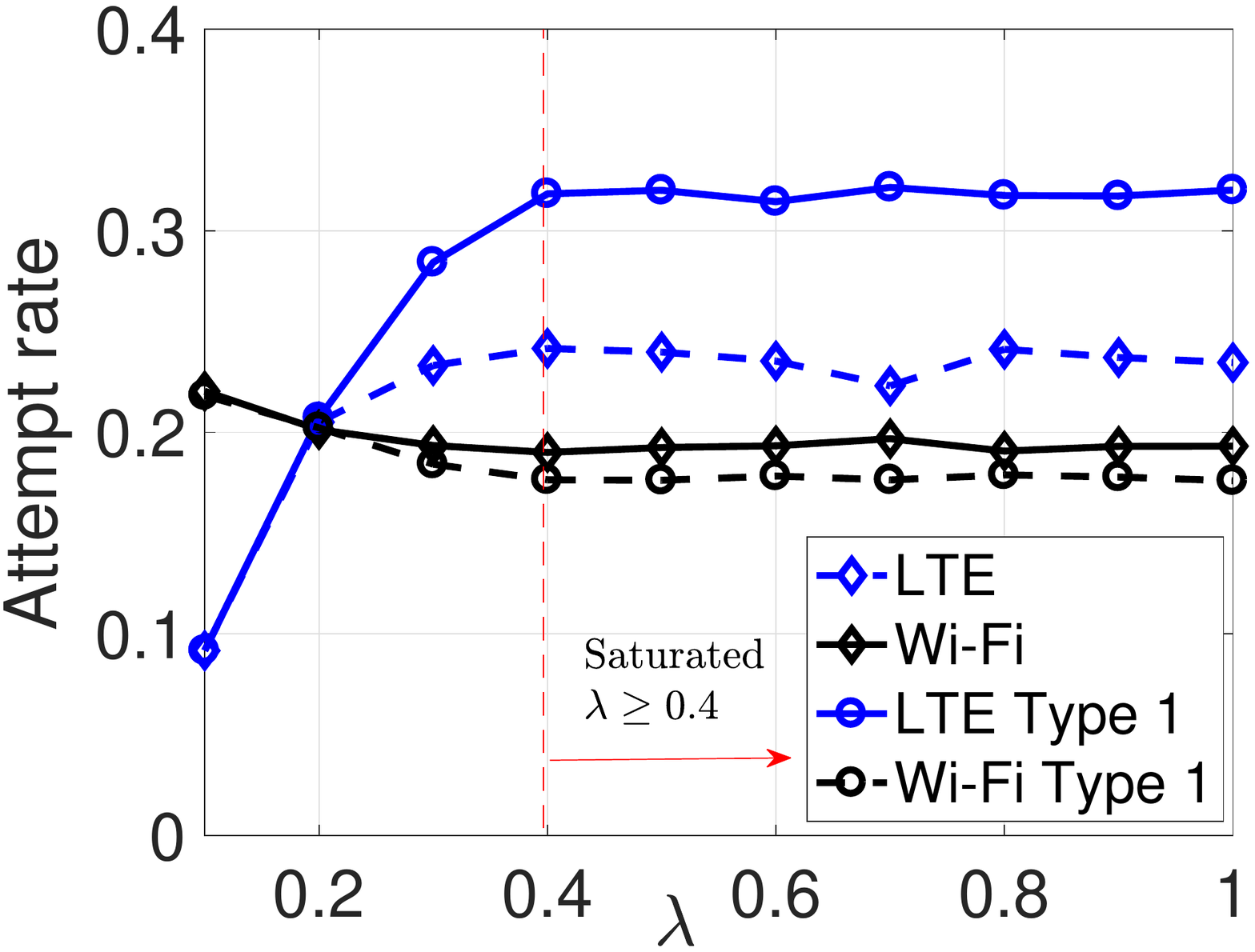}
~& \includegraphics[trim=0.3in 0in 0.0in 0in, width=1.92in]{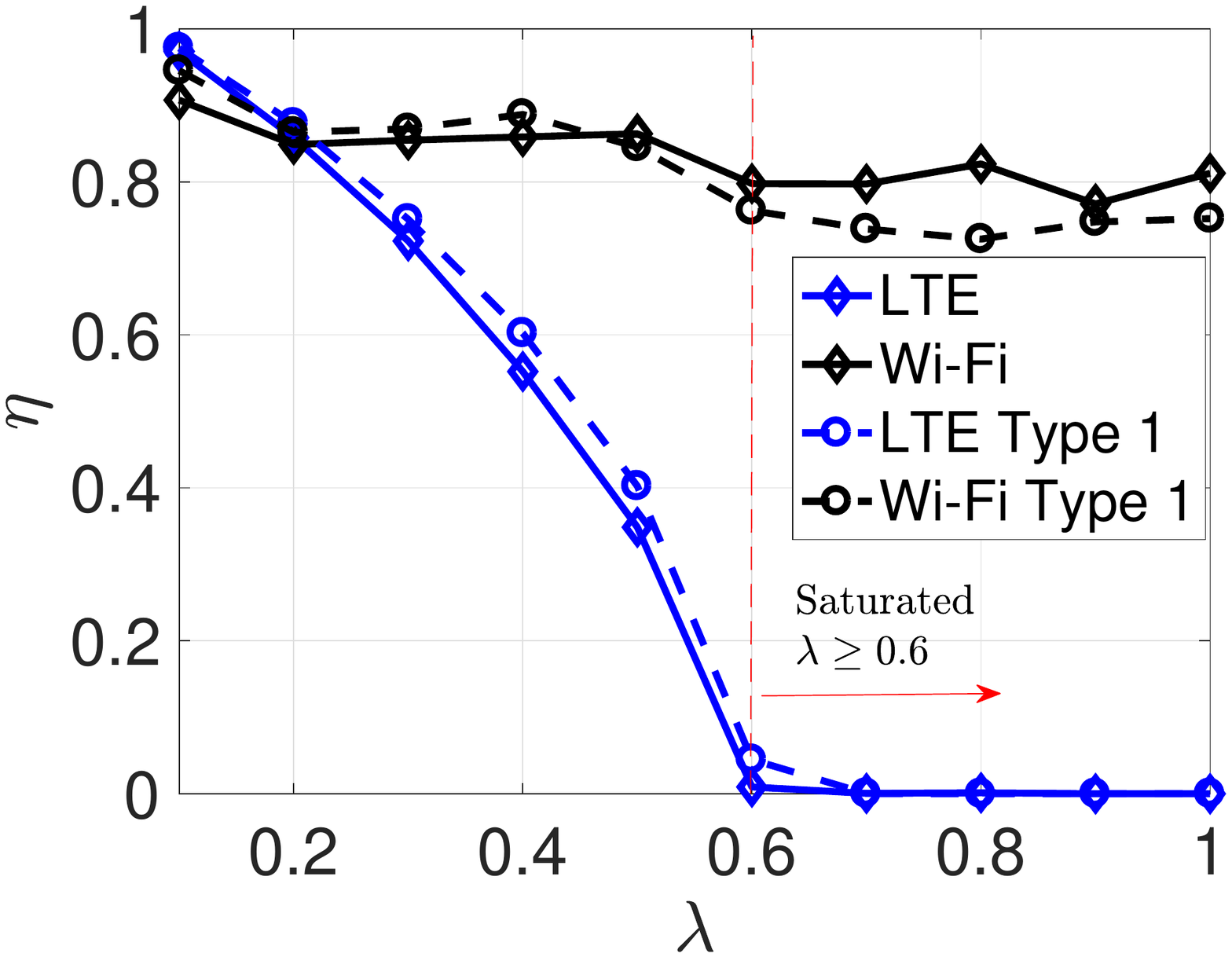} 
~&\includegraphics[trim=0.3in 0in 0.0in 0in, width=1.8in]{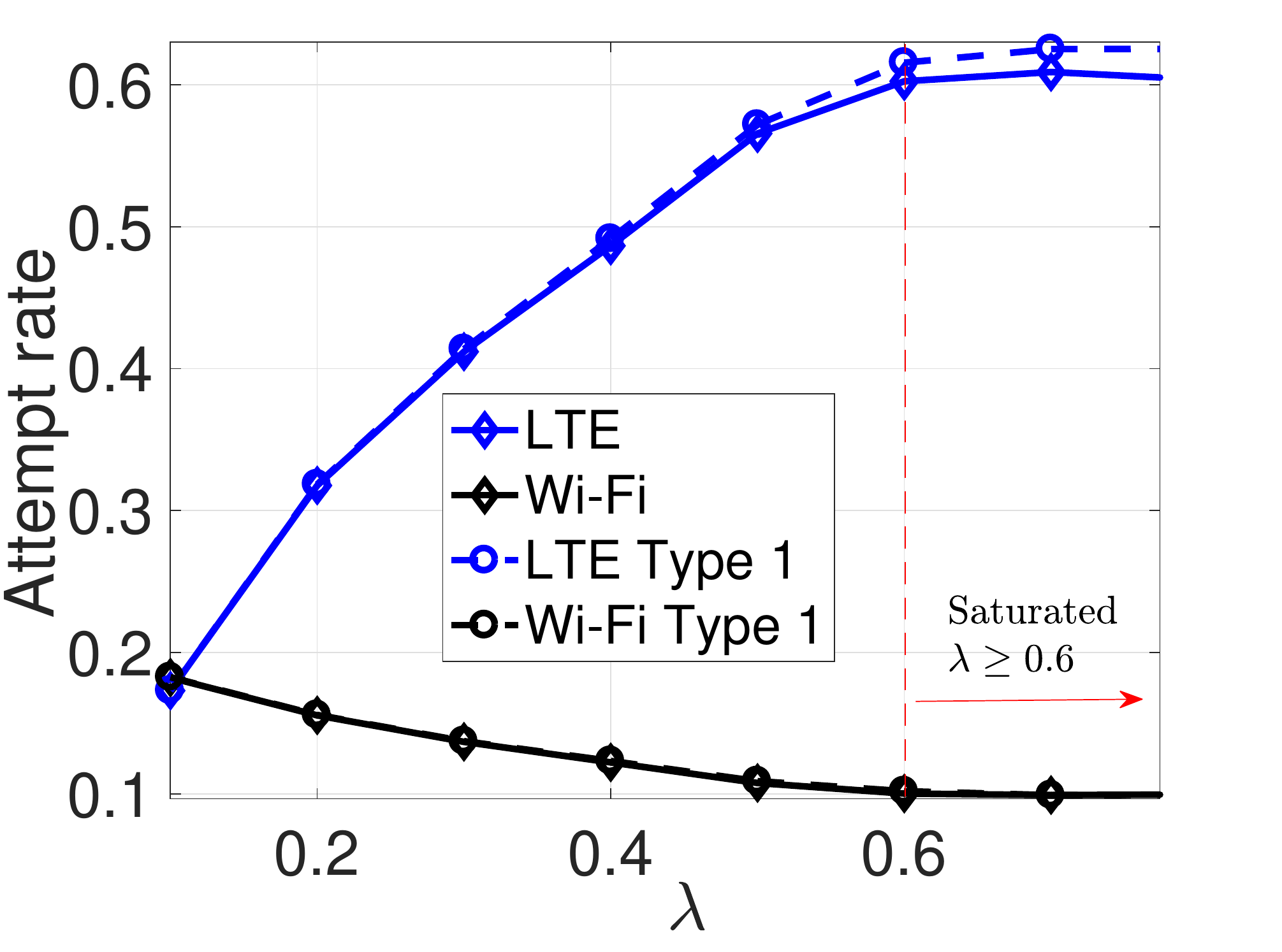} \\
(a) & (b) & (c) & (d) 
\end{tabular}  
\end{center} 
\vspace{-0.1in}
\caption{(a) Attempt rate vs. $\lambda$, and (b)  $\eta$ vs. $\lambda$ when Wi-Fi traffic is saturated. (c) Attempt rate vs. $\lambda$, and (d)  $\eta$ vs. $\lambda$ when Wi-Fi traffic is unsaturated.}
\label{fig:unsat1}
\vspace{-0.1in}
\end{figure*}

\subsection{Effect of LTE Misbehavior on Wi-Fi}

In the first set of experiments, we evaluated the effect of LTE misbehavior on the Wi-Fi channel access opportunities. LTE misbehavior  was implemented by adopting smaller values for the default CW. The LTE chose its backoff uniformly in $\{0,1,\dots,q_m-1\}$, where $q_m$ is the modified CW that is chosen independently of the frame class and transmission round.  
In Fig.~\ref{fig:eval_perf}(a), we show the transmission attempt rate as a function of the normalized reduction in the CW size, denoted by $\nicefrac{q_m}{q}$, where $q$ is the compliant CW ($\nicefrac{q_m}{q}=1$ indicates no misbehavior). We set $\alpha$, the fraction of time that the LTE remains compliant, to 0.5. We considered the coexistence of one LTE with $N_w=1$ and $N_w = 5$ Wi-Fi APs, respectively. 
The Wi-Fi channel access opportunities are shown to degrade when the LTE adopts smaller $q_m$ values whereas the opportunities equalize when $q_m$ approaches $q$.  In addition, the LTE maintains its channel access advantage even when a larger number of Wi-Fi stations compete (note that for $N_W = 5$, the Wi-Fi attempt rate is normalized per AP). The degradation in the Wi-Fi attempt rate  goes up to 50\%. Figure \ref{fig:eval_perf}(b) gives similar intuition when the fraction of time that the LTE misbehaves is varied and $q_m=0.5q$. 

Next, we studied the relation between the number of APs competing with the LTE and the attempt rate. We evaluated the effect of two misbehavior types. In Type 1 misbehavior, the LTE always decreases the CW to $q_m=0.5q$, whereas in Type 2 it used the compliant CW ($q_m = q$), but disregarded the CW exponential growth after collisions.
 We compared the attempt rate under the two misbehavior types with the attempt rate when there is no LTE misbehavior, represented by the labels LTE and Wi-Fi with no type.
 In Fig. \ref{fig:eval_perf}(c), we show the attempt rate as a function of $N_W$. An interesting point here is that the effect of Type 1 misbehavior is more prominent at small $N_W$'s, whereas Type 2 misbehavior has a higher impact at high $N_W$. Overall, Type 1 misbehavior has higher impact than Type 2, as it affects all retransmission rounds. 


In the previous set of experiments,  the LTE and all Wi-Fi APs used the same priority class, i.e., almost similar backoff parameters. In the next set of experiments, we varied the priority class and measured the achieved attempt rate. In Fig.~\ref{fig:eval_perf2}(a), the APs employed a lower priority class that utilizes a smaller CW. We observe that the Wi-Fi performance is almost the same as that of the LTE because reducing the CW for the LTE to $q_m = 0.5q$  equalizes the channel access opportunities for all devices. As expected, the LTE gains are significant when the LTE uses a lower class than Wi-Fi and the LTE also misbehaves. These results are shown in Fig. \ref{fig:eval_perf2}(b) where we see a larger difference in performance relative to Fig. \ref{fig:eval_perf}(c), where the LTE and the APs operate the same class.

\subsection{Receiver Operating Characteristic Curves}

To investigate the efficacy of our misbehavior detection framework, we studied the tradeoff between $P_{fa}$ and $P_d$, for different values of the misbehavior detection threshold $\delta$, through  receiver operating characteristic (ROC) curves.
\subsubsection{Manipulation of the CW $q$}

To measure $P_d$, we implemented a Type 1 misbehavior strategy with $q_m = 0.5q$ when the LTE misbehaved 50\% of the simulation time. To measure $P_{fa}$, we applied our detection framework when the LTE did not misbehave.
Figure~\ref{fig:eval_det2}(a) shows the ROC curve for different lengths of observation set (sizes of set $\Om$) denoted by $J$. 
Indeed, with the increase in the length of the observation set, the ROC approaches the optimal curve indicating that our system can operate with almost sure detection and  almost zero false alarm probability.  In Fig. \ref{fig:eval_det2}(b), we see the effect of the fraction of time the LTE misbehaves. Even at low levels of misbehavior (e.g. $\alpha=0.9$), the misbehavior is detectable. 

\subsubsection{Manipulation of the  defer time $p$} 

We further evaluated the performance of the proposed detection framework when the LTE manipulates the defer time $p$. To simulate this misbehavior, we implemented an LTE that uses the defer time from traffic class $C_1$ (i.e., $p=1$) while transmitting frames that belong to class $C_3$ ($p=3$). Our simulations in Fig. \ref{fig:eval_det2}(c) show
an almost perfect ROC curve for any non-zero false alarm probability, when $N_w = 1$ and $N_w =3$.
The results are justified by the fact that the consistent selection of a smaller defer time skews the estimated distribution of backoff values in a detectable manner. This is a detectable phenomenon for any  $\delta$ that fixes the false alarm probability to a given value.


\subsubsection{CCA threshold manipulation} 

 We further performed another set of experiments to evaluate the manipulation of the CCA threshold. A selection of a  higher CCA threshold increases the number of APs that are ignored by the LTE. To simulate the CCA threshold manipulation scenario, we uniformly deployed multiple APs and one LTE in a square area of $200\times 200$ meters. We set the transmission power of each Wi-Fi AP to 20dBm and modeled the channels between terminals using the free path-loss model (the channel model is not really important here). We set the  carrier frequency to 5 GHz.

We evaluated the performance of our detector when the CCA threshold is set to -63, -68, and -72dBm (the LTE-LAA standard sets the CCA threshold to -73 dBm). To highlight the effect of CCA threshold manipulation, we implemented  a deployment of $N_W=200$ APs. 
Here, we increased the number of APs to ensure that we have a non-negligible number of ignored Wi-Fi transmissions. This is shown in Fig. \ref{fig:eval_det_CCA}(a) where we plot the average number of ignored Wi-Fi transmissions, normalized over the LTE channel access attempts. For instance, when the CCA threshold is set to -72 dBm, the LTE ignores on average one Wi-Fi transmission every ten channel access attempts. On the other hand, the LTE ignores on average one Wi-Fi transmission every other channel access attempt when the threshold is set to -68 dBm. The high number of APs is a relevant scenario in urban areas where there are dense deployments of APs.

In Fig. \ref{fig:eval_det_CCA}(b), we show the ROC for the three CCA thresholds. We observe that when the CCA is lowered by more than 5dBm, the ROC approaches the optimal one. However, our framework does not detect accurately a small change in the CCA. Such misbehavior creates an imperceptible advantage for the LTE in terms of channel access opportunities as the LTE only ignores, on average, one Wi-Fi transmission each ten channel access attempts as shown in Fig. \ref{fig:eval_det_CCA}(a). 

\begin{figure}[t]
\vspace{-1in}
\begin{center}
 \setlength{\tabcolsep}{-0.037in}
 \begin{tabular}{cc}
\includegraphics[trim=0.3in 0in 0.0in 0in, width=1.82in]{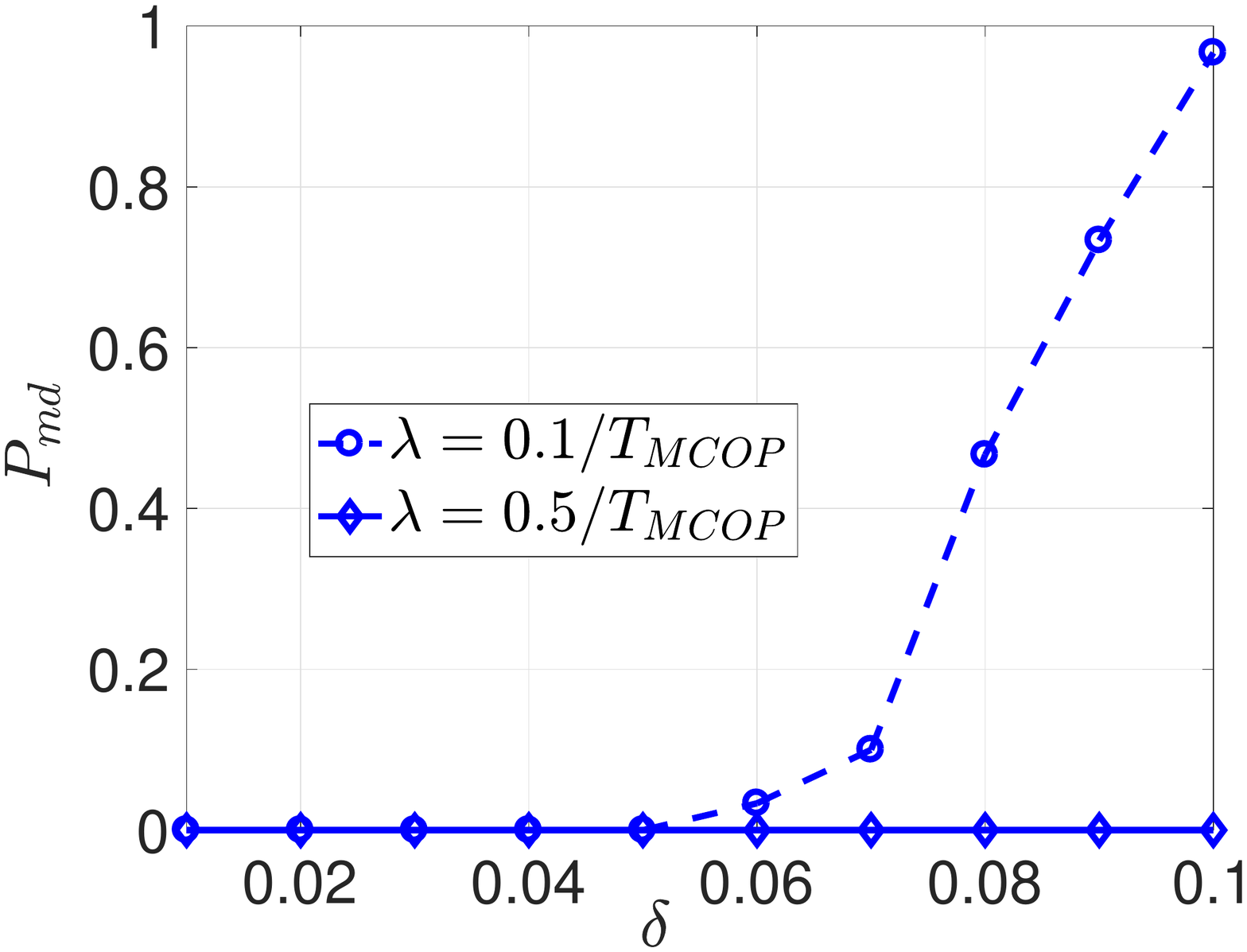} ~&\includegraphics[trim=0.3in 0in 0.0in 0in, width=1.82in]{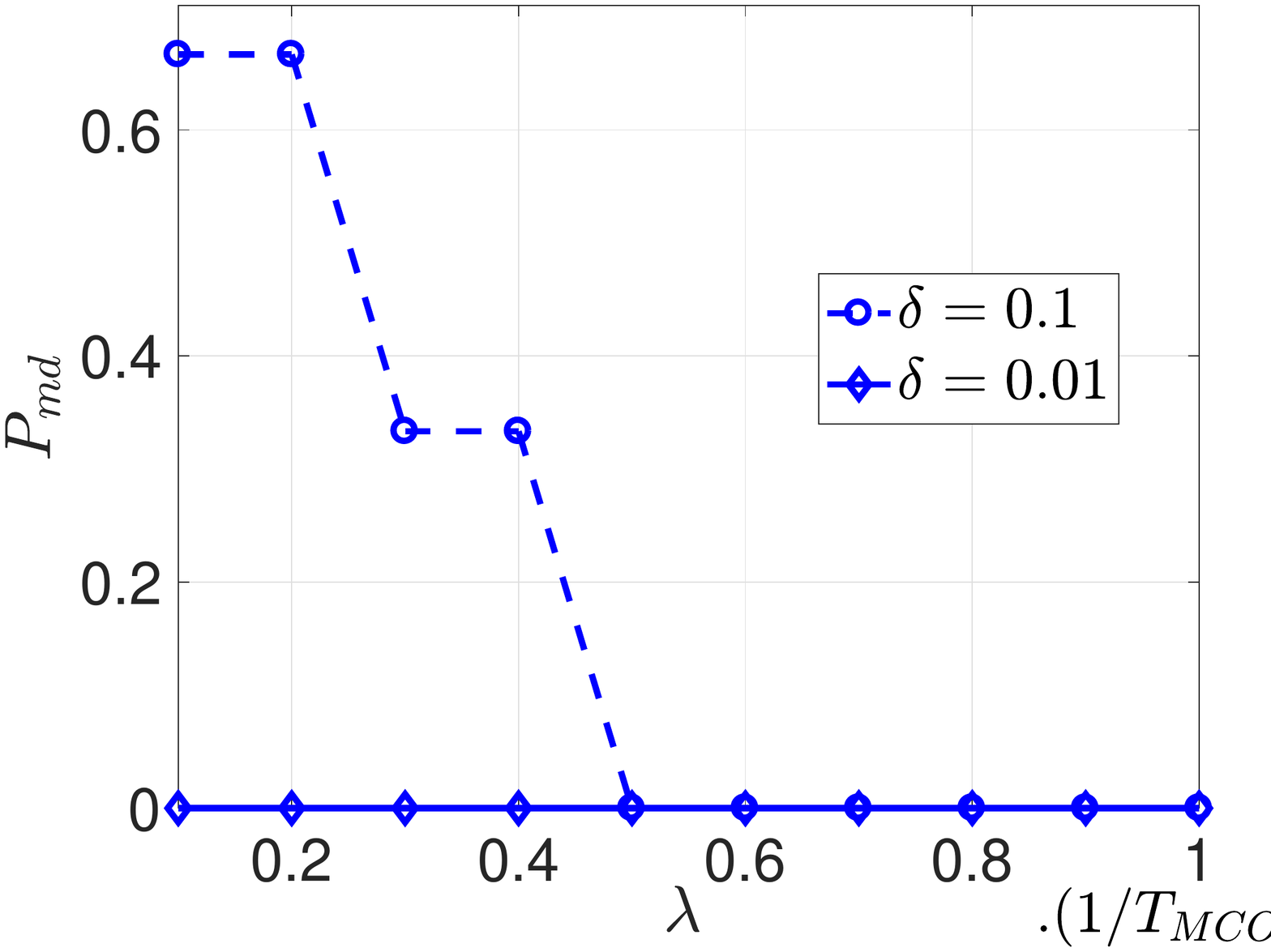} \\
(a) & (b)  \\
\end{tabular}  
\end{center} 
\vspace{-0.15in}
\caption{Misdetection probability ($P_{md}$) vs. : (a)  Threshold $\delta$, and (b) LTE arrival rate $\lambda$. }
\label{fig:unsat2}
\vspace{-0.15in}
\end{figure}

\subsection{Unsaturated traffic}
In the last set of experiments, we studied misbehavior under unsaturated traffic conditions. We implemented a Poisson frame arrival process with an average rate $\lambda$ for each device. We measured the  saturation level by the percentage of time a device's queue stays empty and denoted this parameter by  $\eta$. A device is saturated if $\eta=0$, i.e., it always has a frame to transmit. 

In Fig. \ref{fig:unsat1}, we show the effect of the arrival rate on the attempt rate for different levels of saturation. We implemented a Type 1 misbehavior strategy where the eNB transmits class 3 frames but reduces the contention window to $q_m = 0.5q.$ The eNB misbehaved half the time $(\alpha=0.5)$.  Figure \ref{fig:unsat1}(a) shows 
the saturation levels of both LTE and Wi-Fi with and without LTE misbehavior when the Wi-Fi traffic is always saturated and one eNB competes with five APs. Figure \ref{fig:unsat1}(b) shows the respective attempt rates. Here, the arrival rate is normalized by $\nicefrac{1}{T_{\mbox{MCOP}}}$ which is the maximum service rate (transmissions per second) that meets the medium capacity.
In Fig. \ref{fig:unsat1}(a), we note that $\eta$ always equals zero for the Wi-Fi APs, as they are backlogged by design. For the LTE station, we observe that when the LTE misbehaves, saturation occurs at a higher arrival rate indicating that the LTE gains an advantage in accessing the channel sooner. 

Figure \ref{fig:unsat1}(b) shows that  when the arrival rate is low, misbehavior has no effect on the attempt rate of the Wi-Fi. However, when $\lambda$ increases and the devices approach saturation, the gap between the attempt rate, with and without misbehavior, increases. 
As expected, the attempt rate gap remains constant after saturation is reached (and is consistent with the results shown in Fig. \ref{fig:eval_perf2}). 

 Figures \ref{fig:unsat1}(c) and \ref{fig:unsat1}(d) show the same experiments, but when the APs remain unsaturated while the arrival rate for the LTE increases. The unsaturated condition for the Wi-Fi is also evident in Fig. \ref{fig:unsat1}(c), where the Wi-Fi queue is empty over 70\% of the time for any $\lambda.$ From Fig.~\ref{fig:unsat1}(c), we further observe that the gain from misbehavior is practically diminished. The LTE saturates almost at the same rate $\lambda = 0.6.$ Further, we observe in Fig. \ref{fig:unsat1}(d) that the LTE misbehavior increases the LTE attempt rate in an imperceptible manner.  This makes misbehavior detection less necessary compared to saturated conditions.

Finally, we evaluated the misdetection probability $P_{md}$ under unsaturated traffic conditions.  In Fig. \ref{fig:unsat2}(a), we show $P_{md}$ as a function of  the threshold $\delta$, for Type 1 misbehavior with $q_m=0.5q$ and $\alpha=0.5$. As expected for $\lambda =\frac{0.5}{T_{\mbox{MCOP}}}$ (i.e., under saturated conditions as seen in Fig. \ref{fig:unsat1} (a)), we have almost perfect detection. Under unsaturated conditions ($\lambda =\frac{0.1}{T_{\mbox{MCOP}}}$),  the method of excluding observations that yield idle times larger than expected during a backoff process  enables us to have a reasonable $P_{md}$ with careful selection of $\delta$. Misdetection becomes very small for $\delta\leq 0.05$, however, for this range, the AP is required to collect a large number of observations to avoid false alarms. Generally, this range of $\delta$ is only required whenever the LTE is found operating under unsaturated traffic conditions. Figure \ref{fig:unsat2}(b) shows $P_{md}$ as a function of the arrival rate $\lambda$ at the LTE, for Type 1 misbehavior. As expected,  $P_{md}$ approaches zero as we approach saturation. 

\section{Conclusion}
\label{Conc}
We studied the problem of LTE misbehavior under the LTE-LAA protocol for  coexistent LTE and Wi-Fi systems. We outlined several misbehavior scenarios and  developed a suite of implicit monitoring techniques that enable the Wi-Fi system to estimate the operational parameters of the LTE, without decoding LTE signals. This is a desired property as Wi-Fi devices are not necessarily equipped with LTE receivers.  Our methods rely on operations in the signal domain to identify and classify LTE transmissions. We evaluated these techniques using an experimental setup and verified their efficiency in practical scenarios. 

We further developed a behavior evaluation framework in which a central hub collects all observations from a distributed set of monitoring APs to build a behavior profile for the eNBs and detect misbehavior. We extended our detection method to work reliably for both unsaturated and saturated traffic. We evaluated the performance of our detector via simulations and showed that LTE misbehavior can cause a significant performance degradation for Wi-Fi devices. However, such misbehavior was detectable by our framework with very high probability while achieving a low false alarm probability. Although our framework focuses on the coexistence between LTE and Wi-Fi systems, our ideas can be extended to other coexistence scenarios.

\bibliographystyle{IEEEtran} 
\bibliography{IEEEabrv,./myref,main}

\begin{IEEEbiography}[{\includegraphics[width=1in,height=1.25in]{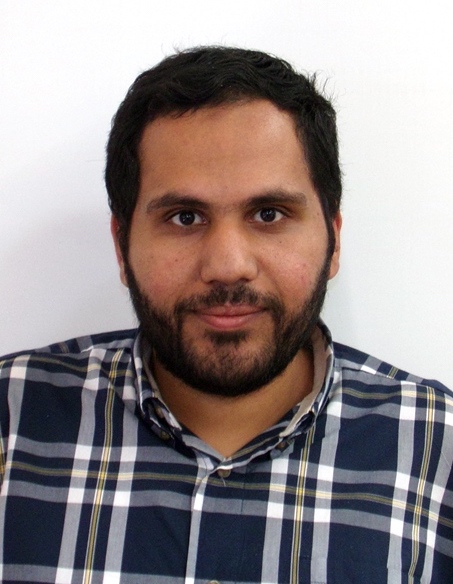}}]{Islam Samy received the B.S. degree in electrical
engineering from Alexandria University in 2011 and
the M.Sc. degree in wireless communication from Nile University in 2014. He is currently pursuing
the Ph.D. degree with the Electrical and Computer Engineering Department, The University of Arizona, where he is also a Graduate Research Assistant. His main research interests include secure and fair resource allocation for heterogeneous coexisting systems, wireless communication, and information theory.
}
\end{IEEEbiography}

\begin{IEEEbiography}[{\includegraphics[width=1in,height=1.25in]{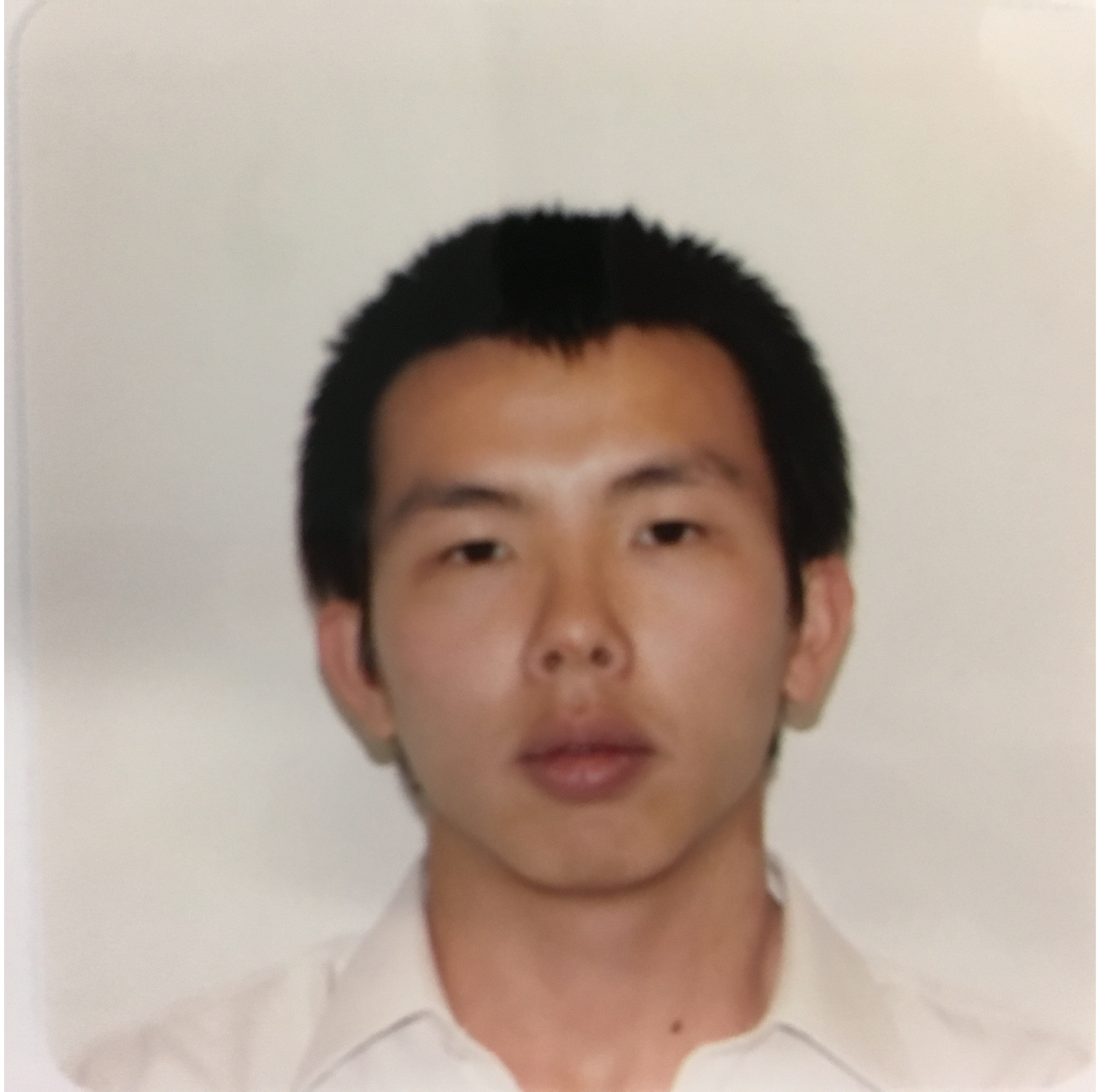}}]{Xiao Han received the BE degree in electrical engineering from Northwest A\&F University, Xianyang, China, in 2016, and MS degree in electrical and computer engineering from University of Arizona, Tucson, United States, in 2019. He is currently working toward the PhD degree in computer science at University of South Florida, Tampa, United States. His research interests include wireless networks and network security.
}
\end{IEEEbiography}

\begin{IEEEbiography}[{\includegraphics[width=1in,height=1.25in]{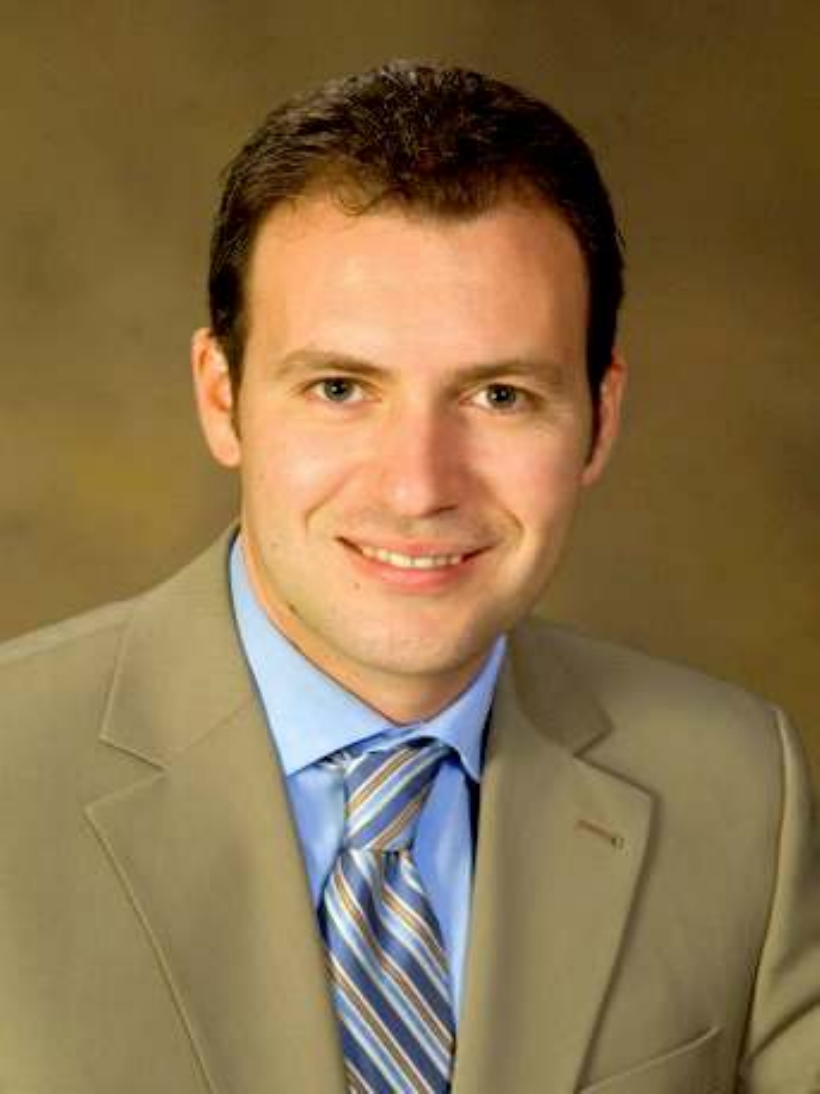}}]{Loukas Lazos is a Professor of Electrical and Computer Engineering at the University of Arizona. Dr. Lazos received his Ph.D. degree in Electrical Engineering from the University of Washington in 2006. His research interests are in the areas of network security, privacy, and wireless communications. His current research focuses on secret-free authentication methods for IoT devices, integrity verification of physical properties, secure and fair channel access for emerging wireless technologies, and fair resource allocation for heterogeneous coexisting technologies, and private information retrieval. Dr. Lazos is the recipient of the US National Science Foundation (NSF) Faculty Early CAREER Development Award (2009) for his research in security of multi-channel wireless networks. He has served as a technical program chair for the IEEE CNS conference, the IEEE GLOBECOM symposium on communications and information systems security and the IEEE DSPAN workshop. He is an associate editor for the IEEE Transactions on Information and Forensics Security journal and the IEEE Transactions on Mobile Computing journal. He has also served and continues to serve on the organization and technical program committees of many international conferences and on expert panels of several government agencies.}
\end{IEEEbiography}

\begin{IEEEbiography}[{\includegraphics[width=1in,height=1.25in]{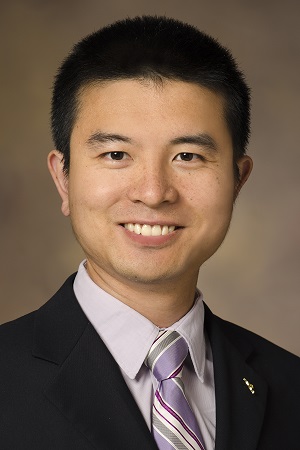}}]{Ming Li is an Associate Professor in the Department of Electrical and Computer Engineering of University of Arizona, and also affiliated with the Computer Science Department. He was an Assistant Professor in the Computer Science Department at Utah State University from 2011 to 2015. He received his Ph.D. in ECE from Worcester Polytechnic Institute, MA, in 2011. His main research interests are wireless and cyber security, with current emphases on cross-layer optimization and machine learning in wireless networks, wireless physical layer security, privacy enhancing technologies, and cyber-physical system security. He received the NSF Early Faculty Development (CAREER) Award in 2014, and the ONR Young Investigator Program (YIP) Award in 2016. He is a senior member of IEEE, and a member of ACM.}
\end{IEEEbiography}

\begin{IEEEbiography}[{\includegraphics[width=1in,height=1.25in]{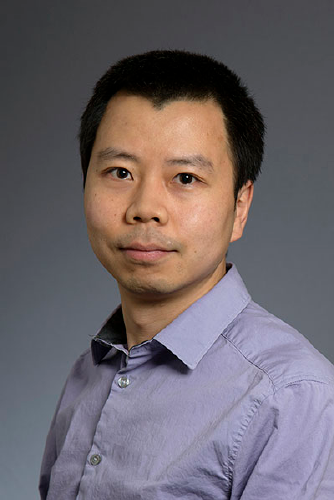}}]{{Yong Xiao}(S'09-M'13-SM'15) is a professor in the School of Electronic Information and Communications at the Huazhong University of Science and Technology (HUST), Wuhan, China. He is also the associate group leader in the network intelligence group of IMT-2030 (6G promoting group) and the vice director of 5G Verticals Innovation Laboratory at HUST. His research interests include machine learning, game theory, and their applications in cloud/fog/mobile edge computing, green communication systems, wireless networks, and Internet-of-Things (IoT).}
\end{IEEEbiography}

\begin{IEEEbiography}[{\includegraphics[width=1in,height=1.25in]{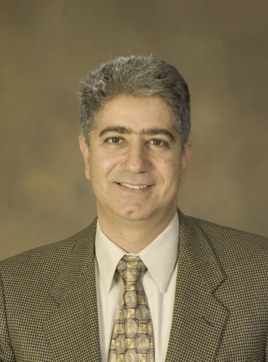}}]{
Marwan Krunz is the Kenneth VonBehren Endowed Professor in ECE and a professor of computer science at the University of Arizona. He directs the Broadband Wireless Access and Applications Center (BWAC), a multi-university NSF/industry center that focuses on next-generation wireless technologies and applications. He also holds a courtesy appointment as a professor at the University Technology Sydney. He previously served as the site director for  the Connection One center. Dr. Krunz’s research emphasis is on resource management, network protocols, and security for wireless systems. He has published more than 300 journal articles and peer-reviewed conference papers, and is an inventor on 12 patents. His latest h-index is 60. He is an IEEE Fellow, an Arizona Engineering Faculty Fellow, and an IEEE Communications Society Distinguished Lecturer (2013-2015). He received the NSF CAREER award. He recently served as the Editor-in-Chief for the IEEE Transactions on Mobile Computing. He also served as editor for numerous IEEE journals. He was TPC chair INFOCOM’04, SECON’05, WoWMoM’06, and Hot Interconnects 9. He was the general vice-chair for WiOpt 2016 and general co-chair for WiSec’12. Dr. Krunz is also an enterpreneur, serving as chief technologist/scientist for two startup companies that focus on 5G and beyond systems and signal intelligence. 
}
\end{IEEEbiography}

\end{document}